\documentclass[fleqn,usenatbib]{mnras}

\usepackage{newtxtext,newtxmath}
\usepackage{multirow}

\usepackage[T1]{fontenc}

\DeclareRobustCommand{\VAN}[3]{#2}
\let\VANthebibliography\thebibliography
\def\thebibliography{\DeclareRobustCommand{\VAN}[3]{##3}\VANthebibliography}

\usepackage{graphicx}	
\usepackage{amsmath}	
\usepackage{xcolor}

\newcommand{\ve}[1]{\boldsymbol{#1}}
\newcommand{\uve}[1]{\boldsymbol{\hat{#1}}}

\newcommand{\grad}{\boldsymbol{\nabla}}

\title[Disk formation with NMHD and jet feedback]{Non-ideal MHD and protostellar feedback effects on disk formation and evolution in numerical simulations of star cluster formation}

\author[N. Filippova et al.]{
Nina Filippova$^{1}$\thanks{E-mail: ninaf@utexas.edu (NF)},
Stella S. R. Offner$^{1}$, 
Michael Y. Grudi\'{c}$^{2}$, and
Philip F. Hopkins$^{3}$
\\
$^{1}$ Department of Astronomy, University of Texas at Austin, TX 78712, USA \\
$^{2}$ Center for Computational Astrophysics, Flatiron Institute, 162 5th Ave, New York, NY 10010, USA \\
$^{3}$ TAPIR, Mailcode 350-17, California Institute of Technology, Pasadena, CA 91125, USA
}

\date{Accepted XXX. Received YYY; in original form ZZZ}

\pubyear{2026}

\begin{document}
\label{firstpage}
\pagerange{\pageref{firstpage}--\pageref{lastpage}}
\maketitle

\begin{abstract}
While recent surveys have resolved hundreds of nearby protostellar disks, numerical simulations assuming ideal magnetohydrodynamics (MHD) have historically struggled to achieve disk formation due to efficient angular momentum removal by magnetic torques. Non-ideal MHD effects, relevant at the low ionization fractions typical of molecular clouds, have been shown to reduce the effectiveness of magnetic braking and promote disk formation. In this work, we present the results from a suite of calculations following the gravitational collapse of 50 $M_{\odot}$ turbulent molecular cloud cores down to the formation and evolution of stellar systems and protostellar disks. We use the radiation-MHD code \textsc{GIZMO} including non-ideal MHD (Ohmic resistivity, ambipolar diffusion, and the Hall effect) and the \textsc{STARFORGE} numerical framework for modeling star formation and stellar feedback. We compare the effects of assuming ideal vs. non-ideal MHD and including sub-grid protostellar jet feedback on disk formation and evolution. Disks form in all of our models but are least massive in the model with ideal MHD and sub-grid jet feedback. Apart from the ideal MHD$+$jets model, we do not observe any significant differences in disk properties between the ideal and non-ideal MHD models; however, ideal MHD disks are embedded in smaller rotating envelopes. Disk sizes are in general agreement with those of observed disks. Jet feedback increases core fragmentation and reduces final stellar masses. Our results suggest that magnetic braking does not efficiently suppress disk formation, regardless of whether ideal or non-ideal MHD is assumed, under the dynamical conditions in which multiple stellar systems form.
\end{abstract}




\section{Introduction}

Stars form during the gravitational collapse of dense clouds of gas and dust, a process that is inhibited by the turbulent, magnetic, and thermal support within the cloud \citep{mckee_ostriker_2007}. Angular momentum conservation during the collapse produces a rotationally-supported disk around the growing protostar, which funnels mass to the protostar as well as provides raw material for planet formation \citep{zhao_2020_review}. Recent surveys by the Very Large Array (VLA) and the Atacama Large Millimeter/submillimeter Array (ALMA) have resolved disks around hundreds of protostars, suggesting that their formation is a regular occurrence during the collapse process \citep{tobin_2020}. Protostellar disk formation is observed in numerical calculations as well. \cite{bate_2018} presented the first population synthesis study of protostellar disks formed in a radiation hydrodynamical simulation of star cluster formation, reporting the statistical properties (such as typical disk sizes) of a diverse population of $>$100 disks formed over $\sim$$10^5$ years of protostellar evolution in a 500 $M_{\odot}$ cloud. However, numerical simulations have historically struggled to achieve disk formation once even relatively modest magnetic fields are included \citep[see, e.g.,][]{joos_2012, santos_2012}, largely due to the efficient removal of angular momentum from the disk that occurs under the assumptions of ideal magnetohydrodynamics (MHD). As observations suggest that molecular clouds are threaded by large-scale magnetic fields on the order of $\sim$10$-$100 $\mu$G \citep{crutcher_2012}, this failure to form disks in magnetized models points to a significant challenge in disk formation theory.

It is clear that angular momentum cannot be perfectly conserved during the collapse. Observations of star-forming clouds measure a typical specific angular momentum of $j \sim 10^{21}$ cm$^2$ s$^{-1}$ \citep[e.g.,][]{goodman_1993}, while stellar cores have a typical specific angular momentum of $j \sim 10^{15}$ cm$^2$ s$^{-1}$ \citep[e.g.,][]{belloche_2013}, a difference of six orders of magnitude that has been referred to as the ``angular momentum problem.” Angular momentum may be removed from the core by magnetic (as well as gravitational) torques \citep[see, e.g., the review by][]{wurster_li_2018}. The presence of charged species $-$ ions, electrons, and charged dust grains $-$ in the star-forming gas causes the fluid motion to become coupled to the magnetic field dynamics. The rotation of the collapsing material drags the magnetic field lines along the direction of rotation, creating a magnetic tension force, and thereby a torque, which counteracts the overall rotation. In the limit of ideal MHD, in which the relative drift between different charged species is neglected, the perfect coupling between the fluid and the field means that even moderate magnetic field strengths can lead to very efficient angular momentum removal and the so-called ``magnetic braking catastrophe" \citep[see, e.g.,][and sources within]{zhao_2020_review, maury_2022}, suppressing disk formation entirely.

Theoretical and numerical studies aimed at resolving the magnetic braking catastrophe have considered adjustments such as introducing misalignment between the magnetic field and the axis of rotation \citep[e.g.,][]{gray_2018, hirano_2020}, adding turbulence \citep[e.g.,][]{santos_2012, joos_2013, lewis_bate_2018}, or including non-ideal MHD effects \citep[e.g.,][]{wurster_2019_low_mass_cluster, zhao_2020_hall, zhao_2021}. Molecular clouds are only partially ionized \citep[with an ionization fraction $x_e = n_e/n_{\rm H}$ ranging from $10^{-4}$ at the cloud boundary down to $10^{-9}$ in the densest cores; see, e.g.,][]{goicoechea_2009}, and so non-ideal MHD effects, such as Ohmic resistivity (electron-ion/neutral drift), ambipolar diffusion (ion-neutral drift), and the Hall effect (ion-electron drift) become increasingly relevant. These effects relax the fluid-field coupling: Ohmic resistivity and ambipolar diffusion act as dissipative terms which weaken the magnetic field, while the Hall effect serves as a dispersive term which modifies the magnetic field line geometry. The influence of each term depends on the local density and temperature, as well as microphysics such as the dust grain size distribution and local cosmic ray ionization rate \citep{zhao_2020_review, zhao_2021, wurster_2021}. The Hall effect has been the most challenging to implement numerically, and is the most recent to be included in such studies.

The classical approach to modeling star formation, pioneered by \cite{larson_1969}, begins with the assumption of an isolated spherical core which collapses due to its own gravity. Early numerical studies of protostellar disk formation adopted this model, typically beginning with a $\sim$1 $M_{\odot}$ uniform-density sphere and investigating the effects of including additional physics, such as rotation, turbulence, magnetic fields, and non-ideal MHD, on disk formation outcomes \citep[see the review by][]{kuffmeier_2024_review}. These simplified models, while suitable for studying the effects of various parameters on core and disk evolution in a well-defined setup, fail to capture the dynamical conditions in which stars form. Observations suggest that the morphology of prestellar cores embedded in dense filaments in Giant Molecular Clouds (GMCs) deviate strongly from spherical symmetry \citep[][]{andre_2014, kainulainen_2017, pineda_2023, hacar_2023}. Additionally, there is increasing evidence for large-scale asymmetric features, or ``streamers" \citep[see the review by][]{pineda_2023}, as well as late-stage infall \citep[e.g.,][]{ginski_2021}, which continue to contribute mass even after the initial envelope mass reservoir has been depleted. And while our familiar solar system contains just one star, observations suggest that this is the exception, rather than the rule. Most stars form in systems consisting of multiple gravitationally-bound stars at sub-parsec separations, with the multiplicity fraction monotonically increasing with the mass of the primary star \citep{offner_2022}. Furthermore, most multiple systems form and evolve into their final configuration by the end of the mass accretion stage \citep{offner_2022}. Several mechanisms have been proposed to explain the formation of multiple stellar systems, such as the migration of existing protostars \citep[e.g.,][]{ostriker_1994, moeckel_bally_2007, bate_2012}, the fragmentation of cloud substructures such as cores and filaments \citep[e.g.,][]{larson_1972, tomisaka_2014, gus_hopkins_2015, gus_2017}, and the fragmentation of the protostellar disk itself \citep[e.g.,][]{adams_1989, shu_1990, bonnell_1994}. Interactions between young stars are common during the first Myr of cluster evolution \citep{pfalzner_2013}. So-called stellar ``flybys" have been shown to truncate disk sizes and induce misalignment between the disk and protostar rotation axes. Once stars form, stellar feedback in the form of radiation, jets, winds, and supernovae affects subsequent star formation and cloud evolution \citep[e.g.,][]{kim_2021, grudic_2022_starforge_feedback}. 

A proper treatment of star (and disk) formation including stellar feedback and dynamical interactions in a turbulent, magnetized plasma must consider complex, non-linear physical processes spanning large dynamical ranges. Considerable improvements in both physical realism as well as computational efficiency over the last decade have made it feasible to attempt to study protostellar disk formation, not only in small ($\sim$1 $M_{\odot}$), rotating, single-protostar/disk models, but also in turbulent, magnetized cloud cores ($\sim$10$-$100 $M_{\odot}$), which form stellar clusters and include various feedback mechanisms. In this paper, we present results from radiation ideal/non-ideal MHD simulations of low-mass cluster formation, using the TURBSPHERE initialization setup of \cite{lane_2022} to generate more realistic turbulent initial conditions from idealized 50 $M_{\odot}$ spherical cores. We use the \textsc{GIZMO} code, which includes non-ideal MHD effects via a modified induction equation valid in the limit of low ionization, to compare disk formation outcomes between models with ideal and non-ideal MHD. We also consider the effects of including sub-grid protostellar jet feedback. While self-consistently launched jets are not observed in most of our models, jet feedback has been shown to lead to an overall reduction in protostellar masses, as well as inject significant energy and momentum into the local star-forming environment \citep[e.g.,][]{gus_2021}. It is very probable that jet feedback affects disk formation and evolution as well.

\section{Methods}

We model the gravitational collapse of a $M_0 = 50 \; M_{\odot}$ turbulent molecular cloud core (hereafter referred to as a cloud for notational convenience) down to the formation and evolution of a star cluster using the 3D radiation MHD code \textsc{GIZMO}, supplemented with modules for non-ideal MHD effects such as Ohmic resistivity, ambipolar diffusion and the Hall effect, as well as additional sink particle formation and feedback prescriptions developed within the \textsc{STARFORGE} numerical framework \citep{hopkins_2015_gizmo, hopkins_raives_2016_gizmo_mhd, hopkins_2017_gizmo_nonideal_mhd_diffusion, grudic_2021_starforge}. We use $N_{\rm gas} = 5 \times 10^6$ equal-mass ($\Delta m = 10^{-5} \; M_{\odot}$) discretized fluid cells to simulate the evolution of the cloud using \textsc{GIZMO}'s Meshless Finite Mass (MFM) solver, in which the 
cells move with the local fluid velocity while maintaining fixed mass in a quasi-Lagrangian manner. This meshless method maintains exact mass, energy and momentum conservation; exhibits superior angular momentum conservation relative to grid methods; does not require artificial diffusion or viscosity terms; and is automatically adaptive in resolution \citep[see][for details]{hopkins_2015_gizmo}. 

The initial cloud size and mass resolution are selected with the aim of resolving protostellar disk behavior within the context of multiple star formation in a realistic turbulent environment. Both parameters are comparable to those used in recent non-ideal MHD 3D SPH studies of cluster and disk formation \citep[e.g., the $50 \; M_{\odot}$, $\Delta m = 10^{-5} \; M_{\odot}$ SPH calculations of][]{wurster_2019_low_mass_cluster}.

Our models include three different sets of MHD physics: ideal MHD (labeled ``ideal"), non-ideal MHD with Ohmic resistivity and ambipolar diffusion (``non-ideal OA"), and non-ideal MHD with Ohmic resistivity, ambipolar diffusion and the Hall effect (``non-ideal OAH"). As the Hall effect has been the most challenging to implement numerically, several previous studies investigating the role of non-ideal MHD in protostellar disk formation did not include the Hall effect; we therefore include calculations both with and without the Hall effect for comparison. For each set of MHD physics, we perform one calculation with no sub-grid \textsc{STARFORGE} protostellar jet feedback and one with jet feedback (appending ``$+$jets" to the model label), resulting in six calculations of cloud evolution. The computational cost of each model ranges from $\sim$240,000 to $\sim$320,000 CPU hours.\footnote{Including non-ideal MHD moderately increases the computational cost; the nonideal Ohmic$+$AD$+$Hall model runs $\sim$20\% slower than the ideal MHD model once the gas density reaches $\rho \gtrsim 10^{-14}$ g cm$^{-3}$. The remaining difference in computational cost between models is mostly due to the formation of tight binaries and multiple systems in several models, requiring short timesteps.} We use the Frontera supercomputer at the Texas Advanced Computing Center (TACC) for our calculations.

\subsection{Initial conditions}\label{sec:ics}
The cloud radius and velocity dispersion are chosen to satisfy the observed Larson size-linewidth relation for molecular clouds (MCs), $\sigma_{\rm 1D} \simeq 0.72 \; R_{\rm pc}^{0.5}$ \citep[e.g.,][]{solomon_1987, finn_2022}, where $R_{\rm pc}$ is the radius in parsecs and $\sigma_{\rm 1D}$ is the 1D turbulent velocity dispersion. The virial parameter is related to the velocity dispersion via
\begin{equation}
    \alpha = \frac{5 \sigma_{\rm 1D}^2 R_0}{G M_0},
\end{equation}
so specifying a turbulent virial parameter $\alpha_{\rm vir} = 1.0$ leads to $R_0 = 0.3077$ pc, for an initial surface density of $\Sigma_0 = M_0/(\pi R_0^2) = 168 \; M_{\odot}/$pc and mass density $\rho_0 = 2.77 \times 10^{-20}$ g cm$^{-3}$. Table~\ref{tab:cloud_params} summarizes the initial cloud parameters, while Table~\ref{tab:sim_params} contains the relevant simulation settings.

\begin{table}
    \centering
    \begin{tabular}{ |l|l|  }
        \hline
        \multicolumn{2}{|c|}{Cloud parameters} \\
        \hline
        $M_0$          & $50 \; M_{\odot}$ \\
        $R_0$          & $0.3077$ pc \\
        $\rho_0$       & $2.77 \times 10^{-20}$ g cm$^{-3}$ \\
        $\Sigma_0$     & $168 \; M_{\odot}$ pc$^{-2}$ \\
        $T_0$          & 20 K \\
        $c_{\rm s, 0}$ & 0.2 km s$^{-1}$ \\
        $\sigma_{\rm 3D, 0}$ & 0.648 km s$^{-1}$ \\
        $\alpha_0$     & 1.0 \\
        $B_0$          & $5.41 \times 10^{-6}$ G \\
        $\mu_0$        & 13.3 \\
        $t_{\rm ff}$   & 0.4 Myr \\
        $t_{\rm cross}$ & 0.465 Myr \\
        \hline
    \end{tabular}
    \caption{Initial cloud parameters: mass ($M_0$), radius ($R_0$), mass density ($\rho_0$), surface density ($\Sigma_0$), temperature ($T_0$), sound speed ($c_{\rm s, 0}$), 3D rm velocity dispersion ($\sigma_{\rm 3D, 0}$), turbulent virial parameter ($\alpha_0$), magnetic field strength ($B_0$), mass-to-magnetic-flux ratio ($\mu_0$), freefall time ($t_{\rm ff}$), and turbulent crossing time ($t_{\rm cross}$).}
    \label{tab:cloud_params}
\end{table}

\begin{table}
    \centering
    \begin{tabular}{ |l|l|  }
        \hline
        \multicolumn{2}{|c|}{Simulation parameters} \\
        \hline
        $L_{\rm box}$    & $10 \; R_0$ \\
        $\rho_{\rm amb}$ & $0.01 \; \rho_0$ \\
        $N_{\rm gas}$    & $5 \times 10^6$ \\
        $\Delta m$       & $1 \times 10^{-5} \; M_{\odot}$ \\
        $\Delta m_{\rm j}$ & $1 \times 10^{-5} \; M_{\odot}$ \\
        $R_{\rm sink}$   & 0.5 AU \\
        $n_{\rm crit}$   & $2 \times 10^{14}$ cm$^{-3}$ \\
        $a_{\rm g}$      & 0.1 $\mu$m \\
        $\rho_{\rm g}$   & 3 g cm$^{-3}$ \\
        $\zeta_{\rm CR}$ & $1.6 \times 10^{-17}$ s$^{-1}$ \\
        \hline
    \end{tabular}
    \caption{Simulation parameters: box size ($L_{\rm box}$), density of the surrounding ambient medium ($\rho_{\rm amb}$), number of gas cells in the cloud ($N_{\rm gas}$), mass resolution ($\Delta m$), mass launched by jets ($\Delta m_{\rm j}$), sink particle radius ($R_{\rm sink}$), critical density for star formation ($n_{\rm crit}$), dust grain radius ($a_{\rm g}$), dust grain density ($\rho_{\rm g}$), and background cosmic ray ionization rate ($\zeta_{\rm CR}$).}
    \label{tab:sim_params}
\end{table}

Since it is currently infeasible to model resolved star formation in a fully Galactic context, MC simulations must assume some initial and boundary conditions, typically informed by theory and observations of MCs and the turbulent interstellar medium (ISM) in general. Popular approaches for setting these conditions for MCs have included a uniform-density sphere with a Gaussian random velocity field with a $\propto k^{-2}$ power spectrum to emulate turbulence and outflow boundary conditions \citep[``\textsc{SPHERE}"; e.g.,][]{bate_2003}, as well a periodic box in which turbulence is first ``stirred" for some period of time and allowed to reach a self-consistent initial turbulent state before self-gravity is switched on \citep[``\textsc{BOX}"; e.g.,][]{maclow_1999}. The \textsc{SPHERE} configuration, however, fails to capture nonlinear, fully-developed turbulence, while the \textsc{BOX} configuration encounters issues once stellar feedback is included, as the periodic boundary conditions force any ejected material to remain within the simulation volume. We choose to use instead the \textsc{TURBSPHERE} initialization setup of \cite{lane_2022}, which combines the \textsc{SPHERE} benefits of a spatially-localized density distribution and outflowing boundary conditions with the \textsc{BOX} advantage of fully-developed turbulence. 

The turbulent initial conditions for the clouds are generated by running the \textsc{TURBSPHERE} setup \citep[in which self-gravity is disabled and the gas is confined by an analytic potential while being continuously stirred to model the turbulent energy cascade down from larger scales; see][for details]{lane_2022} for three turbulent crossing times, $t_{\rm cross} = R_0 / \sigma_{\rm 1D}$. Under this setup, the resulting cloud achieves a quasi-equilibrium state with the desired hallmarks of supersonic turbulence within a few $t_{\rm cross}$, at which point self-gravity is enabled, turbulent driving is disabled, and the star-forming calculations begin. Table~\ref{tab:turb_params} summarizes the turbulent driving parameters used in the \textsc{TURBSPHERE} configuration.

\begin{table}
    \centering
    \begin{tabular}{ |l|l|  }
        \hline
        \multicolumn{2}{|c|}{\textsc{TURBSPHERE} parameters} \\
        \hline
        $v_{\rm drive}$ & 0.5 km s$^{-1}$ \\
        $t_{\rm coher}$ & 0.602 Myr \\
        $\lambda_{\rm min}$ & 0.15 pc \\
        $\lambda_{\rm max}$ & 0.60 pc \\
        seed & 42 \\
        \hline
    \end{tabular}
    \caption{\textsc{TURBSPHERE} turbulent driving parameters: turbulent driving rms velocity ($v_{\rm drive}$), turbulent driving coherence time ($t_{\rm coher}$), minimum wavelength ($\lambda_{\rm min}$), maximum wavelength ($\lambda_{\rm max}$), random turbulent seed (seed).}
    \label{tab:turb_params}
\end{table}

For a uniform magnetic field of magnitude $|\ve{B}| = B$ in a static cloud, the magnetic support against gravitational collapse can be estimated by comparing the mass-to-magnetic flux ratio to a critical value, 
\begin{equation}
    \mu_{\Phi} = \frac{(M/\Phi_B)}{(M/\Phi_B)_{\rm crit}},
\end{equation}
where $\Phi_B = \pi R^2 B$ is the magnetic flux through a spherical cloud of radius $R$, and, following \cite{mouschovias_spitzer_1976}, we define the critical mass-to-flux ratio as
\begin{equation}
    \left(\frac{M}{\Phi_B}\right)_{\rm crit} \approx \frac{0.53}{3\pi} \sqrt{\frac{5}{G}},
\end{equation}
where $G$ is the gravitational constant. In terms of the gravitational potential energy $E_{\rm grav}$ and magnetic energy $E_{\rm mag}$ of a uniform-density spherical cloud,
\begin{equation}\label{eq:mu_energy}
    \mu_{\Phi} \approx \frac{1}{0.53 \sqrt{2}} \sqrt{\frac{|E_{\rm grav}|}{E_{\rm mag}}} \approx 1.33 \sqrt{\frac{|E_{\rm grav}|}{E_{\rm mag}}}.
\end{equation}
We use Eq.~\ref{eq:mu_energy} to estimate $\mu_{\Phi}$ once the cloud is no longer uniformly spherical, e.g., at the conclusion of the \textsc{TURBSPHERE} calculations, by directly calculating $E_{\rm grav}$ and $E_{\rm mag}$ for the gas.\footnote{Previous \textsc{STARFORGE} calculations \citep[e.g.,][]{gus_2020_magnetized_turbulence, gus_2021} define $\mu_{\Phi} \approx 0.4 \sqrt{|E_{\rm grav}|/E_{\rm mag}}$, which is smaller than $\mu_{\Phi}$ in Eq.~\ref{eq:mu_energy} by approximately a factor of $\pi$; the fiducial \textsc{STARFORGE} calculations with $E_{\rm mag} = 0.01 |E_{\rm grav}|$ which are reported as having a mass-to-flux ratio of $\mu_{\Phi} = 4.2$ should in fact have $\mu_{\Phi} = 13.3$ to be consistent with the definition in \cite{mouschovias_spitzer_1976}. Our Eq.~\ref{eq:mu_energy} is consistent with the mass-to-flux ratio as used in similar studies \citep[e.g.,][]{wurster_2019_low_mass_cluster, lebreuilly_2024a, lebreuilly_2024b}.} The initial magnetic field has a uniform magnitude $B_0 = 5.41 \times 10^{-6}$ G in the $z$-direction, corresponding to an initial normalized mass-to-flux ratio $\mu_{\Phi} = 13.3$, or $E_{\rm mag} = 0.01 |E_{\rm grav}|$; however, after $\sim$3 $t_{\rm cross}$, the higher-density gas ($\rho \gtrsim 10^{-20}$ g cm$^{-3}$) has an average magnetic field strength of $B \approx 1.85 \times 10^{-5}$ G ($\mu_{\Phi} \approx 2.4)$ in the ideal MHD case and an average magnetic field strength of $B \approx 1.86\times 10^{-5}$ G ($\mu_{\Phi} \approx 2.5$) in the non-ideal MHD calculations including Ohmic dissipation and ambipolar diffusion due to amplification of the field by a turbulent dynamo. These clouds are magnetically supercritical ($\mu_{\Phi} > 1$), meaning that magnetic support is insufficient to prevent collapse.

\subsection{Non-ideal MHD}\label{methods_nmhd}
Magnetohydrodynamics (MHD) describes the joint dynamics of magnetic fields and gas flows. As long as the fluid approximation holds, an astrophysical plasma can be described using a multifluid approach, where each species (e.g., ions, neutrals, electrons, and charged dust grains) has its own continuity, momentum, and energy equation, coupled by interaction terms describing the collisions between different species. Under several standard assumptions, such as local quasineutrality (i.e., ion and electron number densities are approximately equal; $n_i \approx n_e \approx n$) and negligible electron inertia relative to the ion and neutral inertia (i.e., letting $m_e \rightarrow 0$), the multifluid equations can be simplified to the standard single-fluid MHD description. The ideal MHD approximation further assumes a perfectly-conducting fluid, and neglects the drift velocities between different charged species. Although widely used in astrophysical studies, the assumptions used to derive the ideal MHD equations can break down, particularly in regions with low ionization fractions \citep[such as protostellar disks, where $x_e \lesssim 10^{-10}$; see, e.g., disk models by][]{lesur_2014}.

Non-ideal MHD takes into account collisions between different charged species, and in astrophysical contexts is typically parameterized via Ohmic dissipation (collisions with electrons), the Hall effect (ion-electron drift), and ambipolar diffusion (ion-neutral drift). As the mass density of the fluid is typically dominated by the mass of the neutral gas (i.e., $\rho \approx \rho_n$ and $\rho_i \ll \rho$, where $\rho$, $\rho_n$, and $\rho_i$ are the total, neutral, and ion mass densities), and as collisions between charged and neutral particles dominate the momentum equation, the pressure and momentum contributions from the charged species can typically be ignored. Considering also the long evolutionary timescales of the magnetic field and the flow of neutrals compared to those of the charged species, one can also assume a single continuity equation, dominated by the total mass. This describes the ``strong coupling" approximation, which allows the plasma to be described as a single fluid \citep{shu_1983, pandey_wardle_2008}. Under these assumptions, the main difference between the ideal and non-ideal MHD equations lies in the induction equation:
\begin{align}
    \frac{\partial \ve{B}}{\partial t} &= \frac{\partial \ve{B}}{\partial t} \bigg|_{\rm ideal} + \frac{\partial \ve{B}}{\partial t} \bigg|_{\rm non-ideal} \nonumber \\
    &= \grad \times (\ve{v} \times \ve{B}) \nonumber \\
    &\quad \quad- \grad \times \big\lbrace\eta_{\rm O} \ve{J} + \eta_{\rm H} (\ve{J} \times \hat{\ve{B}}) + \eta_{\rm A} [\hat{\ve{B}} \times (\ve{J} \times \hat{\ve{B}})]\big\rbrace,
\end{align}
where $\ve{J} \equiv \grad \times \ve{B}$, $\hat{\ve{B}} \equiv \ve{B}/|\ve{B}|$, and $\eta_{\rm O}$, $\eta_{\rm H}$, and $\eta_{\rm A}$ are the Ohmic, Hall, and ambipolar resistivities. All of the microphysics governing species properties and interactions is contained in the $\eta$ coefficients, with dependencies
\begin{equation}\label{eq:eta_dependencies}
    \eta \equiv \eta(\rho, T, B, n_j, m_j, eZ_j),
\end{equation}
where $\rho$ and $T$ are the gas density and temperature, $B$ is the magnetic field strength, and $n_j$, $m_j$, and $eZ_j$ are the number density, mass, and electric charge of species $j$.

The numerical implementation of non-ideal MHD in GIZMO has been previously verified in \cite{hopkins_2017_gizmo_nonideal_mhd_diffusion}, which presents the results of numerical tests for the diffusion of a magnetic field under the influence of Ohmic resistivity as well as the growth of the magnetorotational instability \citep[MRI; see, e.g.,][]{sano_stone_hall_mri} with Ohmic resistivity and the Hall effect. We also present the results of the oblique isothermal C-shock test for ambipolar diffusion in Appendix~\ref{appendix_C_shock} and the whistler wave dispersion relation for the Hall effect in Appendix~\ref{appendix_whistler}. Ohmic dissipation and ambipolar diffusion are diffusive terms; diffusion is handled in an operator-split manner from the ideal MHD equation. The Hall effect is a dispersive term, but is handled similarly. The non-ideal MHD coefficients are dynamically calculated based on the local plasma state of the gas. Our implementation closely follows that of \cite{wurster_2016_nicil}; we present the full details here for completion and for ease of comparison with other implementations. The general expressions for the non-ideal coefficients are given by \citep[e.g.,][]{wardle_2007}
\begin{align}
    \eta_{\rm O} &\equiv \frac{c^2}{4\pi \sigma_{\rm O}}, \\
    \eta_{\rm H} &\equiv \frac{c^2}{4\pi \sigma_{\perp}} \frac{\sigma_{\rm H}}{\sigma_{\perp}}, \\
    \eta_{\rm A} &\equiv \frac{c^2}{4\pi\sigma_{\perp}} \frac{\sigma_{\rm P}}{\sigma_{\perp}} - \eta_{\rm O} = \frac{c^2}{4\pi\sigma_{\rm O}} \frac{\sigma_{\rm O} \sigma_{\rm P} - \sigma_{\perp}^2}{\sigma_{\perp}^2} \equiv \frac{c^2}{4\pi \sigma_{\rm O}} \frac{\sigma_{\rm A}}{\sigma_{\perp}^2},
\end{align}
where $\sigma_{\rm O}$, $\sigma_{\rm H}$, and $\sigma_{\rm P}$ are the Ohmic, Hall, and Pedersen conductivities, and $\sigma_{\perp} = \sqrt{\sigma_{\rm H}^2 + \sigma_{\rm P}^2}$ is the total conductivity perpendicular to the magnetic field. The conductivities are given by (e.g., \citealp{wardle_ng_1999}; \citealp{wardle_2007})
\begin{align}
    \sigma_{\rm O} &\equiv \frac{ec}{B} \sum_j n_j |Z_j| \beta_j, \\
    \sigma_{\rm H} &\equiv \frac{ec}{B} \sum_j \frac{n_j Z_j}{1 + \beta_j^2}, \\
    \sigma_{\rm P} &\equiv \frac{ec}{B} \sum_j \frac{n_j |Z_j| \beta_j}{1 + \beta_j^2}.
\end{align}
In the conductivities, $\beta_j$ is the Hall parameter for species $j \in \lbrace i, e, g \rbrace$, which describes the relative magnitude between the magnetic forces and neutral drag. We use a modified form of the Hall parameter for $\beta_e$ and $\beta_i$, as in \cite{wurster_et_al_2016}, so that
\begin{align}
    \beta_e &\equiv \frac{|Z_e| eB}{m_e c} \frac{1}{\nu_{ei} + \nu_{en}}, \\
    \beta_i &\equiv \frac{|Z_i| eB}{m_i c} \frac{1}{\nu_{ie} + \nu_{in}}, \\
    \beta_g &\equiv \frac{|Z_g| eB}{m_g c} \frac{1}{\nu_{gn}},
\end{align}
where $\nu_{jk}$ are the collision frequencies between different species. With these modifications, we recover $\eta_{\rm O}$ from \cite{pandey_wardle_2008} and  \cite{keith_wardle_2014}, under the assumption $\beta_i \ll \beta_e$.

We note that while $\sigma_{\rm O}$ and $\sigma_{\rm P}$ are exclusively positive, $\sigma_{\rm H}$ may take negative values. To prevent $\eta_{\rm A} \lesssim 0$ due to numerical round-off error when $\sigma_{\rm O} \sigma_{P} \approx \sigma_{\perp}^2$, we follow \cite{wurster_2016_nicil} and use a positive-definite formulation for $\eta_{\rm A}$, calculating $\sigma_{\rm A} = \sigma_{\rm O} \sigma_{\rm P} - \sigma_{\perp}^2$ as follows:
\begin{equation}
    \sigma_{\rm A}
    = \left(\frac{ec}{B}\right)^2 \sum_{j>k} \bigg[ \frac{n_j |Z_j| \beta_j}{1+\beta_j^2} \frac{n_k |Z_k| \beta_k}{1 + \beta_k^2} \bigg] \times \left(\frac{Z_k \beta_k}{|Z_k|} - \frac{Z_j \beta_j}{|Z_j|}\right)^2 \bigg].
\end{equation}
Here, $Z_i = +1$, $Z_e = -1$, $Z_n = 0$, and $Z_g$ are the mean ion/electron/neutral/dust grain charges, $m_j$ and $n_j$ the mass and number density of each species, $e$ the electron charge, and $c$ the speed of light. The neutral mass $m_n = \mu_p m_p$ is calculated using the appropriate mean molecular weight $\mu_p$ for the electron abundances, temperature, and molecular fraction determined in the cooling chemistry \citep[for further details about the cooling chemistry see][]{hopkins_2023}. The grain mass is
\begin{equation}
    m_g = \frac{4 \pi}{3} a_g^3 \rho_g,
\end{equation}
where $a_g = 0.1\;\mu$m is the assumed grain radius and $\rho_g = 3$ g cm$^{-3}$ the internal grain material density (typical of both silicate and carbonaceous grains). At the low temperatures and high densities typical of protostellar disks, ions are assumed to be dominated by Mg, so $m_i = 24.3 m_p$ \citep{pollack_et_al_1994}. We use a mean number density $n \approx n_n = \rho / m_n$. The dust grain number density is proportional to the total number density \citep{keith_wardle_2014}:
\begin{equation}
    n_g = \left(\frac{m_n}{m_g}\right) f_{dg} n,
\end{equation}
where the fiducial dust-to-gas ratio is $f_{dg} = 0.01 \; (Z/Z_{\odot})$, but decreases above a sublimation temperature $T_{\rm dust} =1500$ K, adopted from \cite{isella_natta_2005}, and fit to the model by \cite{pollack_et_al_1994}; in practice $f_{dg} \approx 0.01$ throughout the protostellar envelope and disk.

The collision frequencies, $\nu$, are empirically calculated rates. The electron-ion rate is given by \citep{pandey_wardle_2008}
\begin{equation}
    \nu_{ei} = 0.051 n_e \left(\frac{T}{100 \; \text{K}}\right)^{-1.5} \; \text{s}^{-1},
\end{equation}
while the ion-electron rate is given by $\nu_{ie} = \frac{\rho_e}{\rho_i} \nu_{ei}$. The plasma-neutral collision frequencies are given by
\begin{equation}
    \nu_{jn} = \frac{\langle \sigma v \rangle_{jn}}{m_n + m_j} \rho_n,
\end{equation}
where $\langle \sigma v \rangle_{jn}$ is the rate coefficient for the momentum transfer by collisions of particles of type $j$ with neutrals. The neutrals are assumed to be comprised of hydrogen and helium, with mass fractions $X \approx 0.76$ and $Y \approx 0.24$. For electron-neutral collisions, the rate coefficient is then
\begin{equation}
    \langle \sigma v \rangle_{en} = X \langle \sigma v \rangle_{e-\rm H_2} + Y \langle \sigma v \rangle_{e-\rm He},
\end{equation}
where, following \cite{pinto_galli_2008}, we use
\begin{align}
    \langle \sigma v \rangle_{e-\rm H_2} &= 3.16 \times 10^{-11} \left(\frac{v_{\rm rms}}{\text{km s}^{-1}}\right)^{1.3} \text{cm}^3 \; \text{s}^{-1}, \\
    \langle \sigma v \rangle_{e-\rm He} &= 7.08 \times 10^{-11} \left(\frac{v_{\rm rms}}{\text{km s}^{-1}}\right)^{1.0} \text{cm}^3 \; \text{s}^{-1},
\end{align}
with
\begin{equation}
    v_{\rm rms} = \sqrt{v_{\rm d} + \frac{8 k_B T}{\pi \mu_{en}}} \approx \sqrt{\frac{8 k_B T}{\pi m_e}},
\end{equation}
$k_B$ being the Boltzmann constant, where we assume that the electron-neutral drift velocity is $v_d = 0$ and that the electron-neutral reduced mass $\mu_{en}$ is approximately
\begin{equation}
    \mu_{en} = \frac{m_e m_n}{m_e + m_n} \approx m_e.
\end{equation}
Substituting for $v_{\rm rms}$, the electron-neutral collision frequency we use is
\begin{multline}
    \nu_{en} = \bigg[5.15 \times 10^{-9} \left(\frac{T}{100 \; \text{K}}\right)^{0.65} + \\ 1.06 \times 10^{-9} \left(\frac{T}{100 \; \text{K}}\right)^{0.5} \bigg] \left(\frac{\rho_n}{m_n + m_e}\right) \; \text{s}^{-1}.
\end{multline}
The ion-neutral rate coefficient is \citep{pinto_galli_2008}
\begin{multline}
    \langle \sigma v \rangle_{in} = 2.81 \times 10^{-9} Z_i^{1/2} \bigg[ X \left(\frac{p_{\rm H_2}}{\textup{~\AA}^3}\right)^{0.5} \left(\frac{\mu_{i-\rm H_2}}{m_p}\right)^{-0.5} + \\ Y \left(\frac{p_{\rm He}}{\textup{~\AA}^3}\right)^{0.5} \left(\frac{\mu_{i-\rm He}}{m_p}\right)^{-0.5} \bigg] \; \text{cm}^3 \, \text{s}^{-1},
\end{multline}
where the values of the polarizability are $p_{\rm H_2} = 0.804 \textup{~\AA}^3$ and $p_{\rm He} = 0.207 \textup{~\AA}^3$ \citep{osterbrock_1961}, and $\mu_{i-\rm H_2}$, $\mu_{i-\rm He}$ are the ion-hydrogen and ion-helium reduced masses. The ion-neutral collision frequency used is then
\begin{multline}
    \nu_{in} = \bigg[1.91 \times 10^{-9} \left(\frac{\mu_{i-\rm H_2}}{m_p}\right)^{-0.5} + \\ 0.31 \times 10^{-9} \left(\frac{\mu_{i-\rm He}}{m_p}\right)^{-0.5} \bigg] \left(\frac{\rho_n}{m_n + m_i}\right) \; \text{s}^{-1}.
\end{multline}
For grain-neutral collisions, the rate coefficient is given by (\citealp{wardle_ng_1999}; \citealp{pinto_galli_2008})
\begin{equation}
    \nu_{gn} = \frac{4\pi}{3} a_g^2 \delta_{gn}\left(\frac{8 k_B T}{\pi m_n}\right)^{0.5} \left(\frac{\rho_n}{m_n + m_g}\right) \; \text{s}^{-1},
\end{equation}
where $\delta_{gn} \approx 1.3$ is the Epstein coefficient for spherical grains \citep{liu_et_al_2003}.

We assume the dust grains to have a non-evolving 
single grain size with primarily collisional and cosmic-ray charging. The electron and ion number densities vary as (e.g., \citealp{umebayashi_nakano_1980}; \citealp{fujii_et_al_2011})
\begin{align}
    \frac{dn_i}{dt} = \zeta n - k_{ei} n_e n_i - k_{ig} n_i n_g, \\
    \frac{dn_e}{dt} = \zeta n - k_{ei} n_e n_i - k_{eg} n_e n_g,
\end{align}
where $\zeta$ is the ionization rate and $k_{jk}$ are the charge capture rates. We follow \cite{keith_wardle_2014} in assuming that recombination is inefficient, such that charge captures by grains dominate (i.e., $k_{ei} = 0$), and that we have an approximately steady-state system (i.e., $\frac{dn_i}{dt} \approx \frac{dn_e}{dt} \approx 0$). This allows us to calculate the ion and electron number densities as
\begin{align}
    n_i &= \frac{\zeta n}{k_{ig} n_g}, \\
    n_e &= \frac{\zeta n}{k_{eg} n_g}.
\end{align}
For $Z_g < 0$ (as is typically the case), the charge capture rates are \citep{fujii_et_al_2011}
\begin{align}
    k_{ig} &\equiv \pi a_g^2 \left(\frac{8 k_B T}{\pi m_i}\right)^{1/2}(1+\psi), \\
    k_{eg} &\equiv \pi a_g^2 \left(\frac{8 k_B T}{\pi m_e}\right)^{1/2} \exp(-\psi),
\end{align}
where $\psi$ is related to the dust grain charge as $Z_g = -\psi \frac{a_g k_B T}{e^2}$. For a given $n$ and $T$ and assuming charge neutrality, we construct the following equation to solve for $\psi$ (and hence $Z_g$):
\begin{equation}
    \psi = \alpha \left(\exp(\psi) - \frac{(m_i/m_e)^{1/2}}{1+\psi}\right),
\end{equation}
where
\begin{equation}
    \alpha \equiv \frac{\zeta e^2 m_e^{1/2} m_g^2}{(8 \pi)^{1/2} a_g^3 f_{dg} (k_B T)^{3/2} m_n^2 n}.
\end{equation}
At the high densities typical of protostellar disks, we assume that cosmic rays are the dominant ionizing source, so that $\zeta = \zeta_{\rm CR}$. We adopt a uniform background cosmic ray ionization rate $\zeta_{\rm CR} = 1.6 \times 10^{-17}$ s$^{-1}$ (typical of the Solar environment; \citealp{cummings_et_al_2016}).

We emphasize that our implementation is well-suited for modeling non-ideal MHD effects in the cold, dense gas characteristic of the collapsing cloud and the protostellar disk, where singly-ionized Mg dominates the ion population. We do not account for contributions from lighter ions such as ionized H and He, as in, e.g., the NICIL library \citep{wurster_2016_nicil}, which dominate the ion population in highly-photoionized regimes such as the HII regions surrounding massive young stars. In fact, several massive ($\gtrsim$8 $M_{\odot}$) protostars form in our models without sub-grid protostellar jet feedback. However, as we discuss in Section~\ref{section_disk_identification}, we do not expect this limitation of our implementation to affect the formation and evolution of the disks prior to the development of the HII regions.

\subsection{Thermodynamics}\label{sec:thermo}
While the ISM is often assumed to be isothermal for simplicity, many important effects in star formation (such as the dynamics of fragmentation) require an explicit treatment of the thermal structure of the ISM \citep{lee_et_al_2020}. Therefore, for these calculations, we use a gas equation of state with a variable adiabatic index $\gamma$ in order to account for variations in the ratio of para- to ortho-hydrogen, as well as the collisional dissociation of molecular H above $T \simeq 2000$ K. The gas adiabatic index is calculated from a fit to density based on the results of \cite{vaidya_2015}.

Since radiative processes are included, the numerical method must self-consistently co-evolve the gas, dust, and radiation field temperature, as in \cite{hopkins_2020_rad_feedback_methods}. The numerical method operator-splits the adiabatic MHD evolution with a standard implicit cooling algorithm, which solves for equilibrium internal energy, temperature, net cooling/heating rate, mean molecular weight, and ionization state of the gas \citep[treating the adiabatic heating rate from the MHD solver as an additional heating term;][]{hopkins_2017_gizmo_nonideal_mhd_diffusion}. Heating and cooling processes include molecular and fine-structure cooling, cosmic ray heating, dust cooling and heating, photoelectric heating \citep{draine_1978, clark_2012}, metal line cooling, H photoionization, and collisional ionization of H and He. Details of the radiative cooling and thermochemistry modules are given in \cite{hopkins_2023}. In addition to radiative feedback from sink particles (i.e., protostars/stars), we include an external heating source at the boundary of the simulation domain that represents the interstellar radiation field \citep[ISRF; assuming typical solar neighborhood conditions,][]{draine_2011}.

\subsection{Sink particles}\label{sec:sink}
Once dense cores undergo gravitational collapse to form protostars, these protostars begin injecting mass, momentum, and energy into their surroundings via a number of feedback mechanisms, such as radiation and collimated bipolar outflows (i.e., ``jets"). In these simulations, the accretion, dynamics, and feedback of protostars and stars are modeled by using sink particles \citep[e.g.,][]{bate_1995, krumholz_2004, federrath_2014, hubber_2013, bleuler_teyssier_2014}, which are created on-the-fly from gas cells satisfying certain criteria (e.g., a density threshold, here $n_{\rm crit} = 2 \times 10^{14}$ cm$^{-3}$). While the numerical scheme could continue to self-consistently follow collapse down to even higher densities, inserting sink particles at this point lowers the computational cost. At the current mass resolution $\Delta m = 10^{-5} \; M_{\odot}$, each sink particle represents an individual (proto)star.

The sink particle implementation presented in the \textsc{STARFORGE} framework allows gas cells to be accreted by a sink particle if they are within the sink particle radius (here $R_{\rm sink} = 0.5$ AU) and satisfy a number of boundedness/angular momentum/size/density criteria \citep[see][for details]{grudic_2021_starforge}. When a gas cell is accreted, the position, velocity, and internal angular momentum of the sink particle are updated to conserve center of mass, total momentum, and angular momentum.

The luminosity, temperature, and radius of each sink particle are evolved according to the sub-grid protostellar evolution prescription originally implemented in the \textsc{Orion} code by \cite{offner_2009}. This model follows protostellar evolution through a sequence of phases, ending on the main sequence. 

Sink particles act as radiation sources, injecting photons into five radiation bands (H ionizing, FUV, NUV, optical-NIR, and FIR) according to the spectral energy distribution set by the stellar evolution mode. To model radiative feedback, the radiation field is evolved in the  five frequency bins using \textsc{GIZMO's} M1 solver \citep{levermore_1984, hopkins_grudic_2019, hopkins_2020_rad_feedback_methods}, in which gas cells exchange fluxes of radiation across effective faces (i.e., the same mesh-free volume discretization as used for MHD solver). Dust grains can scatter and absorb photons in all five bands, as well as radiate in the FIR band. Lyman continuum photons may be absorbed by HI. Absorbed ionizing photons are promptly re-radiated isotropically in the optical-NIR band; radiation absorbed in all other bands is re-radiated by dust in the FIR band.

As demonstrated by \cite{offner_chaban_2017, gus_2021}, protostellar jets are an important feedback mechanism and dramatically reduce stellar accretion rates. In calculations including sub-grid jet feedback, protostellar jets are modeled using the prescription of \cite{cunningham_2011}: a fraction $f_{\rm w} = 0.3$ of accreted material is launched along the sink angular momentum axis with speed $v_{\rm jet} = f_{\rm K} \sqrt{GM_*/R_*}$, with $f_{\rm K} = 0.3$ \citep[see][for a discussion on the effects of varying the parameters $f_{\rm w}$ and $f_{\rm K}$ on the IMF]{gus_2021}. Jets inject new gas cells 
in pairs near the sink particle, with opposite positions and velocities, conserving center of mass and momentum. Stellar winds and supernovae are also implemented in the numerical model, however these mechanisms are more relevant for high-mass star formation, and so are not included in our calculations of low-mass star formation. Although several high-mass ($\gtrsim$8 $M_{\odot}$) protostars form in our calculations without sub-grid protostellar jet feedback, we do not evolve our star-forming calculations beyond $\sim$1.5 Myr, well before any supernovae are expected, allowing us to safely ignore this feedback mechanism \citep{grudic_2021_starforge}.

\section{Results}
We evolve the six models (three sets of MHD physics, with and without subgrid jet feedback), for approximately $2.5-3.5$ free-fall times ($1-1.4$ Myr), until the gas in the vicinity of the star-forming region has largely been dispersed by feedback or accreted onto the stars. Table~\ref{tab:model_results} summarizes the protostar and protostellar disk formation outcomes for each model.

\begin{table*}
    \centering
    \begin{tabular}{ |l|l|c|c|l|l|  }
        \hline
        \multicolumn{6}{|c|}{Model protostar/protostellar disk formation summary} \\
        \hline
        Model label & MHD & Sub-grid jets & Num. protostars & Disks formed & Disk lifetimes (kyr) \\
        \hline
        \textbf{\texttt{ideal}} & ideal MHD & No & 3 & 1 binary disk & 187 \\
        \textbf{\texttt{ideal+jets}} & ideal MHD & Yes & 4 & 1 quadruple disk\textsuperscript{\textdagger} & 403 \\
        \textbf{\texttt{non-ideal OA}} & Ohmic+ambipolar & No & 1 & 1 single disk & 277 \\
        \textbf{\texttt{non-ideal OA+jets}} & Ohmic+ambipolar & Yes & 5 & 2 single disks & 271 (s$_1$); 361 (s$_2$) \\
        \textbf{\texttt{non-ideal OAH}} & Ohmic+ambipolar+Hall & No & 4 & 2 single $\rightarrow$ 2 binary disks & 30 (s$_1$, s$_2$); 271 (b$_1$, b$_2$) \\
        \textbf{\texttt{non-ideal OAH+jets}} & Ohmic+ambipolar+Hall & Yes & 10 & 1 single $\rightarrow$ 1 binary disk & 151 (s); 397 (b) \\
        \hline
    \end{tabular}
    \caption{Summary of included physics and description of protostar and protostellar disk formation outcomes for the six models. {}\textsuperscript{\textdagger}The quadruple disk formed in the ideal$+$jets model is notably less massive than disks formed in all other models, rarely exceeding $M_{\rm disk} = 10^{-3} M_{\odot}$; see further discussion in Section~\ref{section_disk_identification}.}
    \label{tab:model_results}
\end{table*}

\subsection{Cloud structure}
We begin our comparison of the different models with a description of the global evolution, starting with the \textsc{TURBSPHERE} calculations and then proceeding to the star-forming calculations.
\subsubsection{\textsc{TURBSPHERE} stirring calculations}\label{sec_TURBSPHERE_structure}
\begin{figure*}
	\includegraphics[width=2\columnwidth]{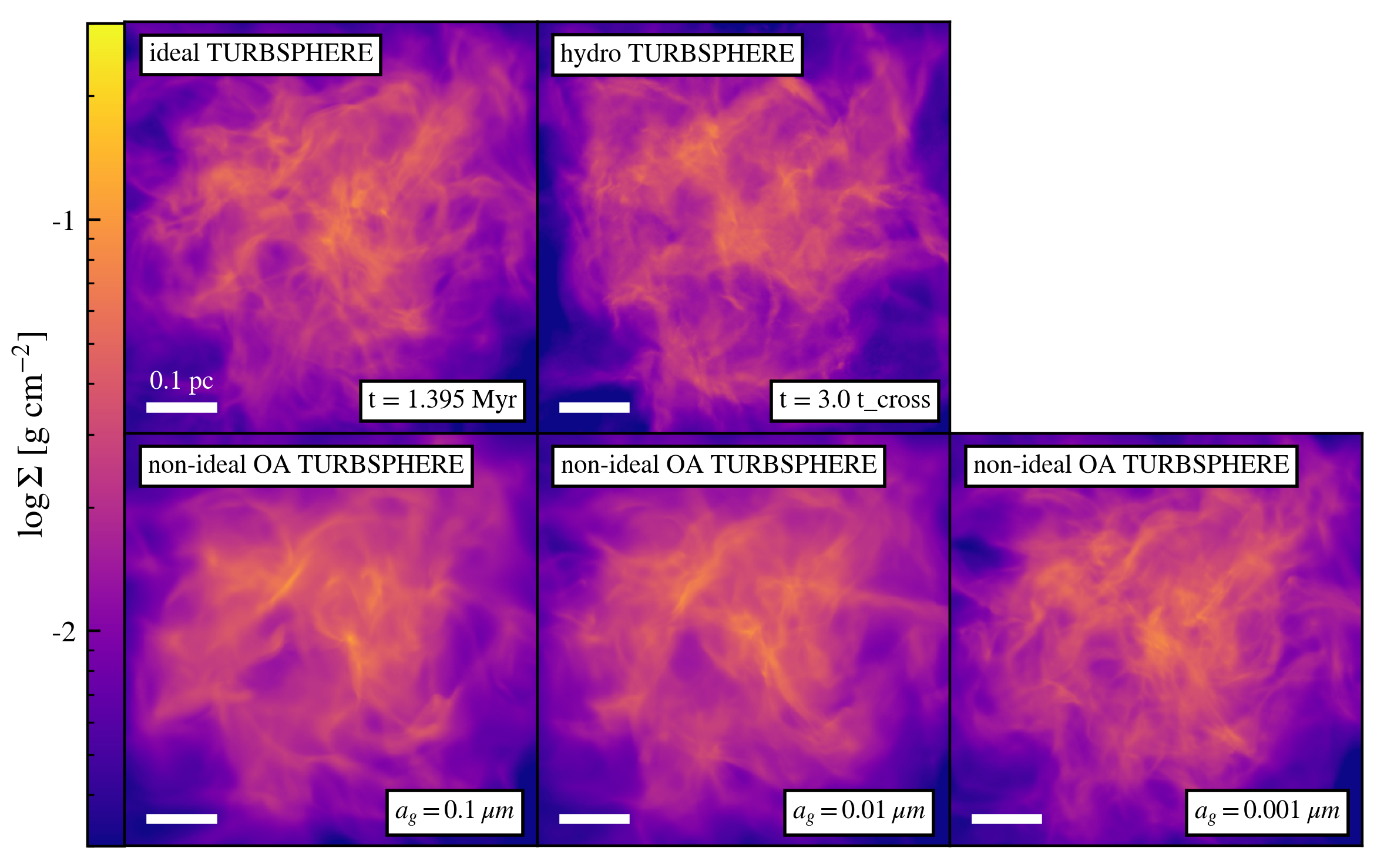}
    \caption{Projected column density at three turbulent crossing times ($t = 1.395$ Myr) during the \textsc{TURBSPHERE} stirring calculations for five different MHD models. Top row: ideal MHD (left) and pure hydrodynamics (middle). Bottom row: Non-ideal MHD (Ohmic$+$AD only) using fixed dust grain sizes $a_g = 0.1$, 0.01, and 0.001 $\mu$m. The snapshots in the leftmost column represent the initial conditions for the star-forming calculations. The snapshots in the bottom row demonstrate the effects of varying the assumed dust grain size on the large-scale shock structure in the non-ideal MHD case.}
    \label{fig:proj_density_turbsphere}
\end{figure*}
\begin{figure}
	\includegraphics[width=\columnwidth]{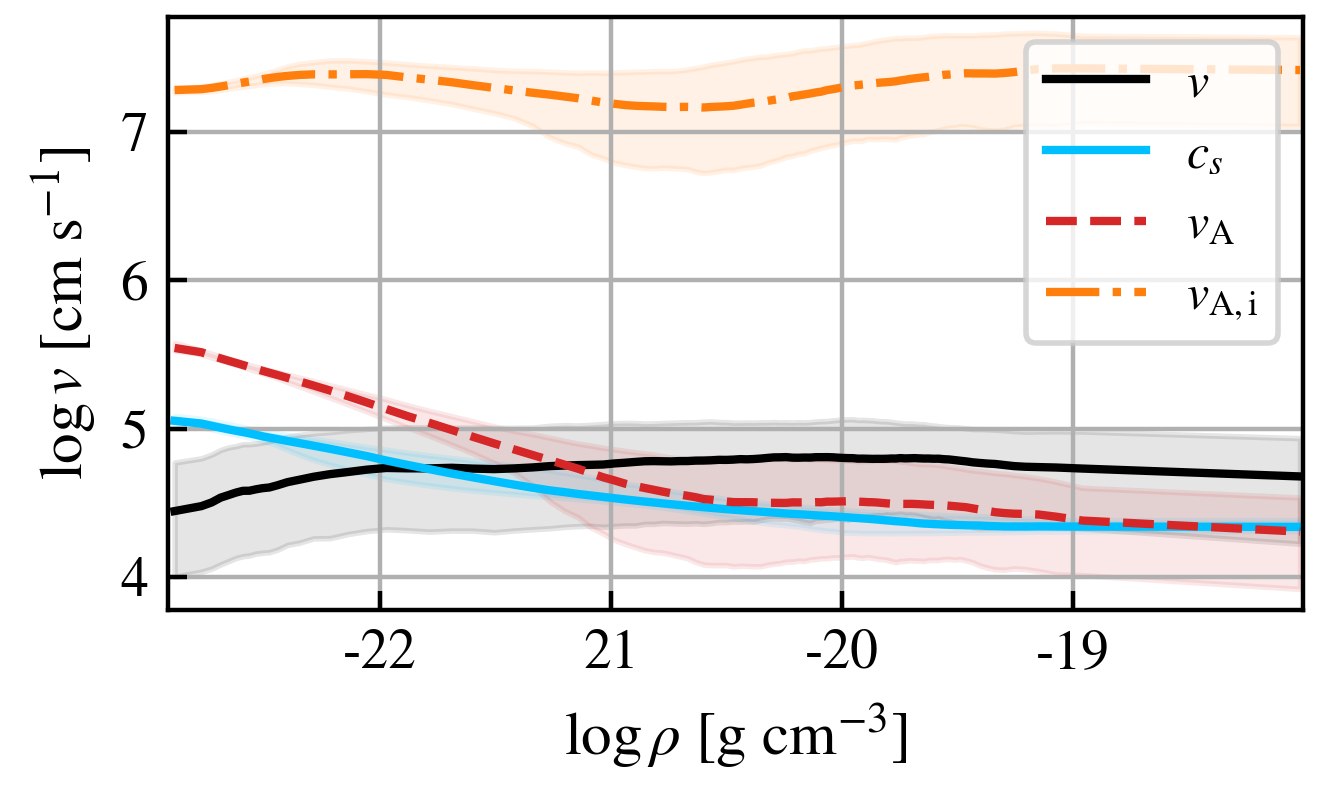}
    \caption{Relevant speeds (magnitude of the gas velocity $v$, gas thermal sound speed $c_s$, bulk gas Alfv\'{e}n speed $v_{\rm A}$, and ion Alfv\'{e}n speed $v_{\rm A, i}$) as functions of gas density for the fiducial non-ideal MHD TURBSPHERE calculations at $t = 3 t_{\rm cross}$. The solid lines show the median value, while the shaded regions enclose the 5th and 95th percentiles.}
    \label{fig:density_profile_TURBSPHERE_velocity}
\end{figure}
\begin{figure}
	\includegraphics[width=\columnwidth]{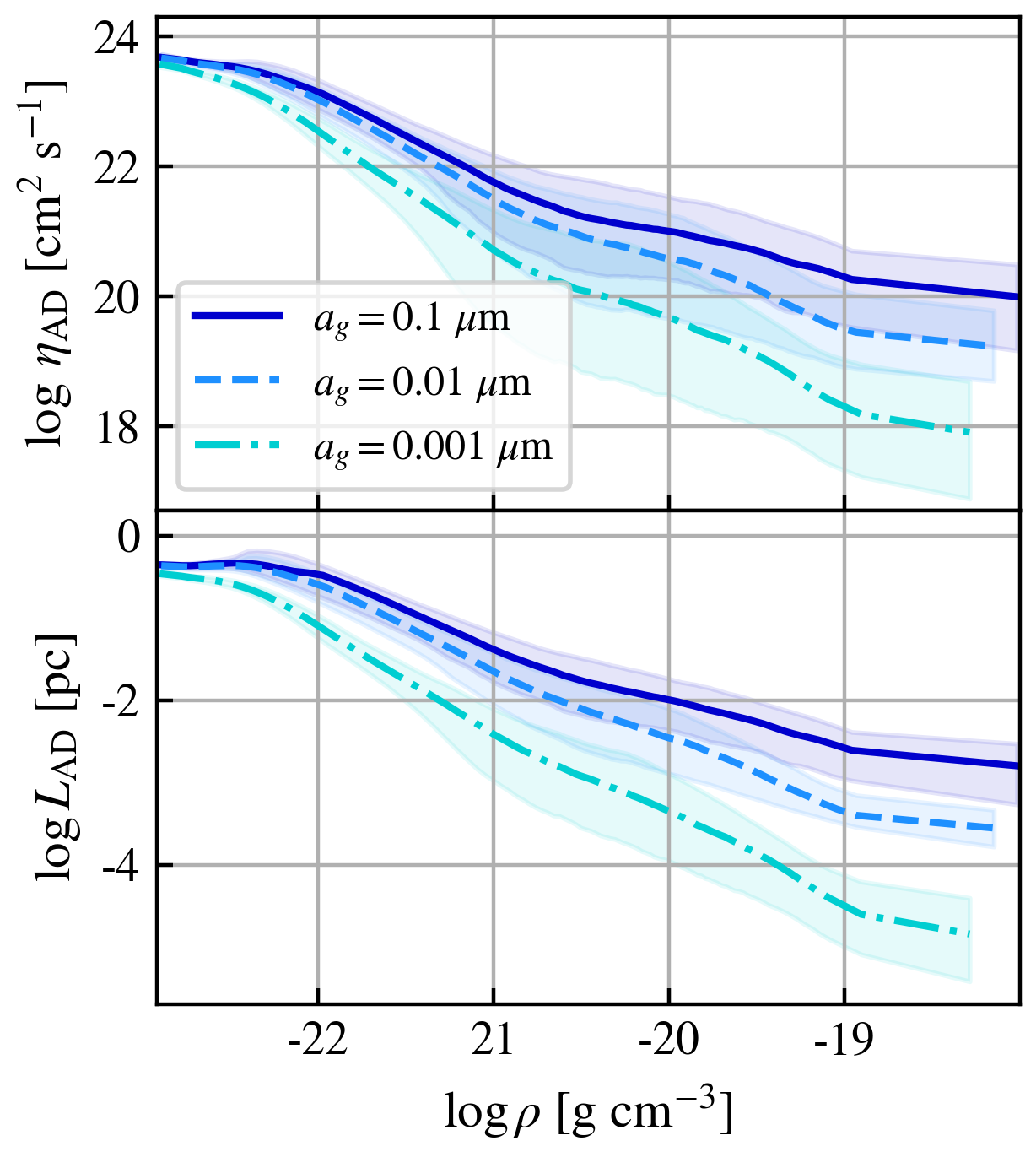}
    \caption{Ambipolar resistivity $\eta_{\rm AD}$ (top) and characteristic ambipolar diffusion length $L_{\rm AD} = \eta_{\rm AD}/v_{\rm A}$ (bottom) as functions of density for the three different non-ideal MHD TURBSPHERE calculations ($a_g = 0.1$, 0.01, and 0.001 $\mu$m) at $t = 3 t_{\rm cross}$. The lines show the median value, while the shaded regions enclose the 5th and 95th percentiles.}
    \label{fig:density_profile_TURBSPHERE_shock_thickness}
\end{figure}
\begin{figure}
	\includegraphics[width=\columnwidth]{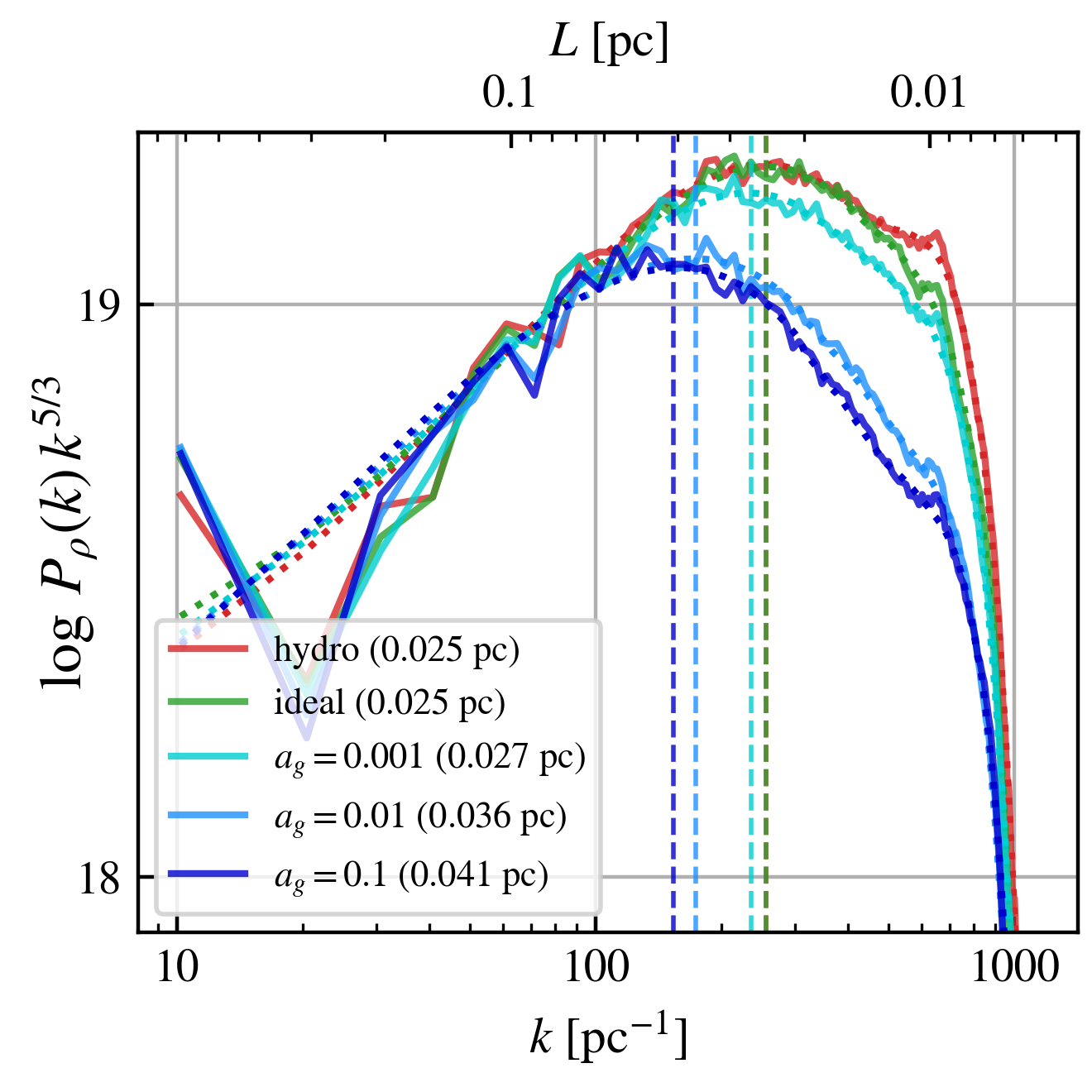}
    \caption{Compensated density power spectra $P_{\rho}(k) k^{5/3}$ comparing the five different models shown in Figure~\ref{fig:proj_density_turbsphere} at $t = 3t_{\rm cross}$. The dotted lines show the data smoothed by a Savitzky-Golay filter; the vertical dashed lines indicate the turnover peaks of the smoothed spectra, with the physical scales listed in the legend. We use a grid resolution $N_{\rm grid} = 128$ over the central $2R_0$ region of the simulation volume to compute the power spectra; the slight peak at $L \approx 0.01$ pc corresponds to the Nyquist frequency at integer wavenumber $k = N/2$.}
    \label{fig:TURBSPHERE_power_spectrum}
\end{figure}
To generate the initial conditions for the ideal and non-ideal MHD models, we evolve the TURBSPHERE stirring calculations for three turbulent crossing times ($t_{\rm cross} = 0.465$ Myr) with two sets of MHD physics: ideal MHD and non-ideal MHD with Ohmic dissipation and ambipolar diffusion only.\footnote{To reduce the computational time, the Hall effect was not included in the non-ideal MHD stirring calculations. However, as ambipolar diffusion is expected to be the dominant non-ideal effect at the lower densities encountered during the stirring calculations, we do not expect these calculations to be substantially altered by the inclusion of the Hall effect.} The influence of non-ideal MHD on the density distribution resulting from the initial turbulent driving phase can be clearly seen in the column density snapshots shown in Figure~\ref{fig:proj_density_turbsphere}. The leftmost column of Figure~\ref{fig:proj_density_turbsphere} shows the initial conditions for the ideal (top) and non-ideal (bottom; with fiducial dust grain size $a_g = 0.1 \; \mu$m) models; for comparison, we also include column density snapshots for a model with hydrodynamics only (top middle), and for non-ideal models with different choices of dust grain size, $a_g = 0.01$ $\mu$m and $a_g = 0.001$ $\mu$m (bottom middle and bottom right, respectively). As expected, the turbulent motions produce a non-homogeneous, filamentary structure in all models. However, while the overall shape and structure of the cloud gas distributions are generally similar, owing to the use of the same initial turbulent seed, it is clear that the non-ideal models have fewer small-scale ($\lesssim$0.1 pc) features: small-scale structure is smoothed out and the densest structures appear broader and more massive with respect to the ideal calculations. Furthermore, the degree of smoothing is correlated with the assumed dust grain size. The non-ideal model with the smallest dust grains ($a_g = 0.001$ $\mu$m) most closely resembles the ideal model, while the fiducial non-ideal model with the largest dust grains ($a_g = 0.1$ $\mu$m) shows fewer small-scale features and appears to have the smoothest density distribution.

The increased smoothing observed in the non-ideal models can be understood by considering the effects of ion-neutral friction, or ambipolar diffusion, on the shocks induced by the turbulent driving. The nature of the shock front depends on both the shock velocity as well as the microphysical properties of the gas. Under ideal MHD conditions, the hydrodynamical variables and the parallel component of the magnetic field undergo a discontinuous transition across a narrow shock front, forming a ``jump" or J-type shock. However, if the fractional ionization is low, as is the case for molecular clouds, the Alfv\'{e}n speed of the ionized component, $v_{\text{A},i} \propto B/\sqrt{\rho_i}$ for ion density $\rho_i$, is much larger than the Alfv\'{e}n speed of the neutral gas, $v_{\text{A}} \propto B/\sqrt{\rho}$, and the shock signal can smoothly propagate upstream through the ionized component. Ion-neutral collisions then subsequently drag and compress the neutrals ahead of the shock front as well. If the neutral fluid remains cool, either because the shock is weak or because radiative cooling efficiently removes the heat generated by the ion-neutral collisions, the neutral fluid is everywhere supersonic in the rest frame of the post-shocked gas and all of the fluid variables are everywhere continuous, resulting in a smooth C-type shock across an extended shock front \citep{draine_1980_C_shock,draine_1983_C_shock}. Figure~\ref{fig:density_profile_TURBSPHERE_velocity} shows the magnitude of the gas velocity $v$, the gas thermal sound speed $c_s$, the bulk neutral gas Alfv\'{e}n speed $v_{\text{A}}$, and the ionized gas Alfv\'{e}n speed $v_{\text{A},i}$ as functions of density in the fiducial (i.e., $a_g = 0.1$ $\mu$m) non-ideal TURBSPHERE model at $t = 3 t_{\rm cross}$. While the gas is supersonic for densities $\rho \gtrsim 10^{-22}$ g cm$^{-3}$ and super-Alfv\'{e}nic for $\rho \gtrsim 10^{-21}$ g cm$^{-3}$, the ionized gas Alfv\'{e}n speed is everywhere several orders of magnitude greater than the gas speed, suggesting appropriate conditions for C-type shocks. 

The expected C-shock thickness can be estimated by considering the characteristic ambipolar diffusion length, $L_{\rm AD} = \eta_{\rm AD} / v_{\rm A}$.\footnote{The shock thickness can also be estimated by considering the compression ratio of neutrals across the shock front, as in \cite{wardle_1990} and \cite{chen_ostriker_2012}. We find good agreement between these estimates and $L_{\rm AD}$ and so use $L_{\rm AD}$ as a measure of the expected C-shock thickness.}  Figure~\ref{fig:density_profile_TURBSPHERE_shock_thickness} shows the effect of the choice of dust grain size on the ambipolar resistivity, $\eta_{\rm AD}$, and the expected shock thickness, $L_{\rm AD}$. The ambipolar resistivity  and characteristic ambipolar diffusion length are in general decreasing functions of density, but both decrease more steeply with density for smaller-sized grains, with $\eta_{\rm AD}$ and $L_{\rm AD}$ differing by approximately one order of magnitude at densities $\rho \sim 10^{-20}$ g cm$^{-3}$ and by more than two orders of magnitude at densities $\rho \gtrsim 10^{-19}$ g cm$^{-3}$ between the $a_g = 0.1$ $\mu$m and $a_g = 0.001$ $\mu$m calculations. We note that, as listed in Table~\ref{tab:cloud_params}, our initial uniform-density sphere has $\rho_0 = 2.77 \times 10^{-20}$ g cm$^{-3}$ and that, after three turbulent crossing times, the mean density of each of the five models shown in Figure~\ref{fig:proj_density_turbsphere} is still approximately $\rho_{\rm mean} \approx 2 \times 10^{-20}$ g cm$^{-3}$.

To compare the estimated shock thickness to the data, in Figure~\ref{fig:TURBSPHERE_power_spectrum} we plot the density power spectrum $P_{\rho}(k)$ compensated by the expected $k^{-5/3}$ power law for Kolmogorov turbulence for the five models shown in Figure~\ref{fig:proj_density_turbsphere}. We also smooth the data by applying a Savitsky-Golay filter and then use the smoothed data to identify the turnover point of each spectra, as indicated by the dotted lines. The non-ideal models with larger dust grains, $a_g = 0.1$ $\mu$m and $a_g = 0.01$ $\mu$m, have turnover points at smaller wavenumbers $k$ and show noticeably less power at smaller physical scales than the hydrodynamical and ideal models as well as the non-ideal model with the smallest dust grains ($a_g = 0.001$ $\mu$m), in agreement with theoretical predictions. These results are consistent with previous non-ideal turbulent box calculations including ambipolar diffusion \citep[e.g.,][see also the discussion in Section~\ref{discussion_nonideal_mhd}]{ntormousi_2016}.

\subsubsection{Star-forming calculations}
\begin{figure}
    \centering\includegraphics[width=0.83\columnwidth]{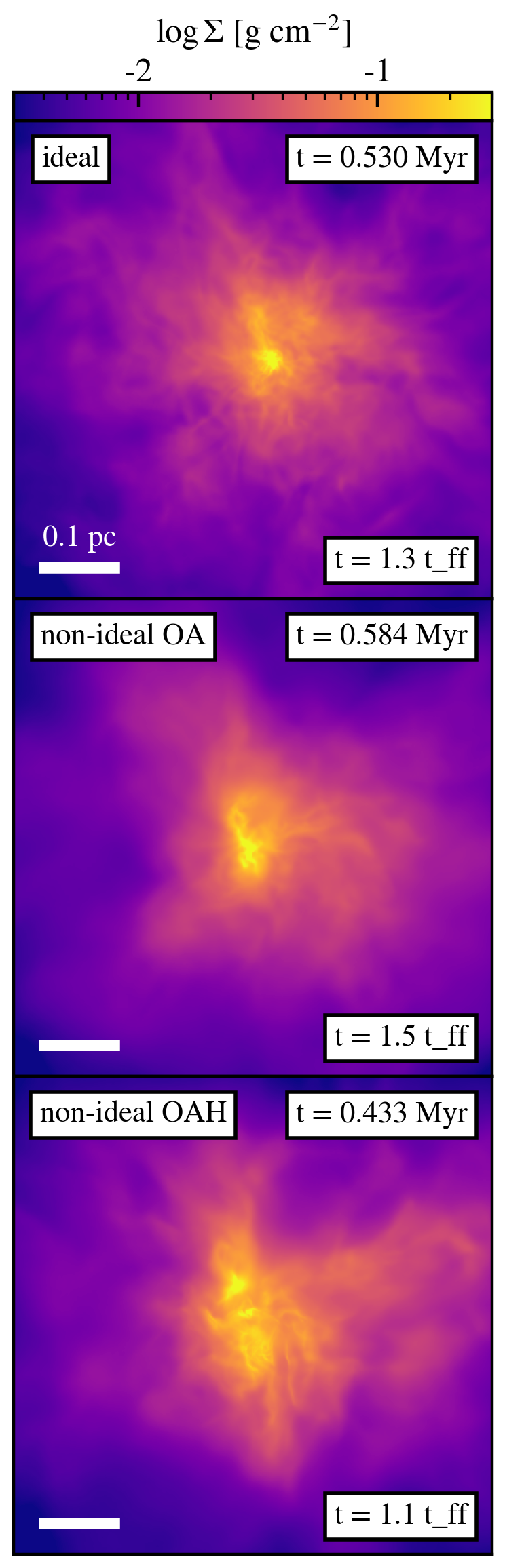}
    \caption{Projected column density just prior to the formation of the first sink particles (i.e., stars) for the different MHD models. Top row: ideal MHD; middle row: non-ideal Ohmic$+$AD; bottom row: non-ideal Ohmic$+$AD$+$Hall.}
    \label{fig:proj_density_pre-sf}
\end{figure}

Figure~\ref{fig:proj_density_pre-sf} shows the gas column density of the ideal and two non-ideal (Ohmic$+$ambipolar, and Ohmic$+$ambipolar$+$Hall) MHD models just prior to the formation of the first sink particles (hereafter referred to as stars). The ideal model appears somewhat clumpier at small ($\lesssim$0.05 pc) scales, possibly as a result of inheriting more small-scale structure from the initial conditions generated by the \textsc{TURBSPHERE} stirring phase; as discussed in Section~\ref{sec_TURBSPHERE_structure}, the initial conditions for the non-ideal star-forming calculations have smoother density distributions due to the influence of ion-neutral friction on the shock structures formed by the turbulent driving. The non-ideal OA model appears to be the most smooth. The densest gas near the centers of the three models shows a variety of asymmetric structures, with a single dense core in the ideal model, and a number of cores connected by filamentary structures in the two non-ideal models. There are also visible differences in the density distributions between the two non-ideal models. For example, the densest regions in the non-ideal OAH model are more spatially separated than in the non-ideal OA model, and there are more extended filamentary structures interspersed with low-density voids in the non-ideal OAH model. These differences can be attributed to the influence of the Hall effect. While Ohmic resistivity and ambipolar diffusion are diffusive effects and tend to relax the bending of pinched magnetic field lines, the Hall effect induces a drift velocity in a direction orthogonal to the bending of the field lines, modifying the magnetic field geometry in a fundamentally different manner \citep[e.g.,][]{zhao_2020_hall}. Differences in the evolution of the magnetic field then lead to the development of different density structures due to the coupling between the gas and the magnetic field.
\begin{figure}
    \centering\includegraphics[width=0.83\columnwidth]{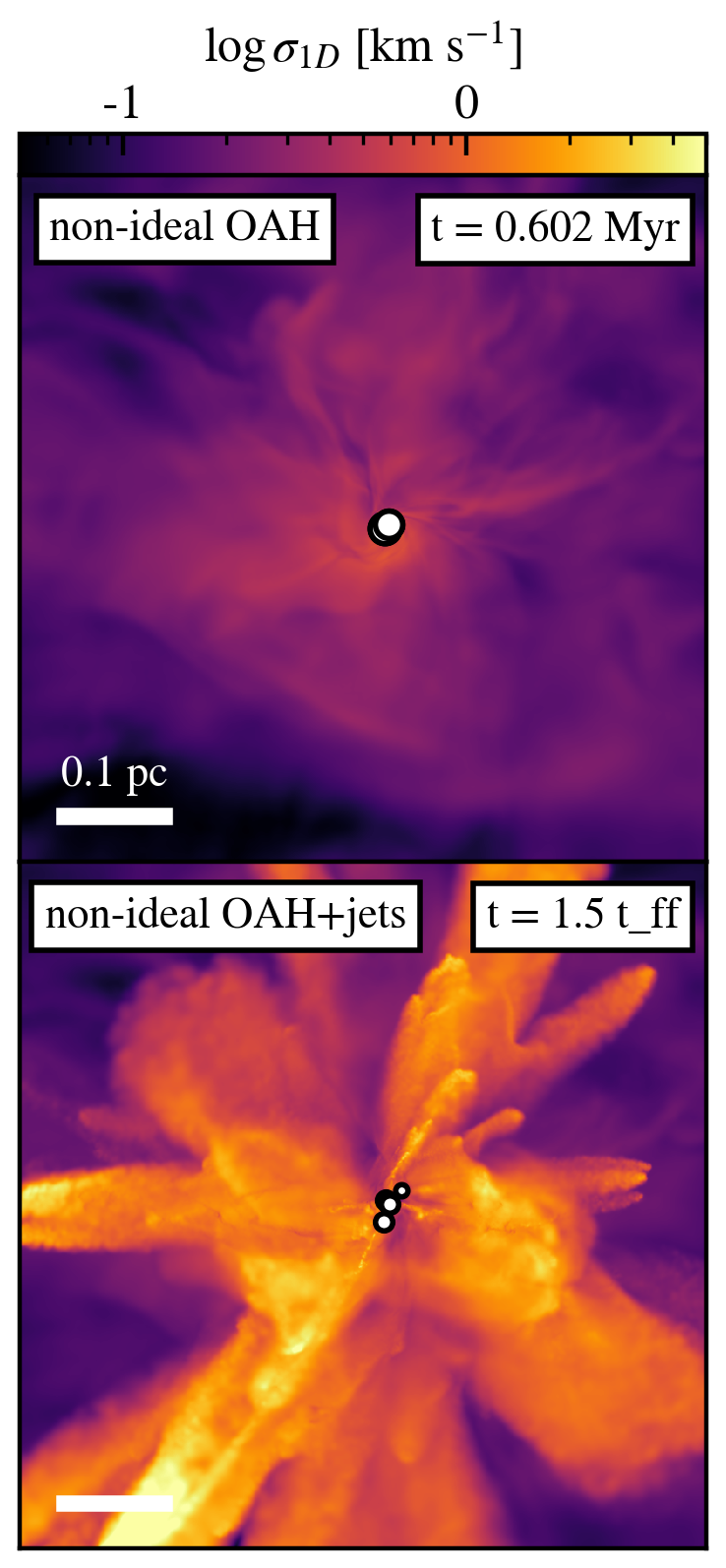}
    \caption{Line-of-sight 1D velocity dispersion ($\sigma_{\rm 1D}$) for the non-ideal OAH (top row) and the non-ideal OAH$+$jets (bottom row) models after stars begin to form, showing the effects of the sub-grid protostellar jet feedback on the global cloud morphology and kinematics. Markers indicate the positions of stars, with marker size proportional to stellar mass.}
    \label{fig:proj_density_feedback}
\end{figure}

Figure~\ref{fig:proj_density_feedback} shows the effects of the sub-grid protostellar jet feedback on the velocity dispersion of the cloud, comparing the non-ideal OAH and non-ideal OAH+jets models at the same point in their evolution, $t = 1.5 t_{\rm ff}$ after the start of the star-forming calculations. At this time, both models have formed multiple stars, shown as white circles in the figure; there are four stars in the non-ideal OAH model and ten stars in the non-ideal OAH+jets model. While the line-of-sight 1D velocity dispersion is at most $\sigma_{\rm 1D} \approx 0.4$ km s$^{-1}$ in the model without jet feedback, the maximum velocity dispersion in the model with jet feedback is greater by an order of magnitude, $\sigma_{\rm 1D} \approx 4$ km s$^{-1}$. The structure of the jets can be clearly seen in the bottom panel of the figure. Jets extend $\gtrsim$0.3 pc from the stars at the cloud center, beyond the initial radius of the cloud, and are oriented along a variety of directions, tracing the range of orientations of the stellar angular momentum axes along which they are launched.
\begin{figure}
    \includegraphics[width=\columnwidth]{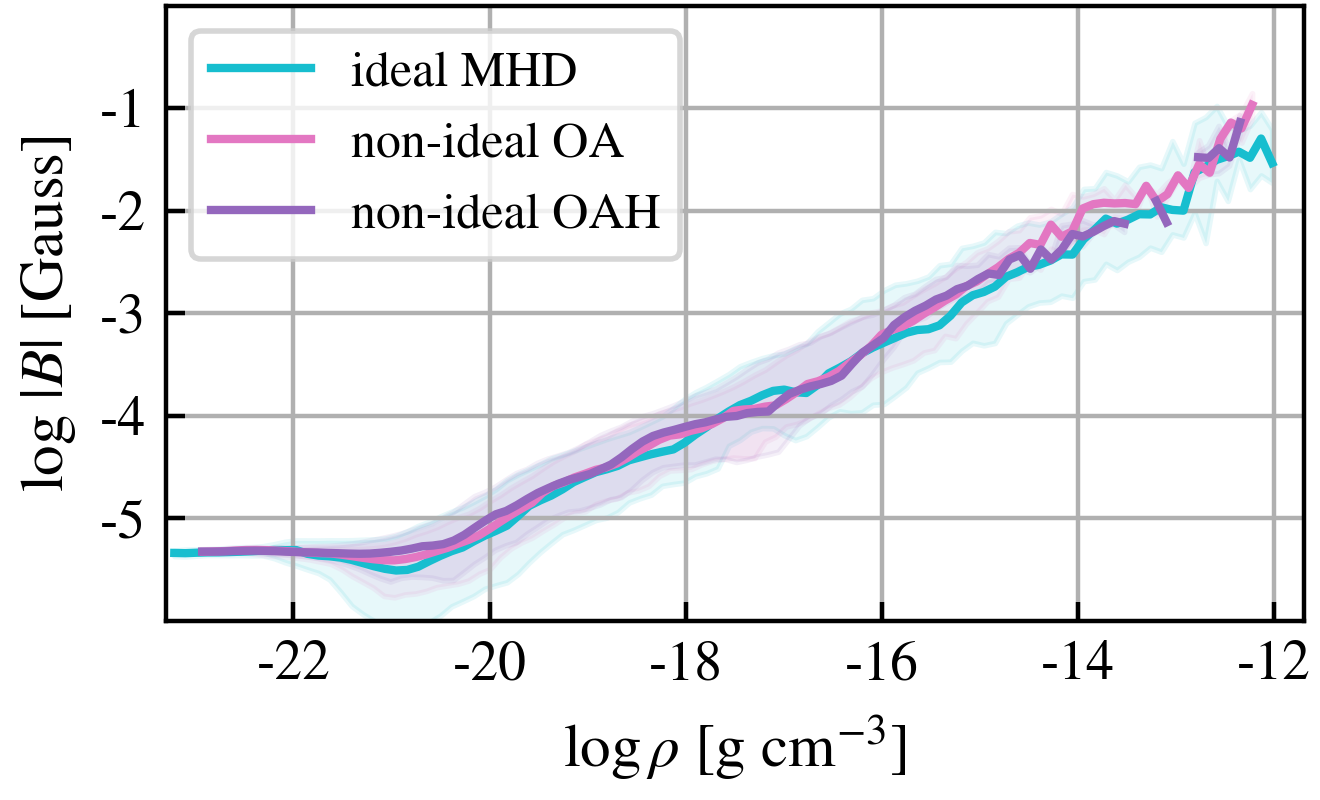}
    \caption{Magnetic field strength as a function of gas density for the ideal model (cyan) and two non-ideal MHD (pink: non-ideal OA; purple: non-ideal OAH) models just after star formation begins. The solid lines show the median value, while the shaded regions enclose the 5th and 95th percentiles.}
    \label{fig:B_vs_rho_pre-sf}
\end{figure}
\begin{figure}
	\includegraphics[width=\columnwidth]{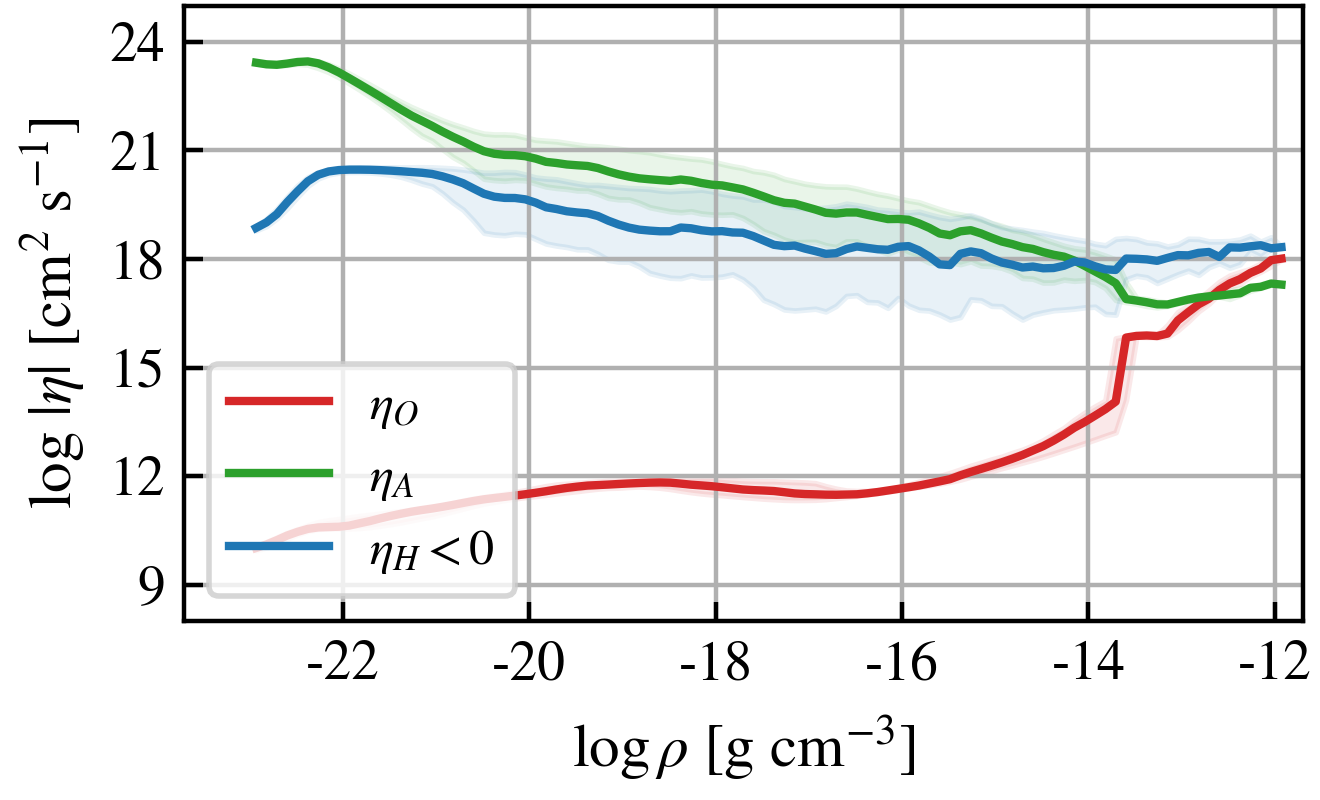}
    \caption{Non-ideal MHD resistivities as functions of gas density for the non-ideal OAH model just after all four stars have formed, $t = 0.50$ Myr. The solid lines show the median value, while the shaded regions enclose the 5th and 95th percentiles.}
    \label{fig:eta_vs_rho_nOAH}
\end{figure}

Figure~\ref{fig:B_vs_rho_pre-sf} shows the magnetic field strength as a function of gas density for the three different sets of MHD physics just after the formation of the first stars. All three models show very similar levels of magnetization, with the magnetic field strength following $B \propto \rho^{1/2}$ between $\rho \gtrsim 10^{-21}$ and $\rho \lesssim 10^{-12}$ g cm$^{-3}$. The median value of the magnetic field strength is slightly higher in the two non-ideal models than in the ideal model for densities $\rho \gtrsim 10^{-16}$ g cm$^{-3}$ but is within the 95th percentile range of the ideal model; see Section~\ref{discussion_nonideal_mhd} for further discussion. The scaling with density and relative similarity between the different MHD models is consistent with earlier studies \citep[e.g.,][]{mocz_2017, wurster_2019_low_mass_cluster, gus_et_al_2022_Bfield_scaling, He_Ricotti_2023, He_Ricotti_2025}.

Figure~\ref{fig:eta_vs_rho_nOAH} shows the values of the non-ideal MHD resistivities as functions of gas density in the non-ideal OAH model just after all four stars have formed at $t = 0.50$ Myr. As expected, ambipolar diffusion is most efficient and Ohmic resistivity is least efficient in the low-density gas \citep[e.g.,][]{mellon_li_2009, krasnopolsky_li_2010, wurster_2021}. Ohmic resistivity becomes increasingly important as gas density increases. The Hall resistivity becomes the dominant non-ideal effect above $\rho \gtrsim 10^{-14}$ g cm$^{-3}$, and is negative throughout our parameter space. While the Hall resistivity may take either positive or negative values \citep[e.g.,][]{wardle_ng_1999}, the transition from negative $\eta_{\rm H}$ to positive $\eta_{\rm H}$ generally occurs at densities greater than $\rho \gtrsim 10^{-12}$ g cm$^{-3}$ for temperatures and magnetizations typical at this stage during the collapse \citep[e.g.,][]{wurster_2016_nicil}.

\subsection{Star formation outcomes}
\begin{figure}
	\includegraphics[width=\columnwidth]{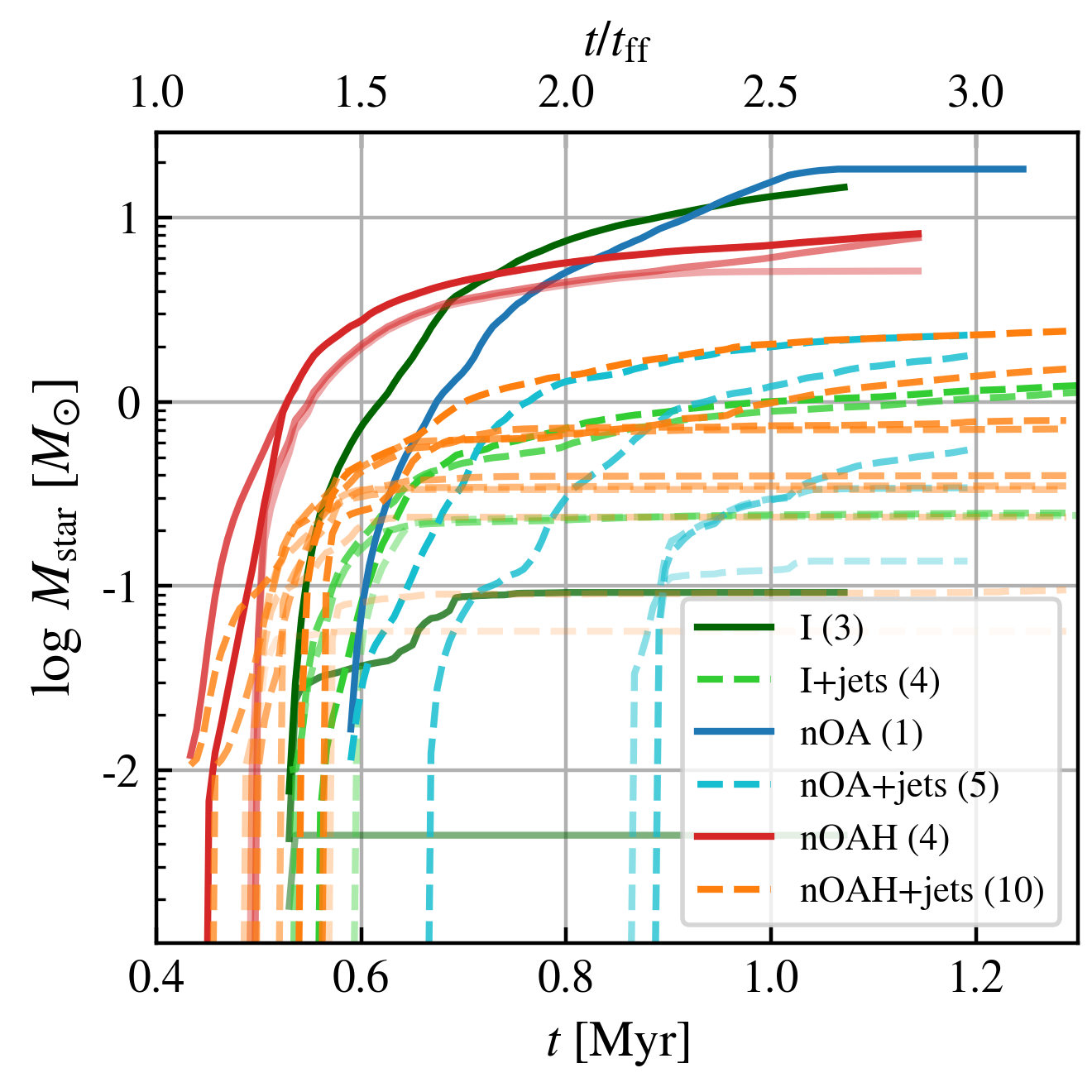}
    \caption{Mass evolution of stars formed in all models. The elapsed time is measured from the beginning of the star-forming calculations, e.g., after completion of the \textsc{TURBSPHERE} stirring calculations.}
    \label{fig:sink_mass}
\end{figure}

Figure~\ref{fig:sink_mass} shows the mass evolution of all stars formed across all six models. Stars formed in models without jet feedback are shown as solid lines, while stars formed in models with jet feedback are shown as dashed lines. Star formation begins roughly $\sim$1$-$1.5 $t_{\rm ff}$ ($\sim$0.4$-$0.6 Myr) after self-gravity is turned on and largely concludes within $\sim$0.2 Myr, with the exception of the non-ideal OA+jets model, which experiences a second burst of star formation another $\sim$0.2 Myr after the first two stars form. These secondary stars form from the fragmentation of a large rotating structure associated with the first stars (discussed further in Section \ref{section_core_filament_fragmentation}). Star formation begins earliest in the two non-ideal OAH(+jets) models and latest in the two non-ideal OA(+jets) models. The total number of stars formed ranges from 1 (non-ideal OA) to 10 (non-ideal OAH$+$jets). The star formation outcomes for each model are summarized in Table~\ref{tab:model_results}.

The calculations with sub-grid protostellar jet feedback are initialized from the final snapshots preceding star formation in the calculations with the same MHD physics but without jets, so all differences between the jet and no-jet models are due to jet feedback. We observe two clear trends once protostellar feedback is included. First, for each set of MHD physics, models with jets consistently form more stars than models without jets. Second, stars in models without jets are generally more massive, where the most massive stars in the models without jets are roughly an order of magnitude larger than those in the models with jet feedback. These trends are consistent with the radiation hydrodynamical studies of \cite{hansen_2012}, who showed that jet feedback increased fragmentation and reduced the mass accretion rate. As discussed in \cite{gus_2021}, protostellar jets change the accretion history of stars not only by removing some fraction of the accreted gas to be ejected as jet feedback but also by disrupting the flow of gas around newly-formed stars.

\subsubsection{Core/filament fragmentation and multiple system formation}\label{section_core_filament_fragmentation}
We observe a variety of dense structures leading to star formation, which are broadly separable into two categories: (asymmetric) cores and filaments; similar geometries of core formation and fragmentation were previously noted by \cite{He_Ricotti_2023}. Some multiple systems form from the fragmentation of a single core, such as the binary system in the ideal MHD model. Here, three stars form within $\sim$6 kyr and $\sim$70$-$220 AU of one another; the lowest-mass member of the triple is almost immediately ejected from the system. In other models, stars form from distinct cores, separated by $\mathcal{O}(10-100$ kyr$)$ in time and $\mathcal{O}(10^3-10^4$ AU$)$ in space, then migrate to become members of bound systems.
\begin{figure}
    \includegraphics[width=\columnwidth]{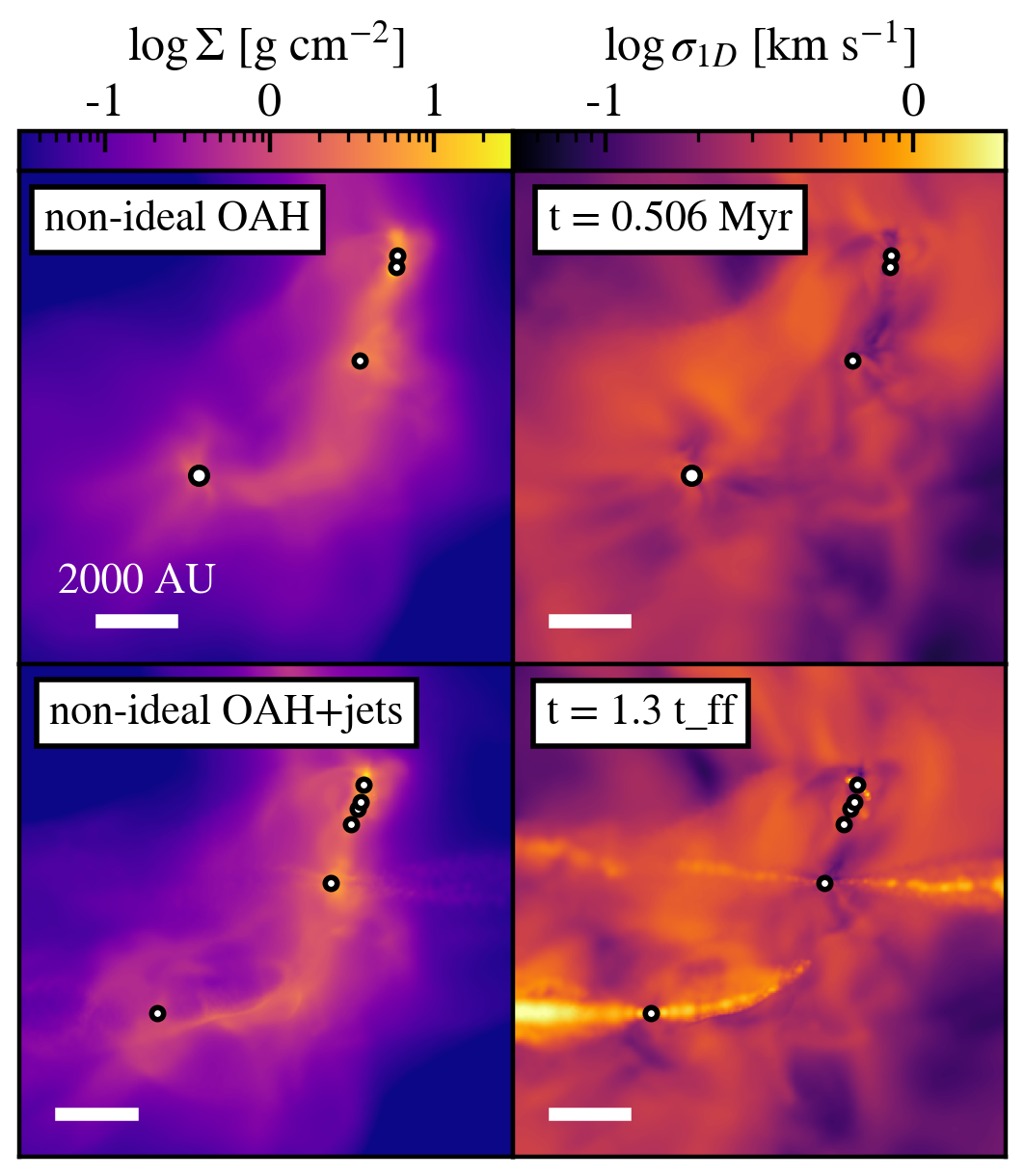}
    \caption{Column density (left column) and line-of-sight 1D velocity dispersion (right column) showing the formation of stars along a filament. Top row: non-ideal OAH; bottom row: non-ideal OAH$+$jets. Markers indicate the positions of stars, with marker size proportional to stellar mass.}
    \label{fig:proj_density_filament_fragmentation}
\end{figure}
Figure~\ref{fig:proj_density_filament_fragmentation} shows stars forming like beads on a string along a filament in the two non-ideal OAH models at the same point in their evolution. The top row shows the model without jets, which has already formed its four total stars (these will later form two interacting binary systems); the bottom row shows the model with jet feedback, which has formed six out of its eventual 10 stars. This figure demonstrates the influence of jet feedback on gas density and structure and thereby the differences in fragmentation outcomes among the models. The jets themselves are not clearly visible in the surface density, although some bow shocks extending in the horizontal direction are visible in the bottom-left panel of 
Figure~\ref{fig:proj_density_filament_fragmentation} and the gas distribution is somewhat more structured. The jets are clearly visible in the velocity dispersion shown in the bottom-right panel, however. 
\begin{figure*}
	\includegraphics[width=2\columnwidth]{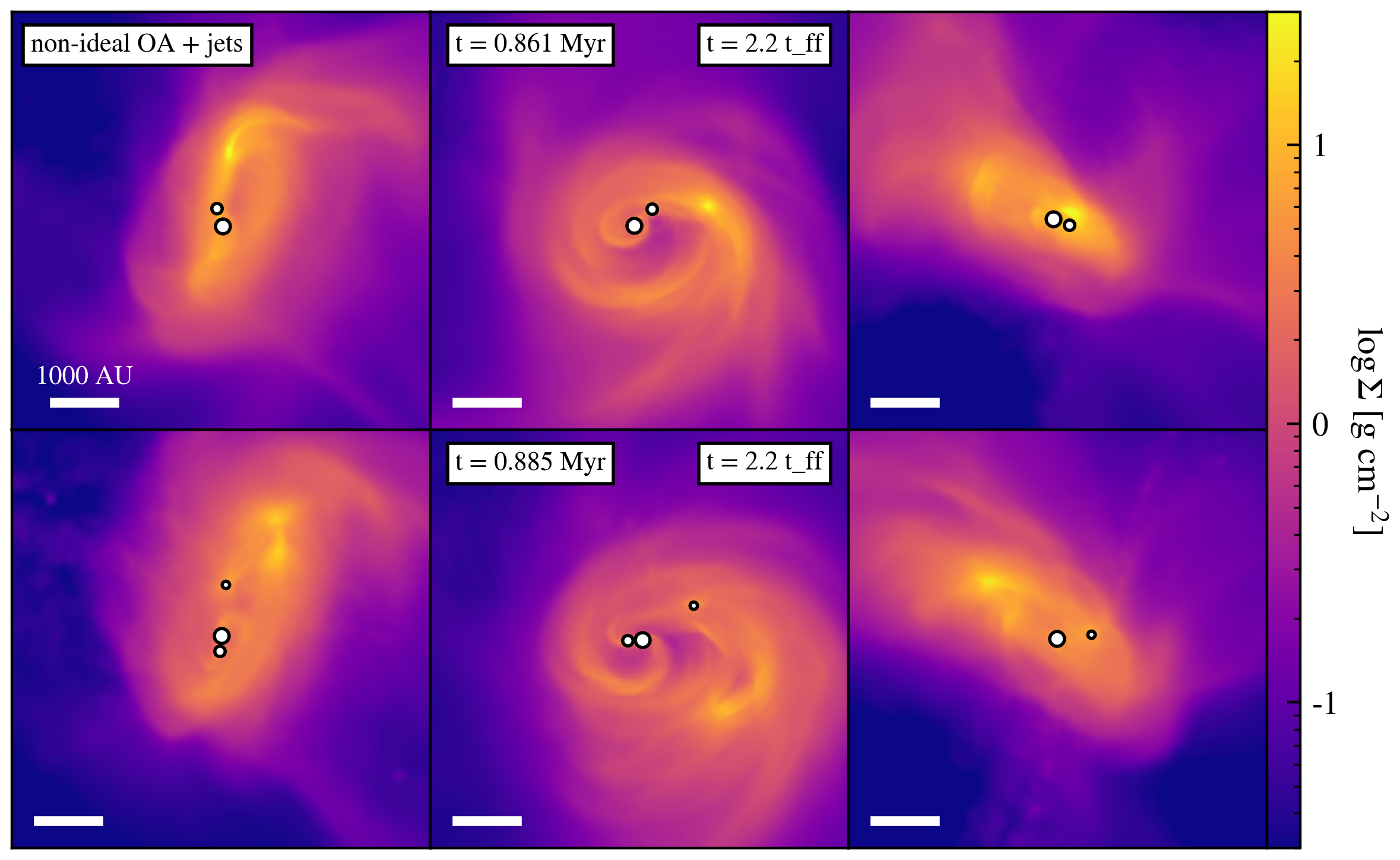}
    \caption{Column density projected along the $x$, $y$, and $z$-axes showing star formation in a rotating core of radius $\sim$2000 AU. Top row: column density just prior to formation of third star. Bottom row: column density just prior to formation of fourth and fifth stars. Markers indicate the positions of stars, with marker size proportional to stellar mass.}
    \label{fig:proj_density_rotating_core_fragmentation}
\end{figure*}

Figure~\ref{fig:proj_density_rotating_core_fragmentation} shows a rotating structure of radius $\sim$2000 AU in the column density snapshots of the non-ideal OA$+$jets model just prior to the formation of the third star (the dense condensation visible in the top row) and the formation of the fourth and fifth stars (the two dense condensations visible in the bottom row). These three stars likely form due to gravitational instability. We note that this structure is not identified as a ``disk” in our analysis, as the gas is below the $n > 10^9$ cm$^{-3}$ density threshold used for disk identification. However, it matches our disk identification criteria if we lower the density threshold to $n > 10^8$ cm$^{-3}$, motivating us to describe this structure as a ``rotating envelope" and to compare the temporal evolution of similar lower-density rotating structures in the other models (see Sections \ref{section_disk_identification} and \ref{section_envelope_identification} for the criteria used to define disks and the extension to lower-density envelope structures). Further analysis of this structure is presented in Appendix \ref{appendix_rotating_envelope}.

\subsection{Disk formation and evolution}\label{section_disk_formation_and_evolution}
We describe our criteria for identifying disks in our calculations and present an overview of the disks formed in our models. We then present the temporal evolution of bulk disk properties, degree of angular momentum and magnetic field alignment, and radial profiles. We also include analysis of the temporal evolution of the lower-density rotating envelope material, which we identify by using a lower density threshold in our disk detection algorithm.
\subsubsection{Disk identification criteria}\label{section_disk_identification}
Following \cite{joos_2012}, we identify disks based on a number of physically-motivated selection criteria:
\begin{itemize}
    \item Disk material must be above a certain density threshold: $n > 10^9$ cm$^{-3}$, where $n$ is the number density;
    \item Disk material must be in Keplerian rotation and near hydrostatic equilibrium: $v_{\phi} > 2 v_r$ and $v_{\phi} > 2 v_z$, where $v_r$, $v_{\phi}$, and $v_z$ are the radial, vertical, and azimuthal velocities, with the rotation axis being the direction of the net angular momentum of the gas within $\sim$6000 AU of the star(s) associated with the disk;
    \item Rotational support must exceed thermal support: $1/2 \rho v_{\phi}^2 > 2 P_{\rm th}$, where $\rho$ is the gas density and $P_{\rm th}$ is the thermal pressure; this check is intended to exclude the rotating central adiabatic core.
\end{itemize}
In each simulation snapshot, we first identify all disks associated with individual stars in order of least to most massive. We then identify all circummultiple disks associated with binary and higher-order systems. For ``isolated” stars and stellar systems (i.e., separated from other stars by at least $6000$ AU), we include all gas within a distance $r_{\rm max} = 6000$ AU of the central star(s) when searching for disks. For stars with neighbors closer than $6000$ AU, we set $r_{\rm max} = 0.65 r_{\rm nearest}$, where $r_{\rm nearest}$ is the distance to the closest star or center of mass of a bound stellar system; this prevents material visibly associated with a neighboring disk from being identified as disk material.

Most disks persist for several hundred kyr; however, a few transient disks appear for a snapshot or two ($\sim$6$-$12 kyr) before being disrupted by merging stellar systems and/or jet feedback from nearby stars; we exclude these short-lived disks from further analysis of the disk property evolution.
\begin{figure*}
	\includegraphics[width=2\columnwidth]{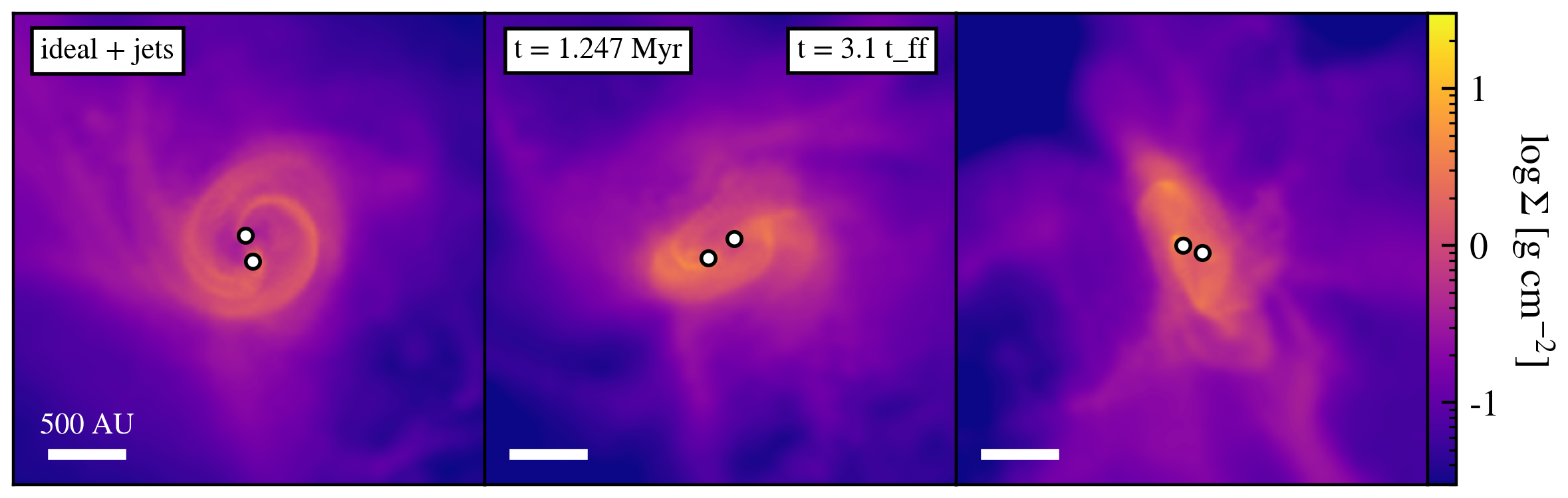}
    \caption{Column density projected along the $x$, $y$, and $z$-axes showing the low-mass quadruple disk that forms in the ideal$+$jets model. While only two markers representing stars are visible in each projection, there are four stars in two tight binaries in this quadruple system.}
    \label{fig:proj_density_Ij_disk}
\end{figure*}
\begin{figure*}
	\includegraphics[width=2\columnwidth]{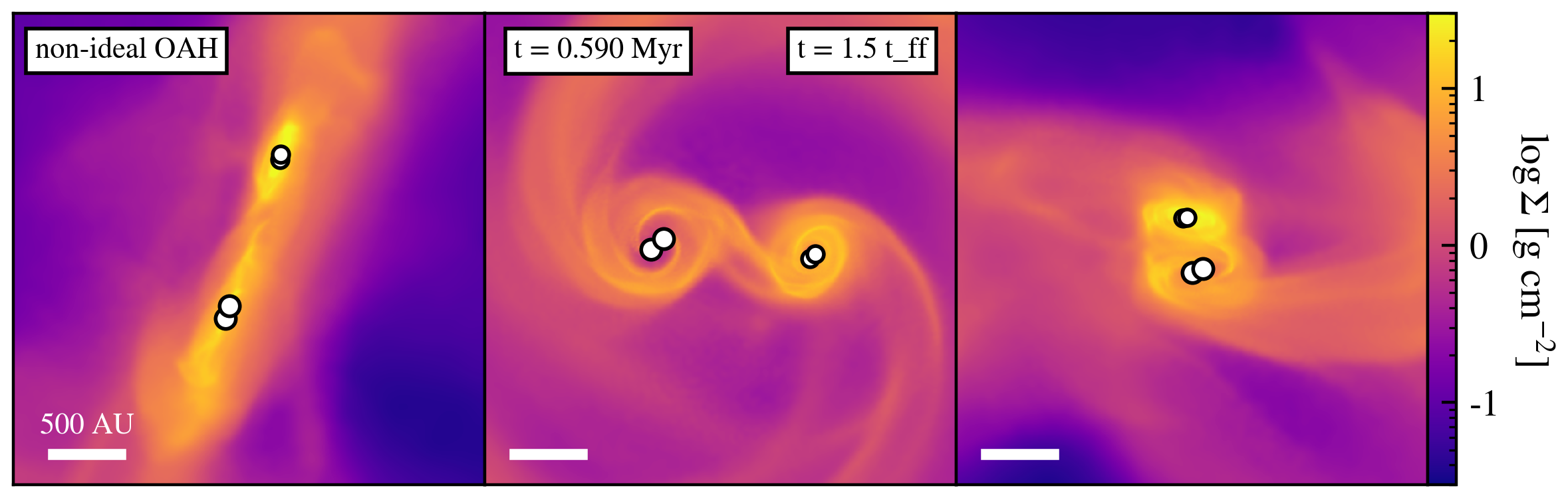}
    \caption{Column density projected along the $x$, $y$, and $z$-axes showing the two interacting binary systems that form in the non-ideal OAH model.}
    \label{fig:proj_density_nOAH_binaries}
\end{figure*}

Using the above criteria, disks form in all of our models. However, the quadruple ``disk" identified in the ideal$+$jets model is notably less massive ($M_{\rm disk} \lesssim 10^{-3} \; M_{\odot}$) than the disks formed in the other models and contains less than 100 gas cells (often less than 10) in nearly all snapshots. We include this disk in our analysis of bulk disk properties, such as mass and radius, with the caveat that it is likely unresolved, but exclude it from more detailed analysis of disk radial structure in Section \ref{section_disk_radial_profiles}.
Figure~\ref{fig:proj_density_Ij_disk} shows the column density of this disk.

We observe a diverse set of disks and disk-hosting stellar systems. The non-ideal OA model forms a single massive ($\sim$18 $M_{\odot}$) star and hosts the most massive disk. The photoionizing radiation from this star eventually forms an HII region, which rapidly expands and clears out the remaining gas surrounding this star. However, this occurs $\sim$130 kyr after the disk disappears, having largely been accreted onto the star. The ideal and non-ideal OAH models without jet feedback also form massive stars (ideal: $\sim$10 $M_{\odot}$; non-ideal OAH: $\sim$8 $M_{\odot}$), but no HII regions form prior to the end of the calculations. HII regions therefore do not affect disk evolution in any of our models. The two single disks that form in the non-ideal OA$+$jets model exhibit the most variability in disk mass, possibly as a result of jet feedback from the secondary stars that form in the fragmenting rotating structure surrounding the first two stars, as seen in Figure~\ref{fig:proj_density_rotating_core_fragmentation}. The non-ideal OAH model forms four stars, which become two interacting binary systems separated by $\sim$1000 AU, as seen in Figure~\ref{fig:proj_density_nOAH_binaries}. The stellar masses are very similar, with the stars belonging to a given binary being closest in mass, and the circumbinary disk masses are also quite comparable. As seen in the figure, we observe a bridge of material connecting the two binary disks throughout their evolution. These two binary disks are fed material by two external arcing streams of material, which may be larger spiral arms or ``streamers", seen in observations as asymmetries in protostellar envelopes with total lengths ranging from $\sim$500 AU to over $10^4$ AU \citep{pineda_2023}. The non-ideal OAH$+$jets model forms a single disk at early times. This model forms the most stars and thus has the most chaotic, disruptive jet feedback during this period. A number of other circumstellar and circummultiple disks form during the first $\sim$100 kyr following the onset star formation but persist for only a single snapshot or two before being disrupted by stellar interactions and jet feedback. However, about 200 kyr after the single disk disappears, a circumbinary disk forms and persists for at least another $\sim$400 kyr. This binary system includes the star associated with the earlier single disk, and accretes material not from the natal core but from late-stage infalling gas which has been kicked up by jet feedback.

\subsubsection{Envelope identification}\label{section_envelope_identification}
We adopt the same density threshold ($n > 10^9$ cm$^{-3}$) for disk identification as in \cite{joos_2012}, who use a density threshold to avoid including large spiral arms connected to the disk and to obtain more realistic estimates of the disk shape. However, our projected column density plots tend to show that the disks in the non-ideal MHD models are embedded in large, rotating, lower-density structures extending out to several thousands of AU, which feed into the disks via extended structures resembling spiral arms or streamers; see Fig.~\ref{fig:proj_density_nOAH_binaries} for an example. The projected column density plots of the ideal MHD disks, however, show that these disks tend to be embedded in a more tenuous and filamentary reservoir of material; see, e.g., Fig~\ref{fig:proj_density_Ij_disk}. Motivated by these observations, we lower the density threshold to $n > 10^8$ cm$^{-3}$ and then repeat the disk detection algorithm described in Section~\ref{section_disk_identification} to identify the lower-density rotating material associated with each protostar or protostellar system. Because all of the disks we identify for each model belong to a single bound hierarchical system, we are able to define a common ``rotating envelope" for each model by merging the lower-density rotating material identified for each subsystem. We henceforth call this material the ``envelope," and note that it differs from the disks only by including more of the extended, lower-density material surrounding the disks. The fragmenting rotating envelope discussed in Appendix \ref{appendix_rotating_envelope} is an example of one of these structures.

\subsubsection{Evolution of disk bulk properties}\label{section_bulk_properties}
We examine the evolution of several bulk properties of the disks, such as the mass, radius, disk-to-stellar mass ratio, and aspect ratio. We also compare the evolution of several bulk properties of the lower-density envelope, as defined in Section \ref{section_envelope_identification}.
\begin{figure}
    \includegraphics[width=\columnwidth]{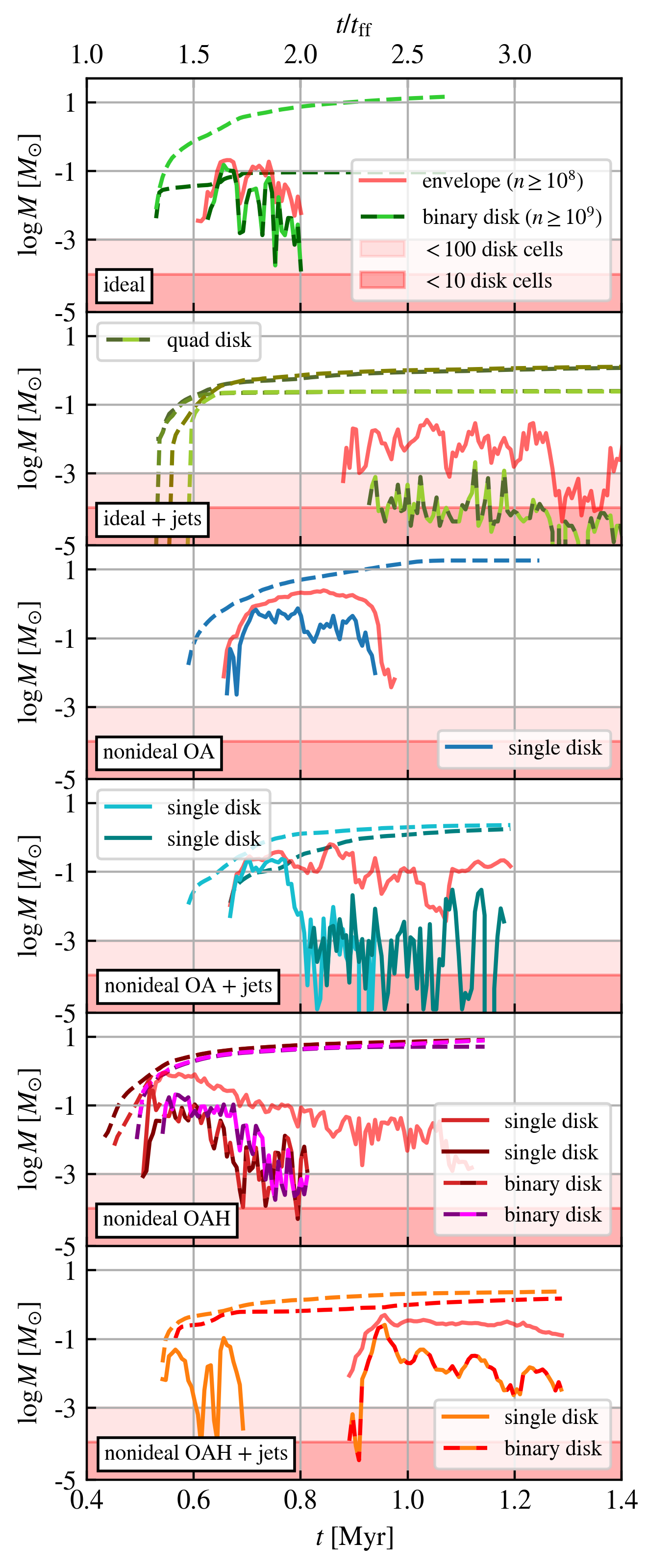}
    \caption{Disk/envelope mass evolution. The elapsed time is measured from the beginning of the star-forming calculations (when self-gravity is enabled). The dotted lines show the masses of the stars associated with each disk. Dashed lines combining multiple colors represent circumbinary and higher-order disks. Disks persisting for less than $\sim$12 kyr are not depicted. Red solid lines in each panel shows the evolution of the envelope mass, i.e., rotating gas with $n > 10^8$ cm$^{-3}$. Red shaded regions indicate when disks contain less than 100 gas cells and may be poorly resolved.} 
    \label{fig:disk_mass}
\end{figure}
\begin{figure}
    \includegraphics[width=\columnwidth]{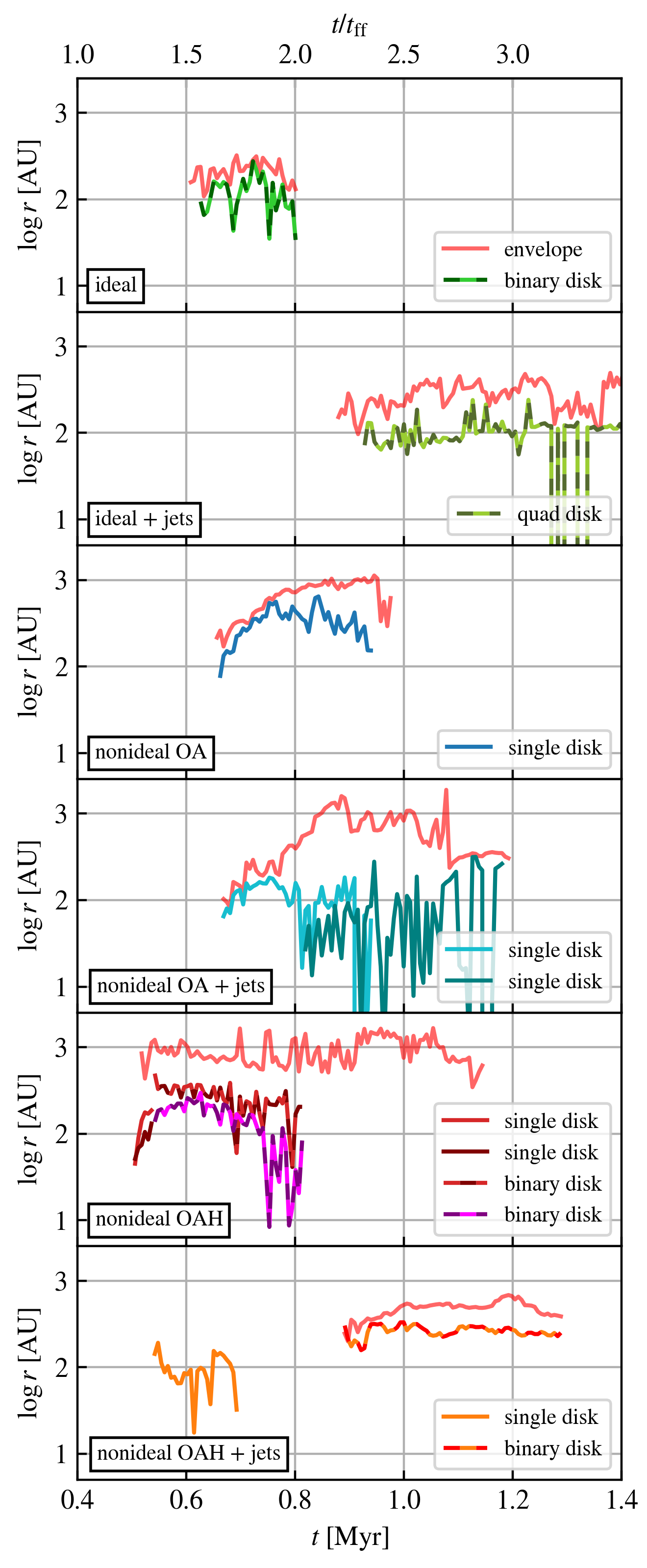}
    \caption{Same as Fig.~\ref{fig:disk_mass}, but showing the evolution of the disk/envelope radius, taken to be the radius containing 63.2 percent of the total disk/envelope mass.}
    \label{fig:disk_radius}
\end{figure}
\begin{figure}
    \includegraphics[width=\columnwidth]{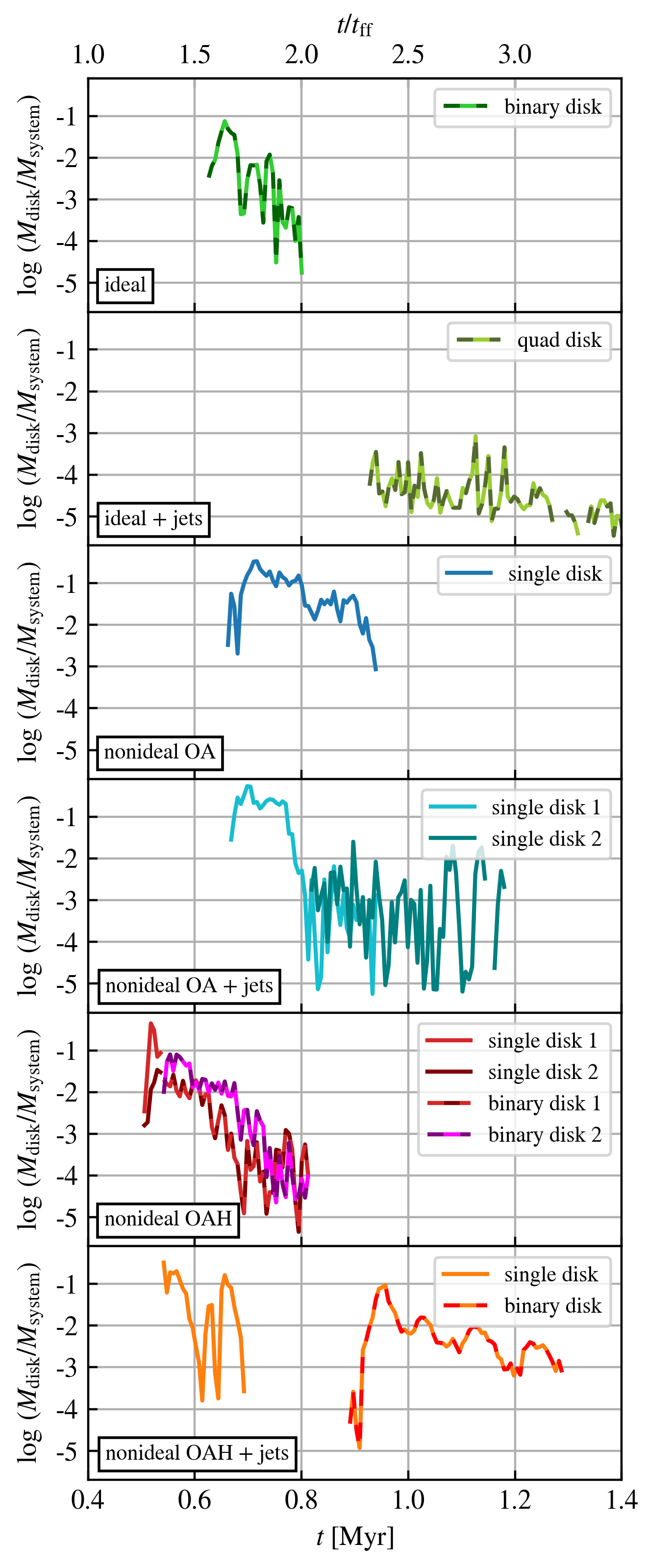}
    \caption{Same as Fig.~\ref{fig:disk_mass}, but showing the evolution of the ratio of disk to stellar mass.}
    \label{fig:disk_mass_ratio}
\end{figure}
\begin{figure}
    \includegraphics[width=\columnwidth]{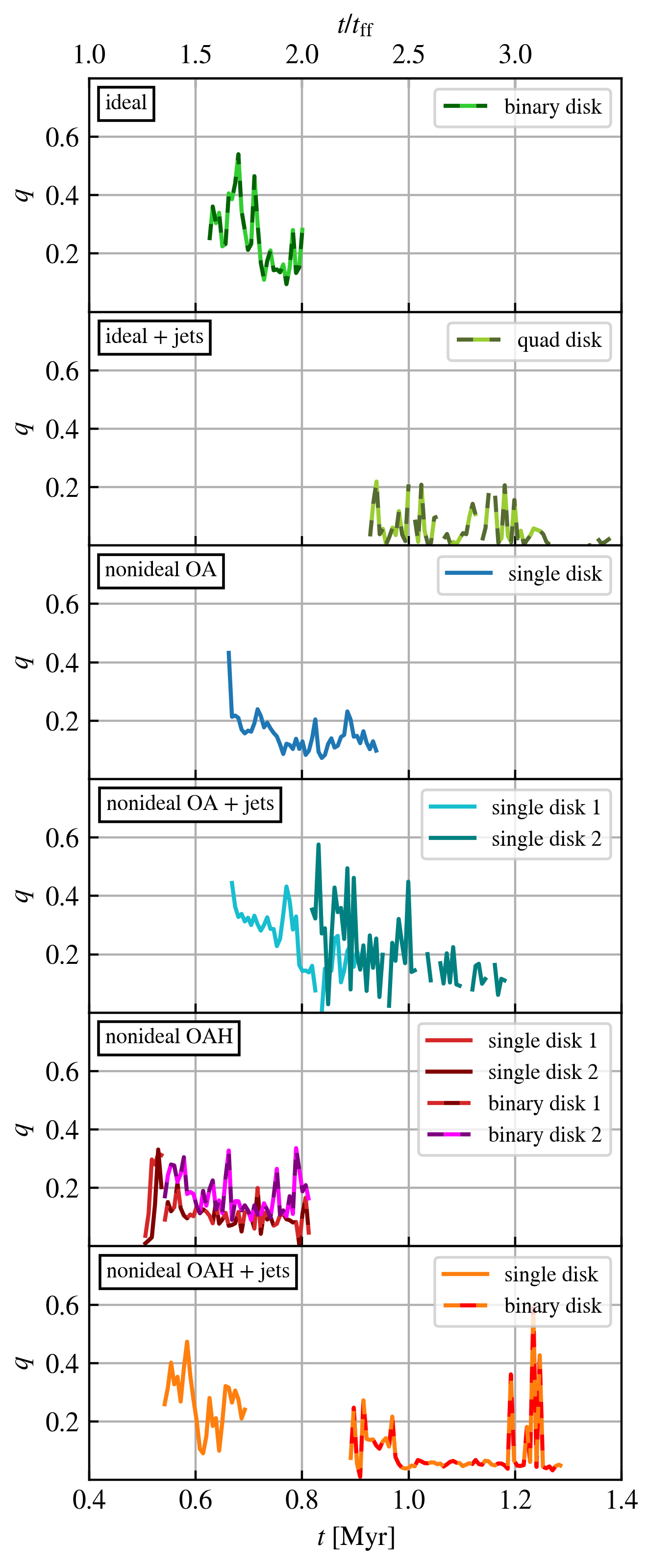}
    \caption{Same as Fig.~\ref{fig:disk_mass} but showing the evolution of the disk aspect ratio.}
    \label{fig:disk_aspect_ratio}
\end{figure}

We define the total disk mass to be the sum of the mass of all gas identified as belonging to the disk (the total envelope mass is likewise defined to be the sum of the mass of all the gas identified as belonging to the envelope). Following \cite{bate_2018}, we then define the disk (envelope) radius to be the radius containing 63.2 percent of the total mass. This is based on a truncated power-law surface density profile derived from models of viscously evolving disks \citep{lyden-bell_pringle_1974, hartmann_1998}, often used to fit observed disks \citep[e.g.,][]{andrews_2010, tazzari_2017}:
\begin{equation}
    \Sigma(r) = \Sigma_c \left(\frac{r}{r_c}\right)^{-\gamma} \exp \bigg[ -\left(\frac{r}{r_c}\right)^{(2-\gamma)}\bigg],
\end{equation}
where $\Sigma_c$ is a normalization factor, $r_c$ is the exponential cut-off radius, and $\gamma$ is the power-law slope.\footnote{Another approach defines the disk radius as the median value of the maximal disk extent in 50 equal-width azimuthal slices \citep[see, e.g., ][]{lebreuilly_2024a}. We find that both criteria give similar disk radii and so we adopt the criterion based on the enclosed disk mass for our analysis.}

Figure~\ref{fig:disk_mass} and Figure~\ref{fig:disk_radius} show the evolution of the disk masses and radii for all disks that persist for $\gtrsim$12 kyr in all models. We also include the evolution of the envelope masses and radii. We find that most disks form within $\sim$100 kyr of the formation of the host stars (see the dashed lines representing stellar masses in Figure~\ref{fig:disk_mass}) and then persist for $\sim$150$-$400 kyr (see Table~\ref{tab:model_results}); the quadruple disk formed in the ideal+jets model (plotted in the second panel) and the binary disk formed in the non-ideal OAH+jets model (plotted in the bottom panel), which form roughly $\sim$0.4 Myr after the onset of star formation, are notable exceptions that are discussed further in Section~\ref{sec:effects_of_jet_feedback}. The disks are typically most massive soon after formation and become less massive with time. As discussed in Section~\ref{section_disk_identification}, the quadruple disk formed in the ideal$+$jets model (second panel) is notably less massive ($M_{\rm disk} \lesssim 10^{-3} M_{\odot}$) than the disks formed in the other models and contains less than 100 gas cells (often less than 10) throughout most of its evolution. The mass of the binary disk formed in the ideal model (top panel) is comparable to the disk masses of the non-ideal models, although the ideal binary disk lifetime (187 kyr) is $\sim$200 kyr shorter than the lifetimes of the non-ideal binary disks, such as those formed in the non-ideal OAH (fifth panel; 361 kyr) and the non-ideal OAH$+$jets (bottom panel; 397 kyr) models.

The differences between the models assuming ideal vs. non-ideal MHD, as well as those due to the inclusion of jet feedback, become much more pronounced when considering the mass evolution of the larger, lower-density envelopes. The envelopes generally become detectable at or shortly prior to ($\lesssim$10 kyr) the time of disk formation. The initial jet feedback from the 10 protostars, which form in the non-ideal MHD$+$jets model is so chaotic that we are unable to define a coherent envelope until the formation of the binary disk at $t \sim 0.9$ Myr; this does not, however, prevent us from making the following observations.

Comparing the three models without subgrid jet feedback (first, third, and fifth panels), we clearly observe that the envelope in the ideal model is less massive and exists for a significantly shorter period of time than the envelopes in the non-ideal OA and the non-ideal OAH models. The ideal envelope is present for 193 kyr; the non-ideal OA envelope is present for 320 kyr; and the non-ideal OAH envelope is present for 602 kyr. The ideal envelope is also only slightly more massive than the ideal binary disk, suggesting that this disk lacks an extensive envelope of material to accrete from, whereas the non-ideal OA and non-ideal OAH envelopes are typically one to two orders of magnitude more massive than the enclosed disks and thus provide large reservoirs of material to replenish the disk and extend the disk lifetime. These results indicate that magnetic braking is more efficient in the ideal MHD case and that non-ideal MHD effects are important on these scales.

Comparing the three models with jet feedback (second, fourth, and sixth panels), we first observe that jet feedback tends to extend the envelope lifetime and stabilize the envelope mass, likely due to the jet feedback replenishing the mass reservoir available to be accreted by the envelope and disk. The envelope in the ideal$+$jets model is detectable for 602 kyr (vs. 193 kyr for the ideal model); the envelope in the non-ideal OA$+$jets model is present for 524 kyr (vs. 320 kyr for the non-ideal OA model); and while the non-ideal OAH$+$jets envelope is detectable for only 397 kyr (vs. 602 kyr for the non-ideal OAH model), this is partly due to the fact that early jet feedback disrupts the envelope during the initial star-forming period. However, when comparing the two non-ideal OAH models, with and without jet feedback, we see that the envelope mass steadily declines from $\sim$1.0 to $\sim$0.001 $M_{\odot}$ in the model without jet feedback, but remains rather constant at $\sim$0.2$-$0.3 $M_{\odot}$ in the model with jet feedback. When comparing the envelope mass among the models with jet feedback, we clearly see that the envelope in the ideal$+$jets model is generally one to two orders of magnitude less massive than the envelopes in the non-ideal OA$+$jets and the non-ideal OAH$+$jets models. 

We do not focus on binary properties in this work, but we note that Figure~\ref{fig:disk_mass} contains examples of binaries with a range of mass ratios between the two stellar companions. The masses of stars hosting disks are shown as dashed lines in this figure. For example, the binary system hosting the circumbinary disk formed in the ideal model shown in the top panel of Figure~\ref{fig:disk_mass} has a very uneven mass ratio, with the more massive star reaching $\sim$10$M_{\odot}$ by the time the disk has been accreted onto the binary system, while the smaller star stagnates at only $\sim$0.1$M_{\odot}$. This binary system initially forms as a triple system, but the third star is ejected within $\sim$6 kyr after the formation of the system. The ideal+jets model shown in the second panel forms a hierarchical quadruple system composed of two tight binaries (see Figure~\ref{fig:proj_density_Ij_disk}), and the masses of the two stars in each of the binaries are nearly identical. The non-ideal OAH model shown in the fifth panel also forms two interacting tight binaries (see Figure~\ref{fig:proj_density_nOAH_binaries}), which may be viewed as a bound quadruple system, and again the masses of the members of each binary are very similar, though they begin to deviate after about $t \sim 1$ Myr, after all disks have been accreted. The binary formed at about $t \sim 0.9$ Myr in the non-ideal OAH+jets model shown in the bottom panel presents an intermediate case. The mass ratio is only about $M_2/M_1 \sim 0.4$ when the binary disk first forms, where $M_1$ and $M_2$ are the masses of the primary and secondary stars, but increases to $M_2/M_1 \sim 0.65$ by the end of the disk lifetime. Although this is a small sample of disks, we find a diverse set of binary mass ratios.

As shown in Figure~\ref{fig:disk_radius}, disk radii range from a few 10 to a few 100 AU, broadly consistent with observations of circumstellar and circummultiple disks (see discussion in Section \ref{section_observations}). The overall trend in disk radius generally remains flat, but there are some variations in disk size, which are due to a combination of interactions and perturbations with other stars and disks as well as accretion variability. Comparing the envelope radii shows that the ideal envelopes tend to be quite compact: the ideal and ideal$+$jets models have average envelope radii of $\sim$200 AU and $\sim$300 AU, respectively, while the average envelope radii of the four non-ideal models range from $\sim$500 to $\sim$1000 AU. The ideal envelopes are thus both significantly less massive and more compact than the non-ideal envelopes.

Figure~\ref{fig:disk_mass_ratio} shows the evolution of the disk-to-stellar-mass ratio. At early times, the disk-to-stellar mass ratio exceeds $M_{\rm disk}/M_{\rm star} \gtrsim 0.1$ for many of the disks; disks are most likely to be gravitationally unstable and develop strong episodes of spiral activity during this early evolutionary stage \citep{kratter_lodato_2016}. However, the ratio then decreases by many orders of magnitude over the disk lifetime as the star continues to grow while the disk becomes less massive. No disk fragmentation is observed (see also Section \ref{section_disk_stability}). 

To compute the disk aspect ratio $q$, we center the coordinates of all disk cells about the disk center of mass, compute the covariance matrix of the centered coordinates, and obtain the eigenvalues $\lambda_j$ of the covariance matrix. The disk aspect ratio is then defined to be $q = \min_j(\sqrt{\lambda_j})/\max_j(\sqrt{\lambda_j})$. Figure~\ref{fig:disk_aspect_ratio} shows the evolution of the aspect ratio of all disks persisting for $\gtrsim$12 kyr.  We see that most of the disks have an aspect ratio $q \lesssim 0.2$ for most of their evolution, and that the aspect ratio tends to decrease with disk age.

We note that the variability in the disk masses and radii is physical and not simply due to statistical fluctuations in the number of included cells. Disk morphologies visibly vary over multiple timesteps, and fluctuations in disk masses closely follow changes to the disk specific angular momentum. Disk masses and radii at early times (e.g., during the Class 0/I phases) are controlled by the infalling angular momentum from the envelope and/or feeding filaments, and as it is not guaranteed that the angular momentum vector of the infalling material will remain constant, it is likely that disk sizes do not grow monotonically with time \citep[see, e.g., the review by][]{kratter_lodato_2016}.

\subsubsection{Angular momentum and magnetic field misalignment}
\begin{figure}
    \includegraphics[width=\columnwidth]{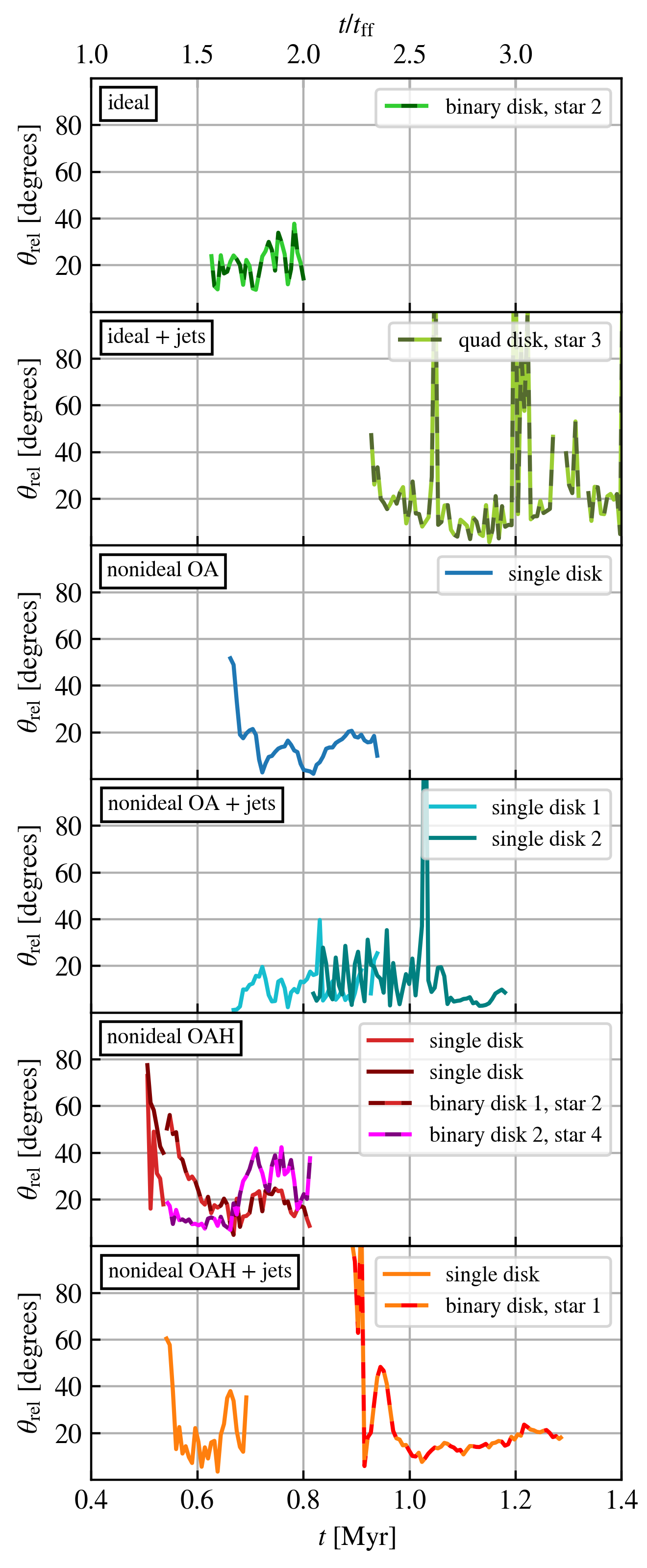}
    \caption{Evolution of the misalignment angle between the disk and sink particle specific angular momentum vectors, where for multiple systems we consider only the angular momentum of the most massive sink particle. The sink particle angular momentum vector is not quite the same as the stellar angular momentum vector due to unresolved processes but represents our best available estimate.}
    \label{fig:disk_sink_misalignment}
\end{figure}
\begin{figure}
    \includegraphics[width=\columnwidth]{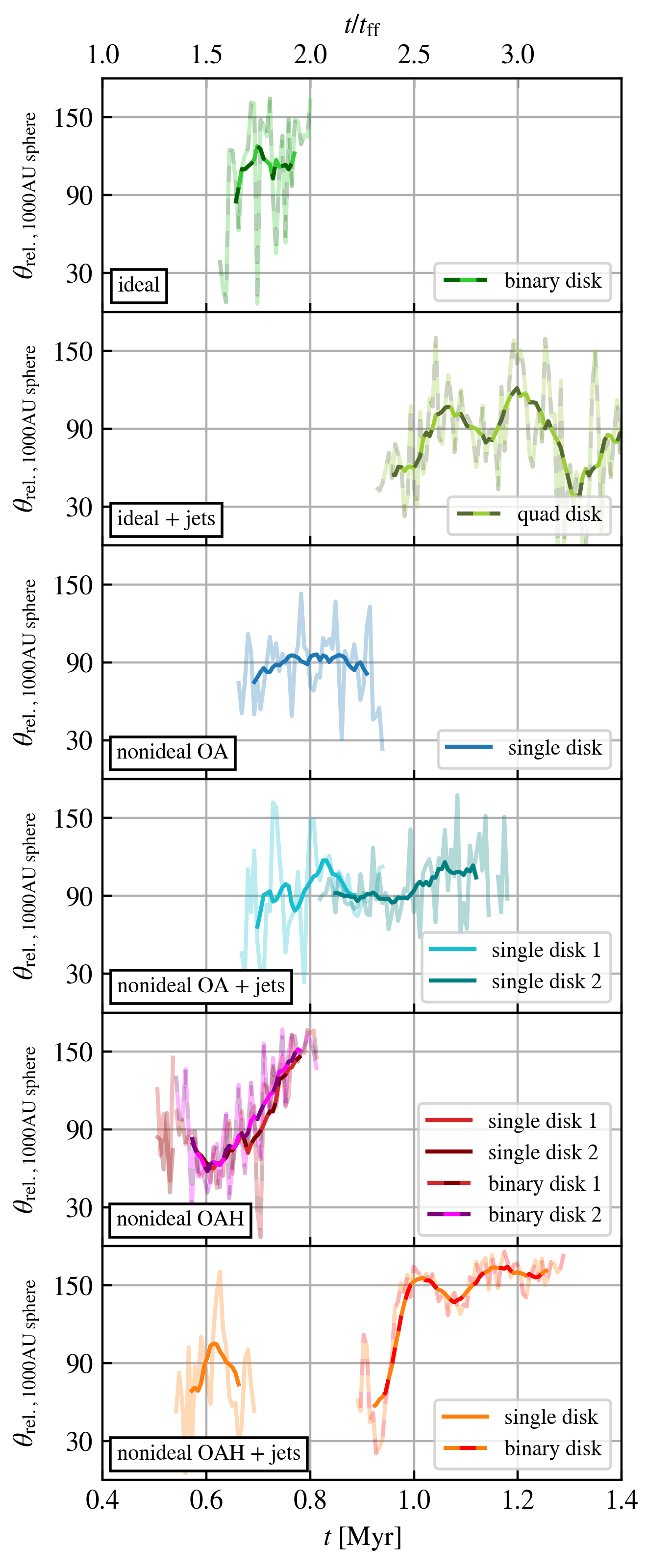}
    \caption{Evolution of the misalignment angle between the disk angular momentum vector and the mean local magnetic field direction for all disks formed. The mean local magnetic field is averaged over all gas within a 1000 AU radius centered at the disk center of mass. The transparent lines show the misalignment angle for all snapshots, while the opaque lines show the moving average computed over intervals of 10 snapshots ($\sim$60 kyr).}
    \label{fig:disk_B_misalignment}
\end{figure}
Figure~\ref{fig:disk_sink_misalignment} shows the relative angle between the sink particle and disk specific angular momentum vectors\footnote{The sink particle angular momentum includes the angular momentum of all gas cells accreted by the sink particle. The sink particle angular momentum is not, strictly speaking, the same as the stellar angular momentum as there are unresolved processes which may affect the angular momentum transport between the gas and the star; however, it represents our best estimate of the stellar angular momentum direction. We clarify that this is not orbital angular momentum.}, used as a proxy for the disk and stellar rotation axes, where for circummultiple disks, we use the angular momentum vector of the most massive star associated with the disk. We see that the angular momentum vectors tend to be generally well-aligned ($\theta_{\rm rel} \lesssim 20$ degrees) for most of the disk evolution. Several disks are initially quite misaligned with their stars ($\theta_{\rm rel} \sim$60$-$80 degrees), and then become more aligned with time; this is expected as disks continue to transfer angular momentum to the central star(s). Stars that begin with wide separations are likely to be initially misaligned and then align as their separation decreases, since they will be accreting gas from the same reservoir that has same net angular momentum direction \citep[e.g.,][]{GuszejnovRaju2023}. Sharp excursions away from alignment tend to be produced by dynamical interactions, which perturb the disk.

Figure~\ref{fig:disk_B_misalignment} shows the evolution of the angle between the disk angular momentum vector and the direction of the mean local magnetic field, where the mean local magnetic field is the field averaged over a 1000 AU sphere centered at the disk center of mass. We note that most of the variability in Figure~\ref{fig:disk_B_misalignment} is due to changes in the orientation of the local magnetic field; the disk angular momentum vector changes orientation on much longer timescales. To reduce the scatter, we also plot the moving average, computed over intervals of 10 snapshots (approximately 60 kyr). The disks formed in the ideal MHD models show the most variability in the alignment with the local magnetic field, both from snapshot to snapshot as well as over the long-term average. The disks formed in the models with non-ideal MHD including only Ohmic dissipation and ambipolar diffusion show a tendency to remain aligned perpendicular to the local magnetic field, albeit with some significant variability ($\sim$$\pm 60^{\circ}$). The long-lived binary disks formed in the non-ideal MHD models with the Hall effect included, meanwhile, show a tendency for increasing misalignment with the local magnetic field, reaching angles $\gtrsim$$150^{\circ}$. We discuss this observation further in Section~\ref{discussion_nonideal_mhd}.

\subsubsection{Which stars host disks?}
\begin{figure}
    \includegraphics[width=\columnwidth]{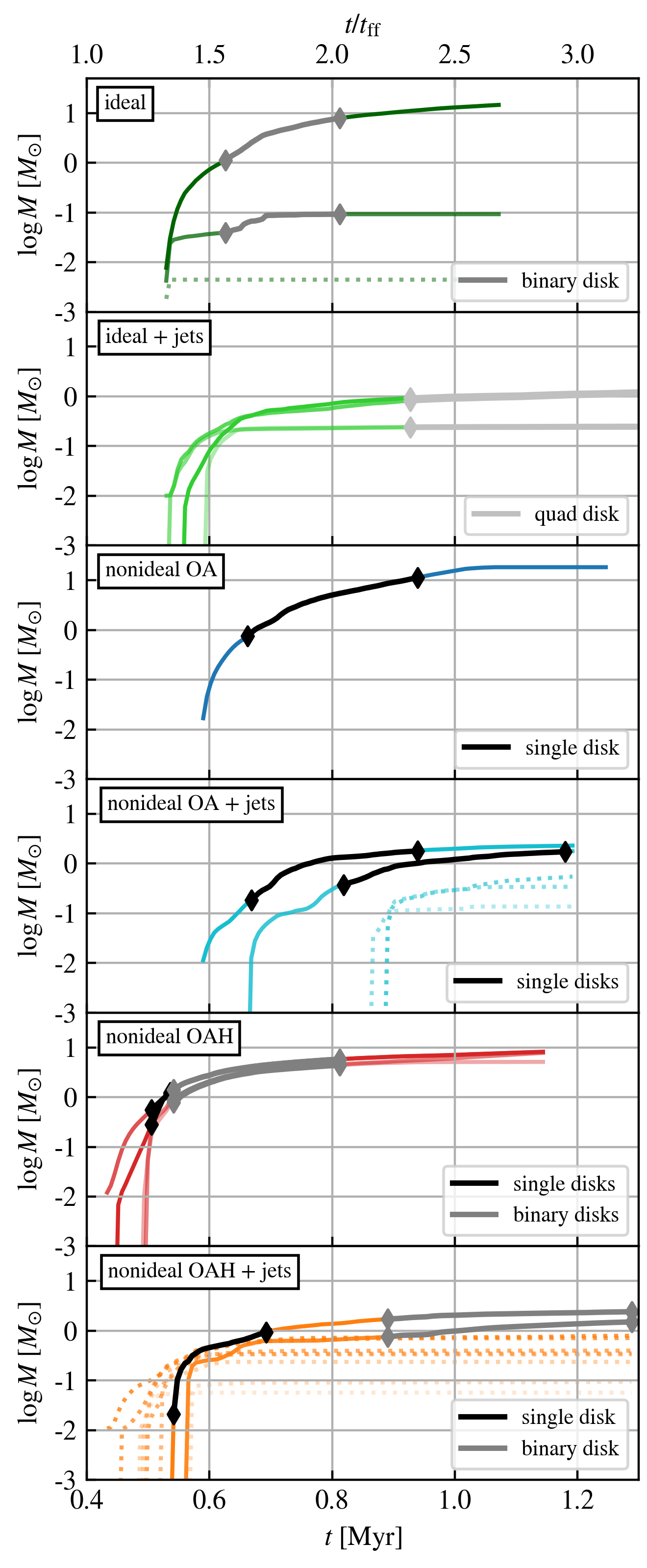}
    \caption{Mass evolution of all stars in all models, highlighting disk presence. Solid colored lines correspond to stars associated with a disk at some point during the calculations; dotted lines correspond to stars with no disks. Black and gray line sections indicate the presence of disks. Diamond markers indicate the first and last snapshots in which a given disk exists.}
    \label{fig:sinks_with_disks}
\end{figure}

Disks form in all models, but not all stars are associated with disks. Figure~\ref{fig:sinks_with_disks} again shows the mass evolution of all stars across all models but now emphasizing the stars hosting disks. We see that, by the end of the simulation, the stars with disks are consistently the most massive stars in that model. The smallest stars, meanwhile, are typically either components of close binaries or else are dynamically ejected into the low-density gas, where there is little material available for disk formation. It is also possible that these stars host disks that fall below our disk mass criterion (i.e., more than 100 gas cells belonging to the disk, or $M_{\rm disk} \geq 0.001 M_{\odot}$). Interestingly, most of the stars that eventually have sizable disks do not have disks, as identified by our disk criteria, until $\sim$0.1 Myr after their formation. This is partially because the early accreting material is turbulent and because these young protostars have masses below $0.1 M_{\odot}$, at which point the amount of bound mass is small.

The disk lifetimes, ranging from $\lesssim$200 to $\gtrsim$400 kyr, are consistent with estimated Class 0/I lifetimes, during which protostars are embedded in their natal envelopes \citep{dunham_2014}.

\subsubsection{Disk stability}\label{section_disk_stability}
A protostellar disk may become gravitationally unstable and fragment to form companion stars and/or planets. The standard reference for estimating the onset of gravitational instability in differentially-rotating, self-gravitating disks is the Toomre $Q$ parameter \citep{toomre_1964}:
\begin{equation}
    Q = \frac{c_s \kappa}{\pi G \Sigma},
\end{equation}
where $c_s$ is the sound speed, $\kappa$ the epicyclic frequency (equal to the angular velocity in a Keplerian disk), $G$ the gravitational constant, and $\Sigma$ the surface density. The disk is stable against fragmentation for $Q > 1$. As $Q \rightarrow 1$, self-gravity becomes increasingly important and the disk becomes unstable due to the growth of spiral density waves.

\begin{figure}
	\includegraphics[width=\columnwidth]{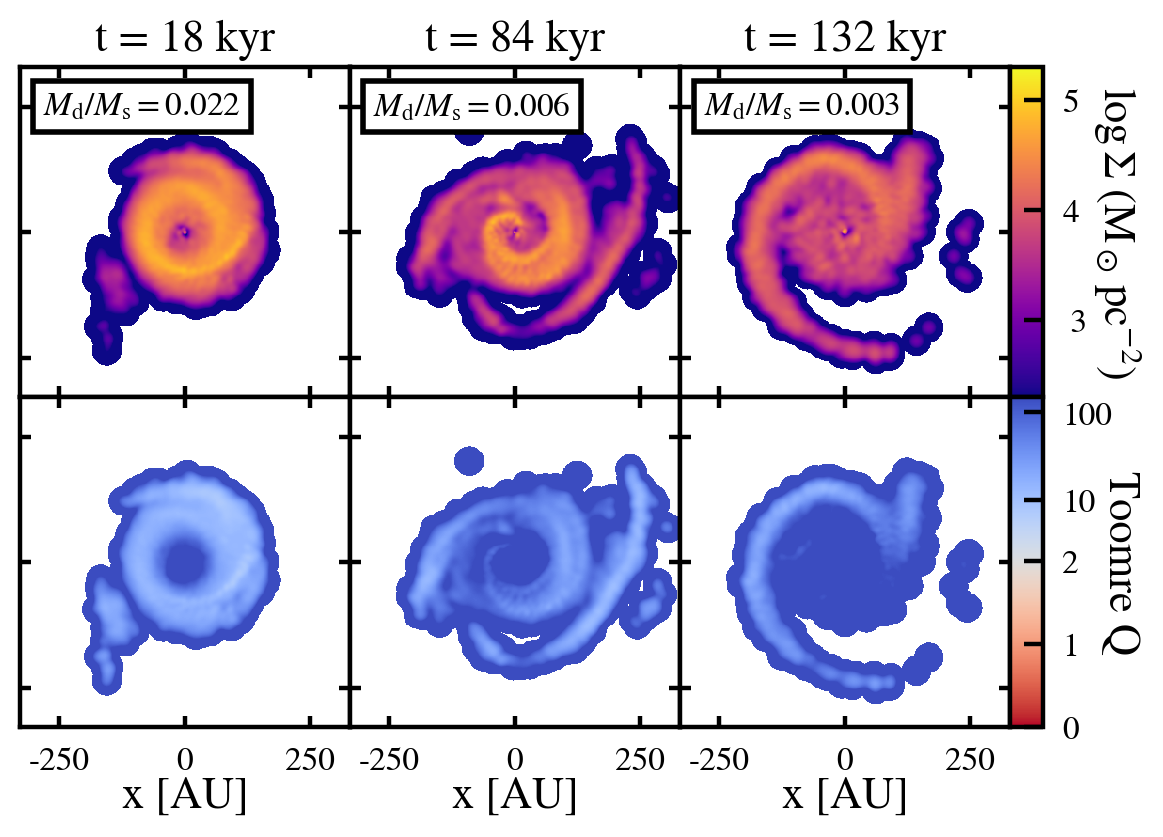}
    \caption{Projected surface density (top row) and Toomre Q (bottom row) for the binary disk formed in the ideal MHD model. We project along the direction of the net angular momentum of the gas, and use a linear colormap normalization between $Q=0$ and $Q=2$ and a logarithmic normalization for $Q > 2$. The disk-to-stellar mass ratio $M_{\rm d}/M_{\rm s}$ is reported in the upper left corner of each surface density panel.}
    \label{fig:toomre_Q_ideal_binary}
\end{figure}

\begin{figure}
	\includegraphics[width=\columnwidth]{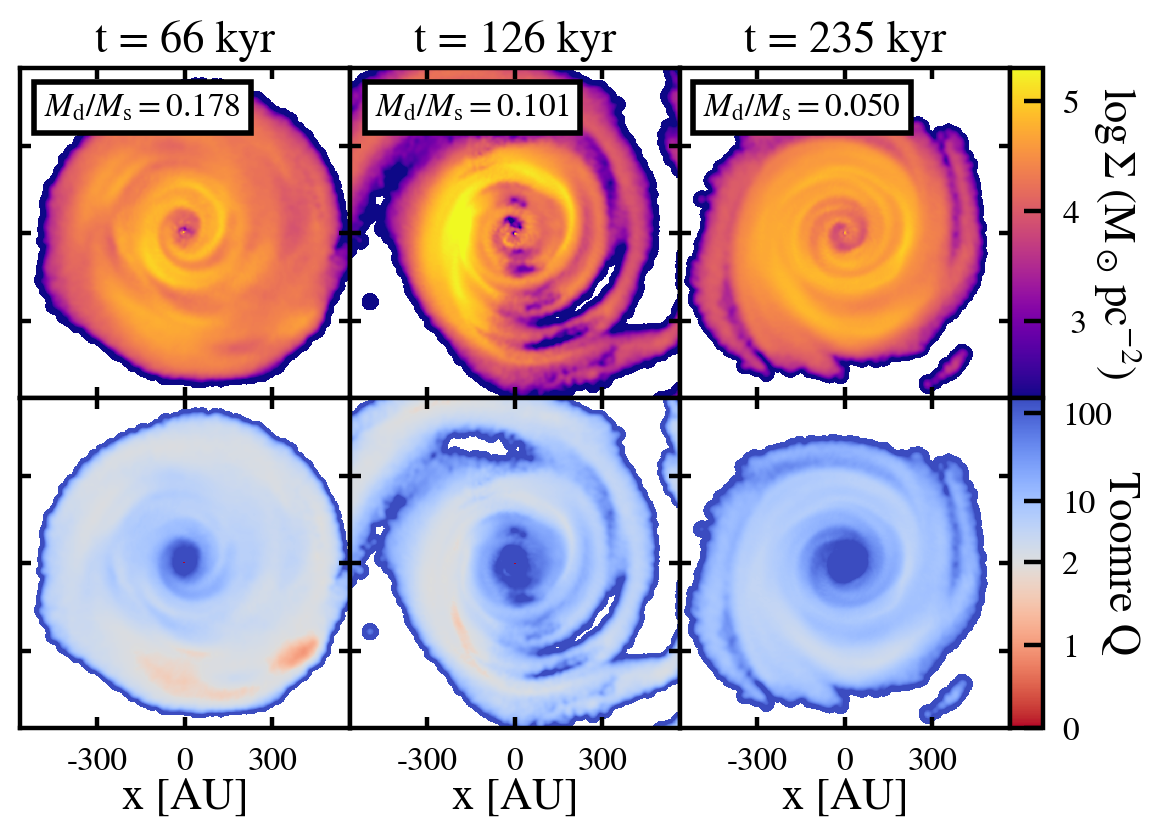}
    \caption{Projected surface density (top row) and Toomre Q (bottom row) for the single disk formed in the nonideal OA model.}
    \label{fig:toomre_Q_nOA_single}
\end{figure}

\begin{figure}
	\includegraphics[width=\columnwidth]{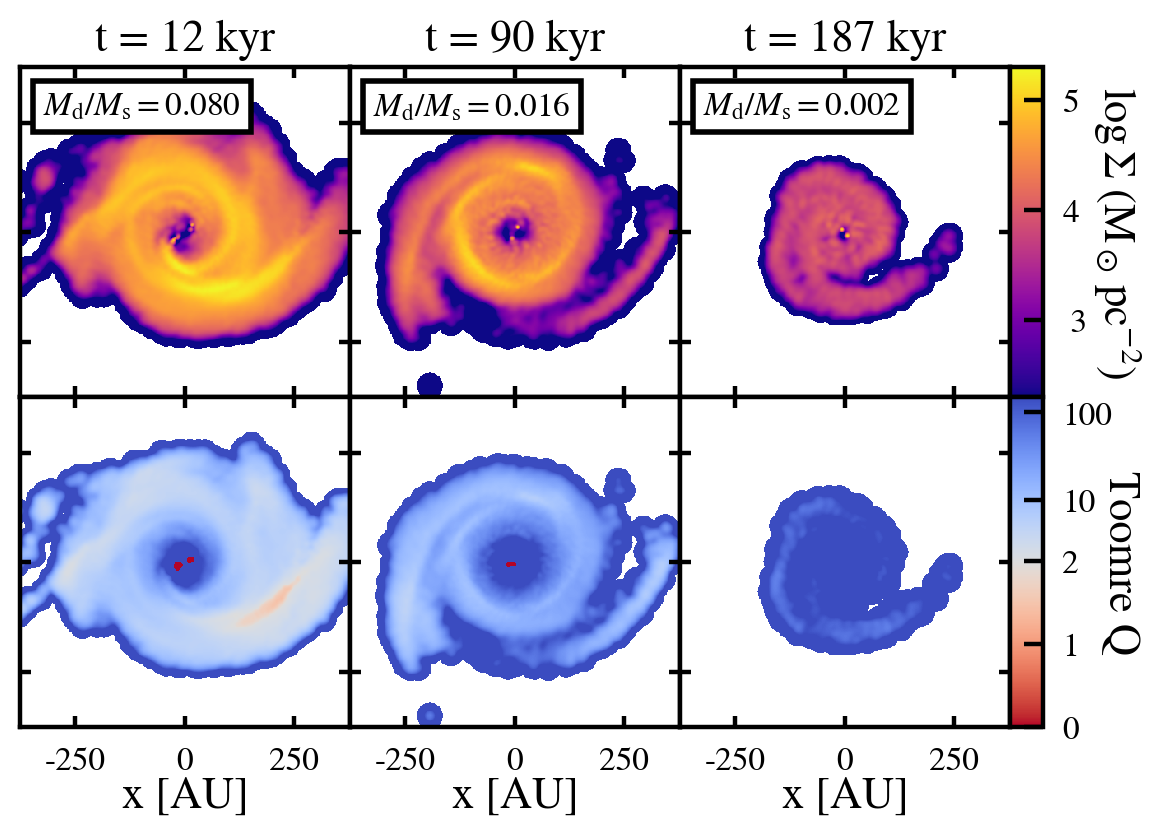}
    \caption{Projected surface density (top row) and Toomre Q (bottom row) for the more-massive binary disk (labeled ``binary disk 2" in Figures~\ref{fig:disk_mass}--\ref{fig:disk_B_misalignment}) formed in the nonideal OAH model.}
    \label{fig:toomre_Q_nOAH_binary_2}
\end{figure}

\begin{figure}
	\includegraphics[width=\columnwidth]{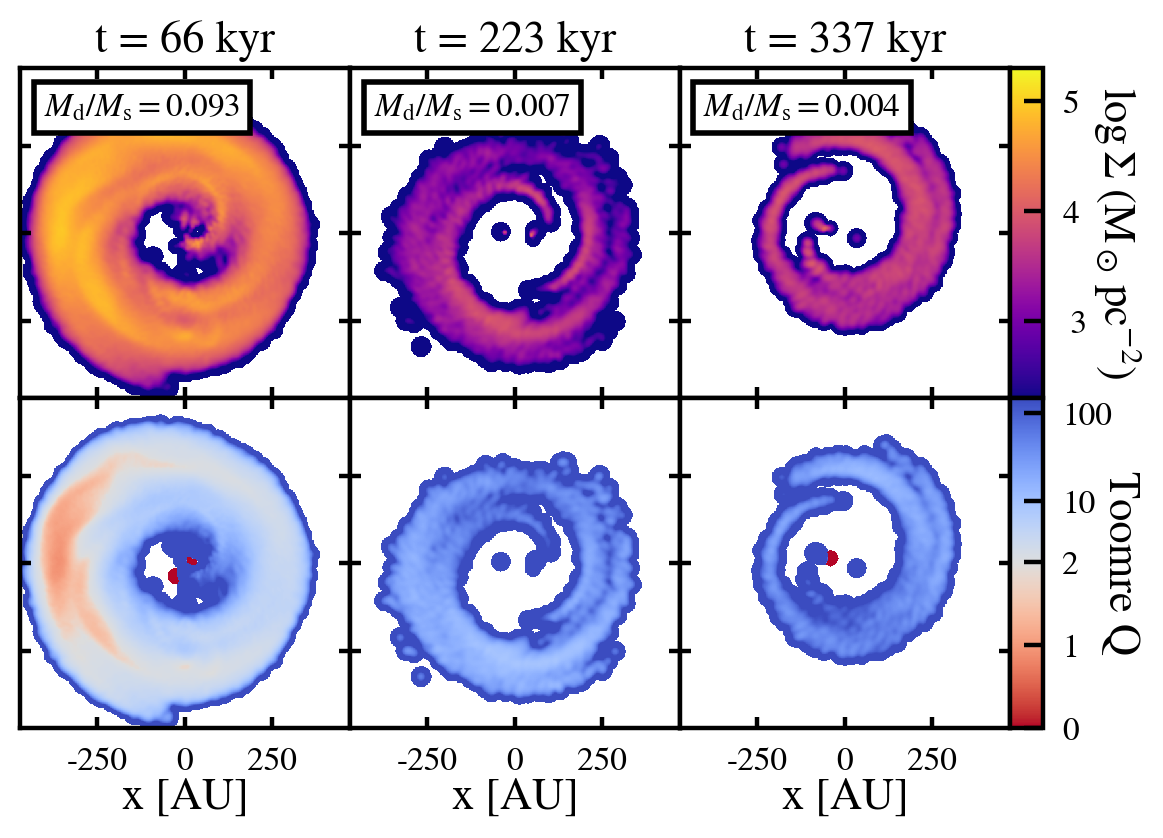}
    \caption{Projected surface density (top row) and Toomre Q (bottom row) for the binary disk formed in the nonideal OAH+jets model.}
    \label{fig:toomre_Q_nOAHj_binary}
\end{figure}

We calculate the Toomre $Q$ for the disks formed in our models and find that the disks are generally stable against gravitational fragmentation. Figures~\ref{fig:toomre_Q_ideal_binary}--\ref{fig:toomre_Q_nOAHj_binary} show the surface density and evolution of $Q$ for four of the most massive and longest-lived disks in our models: the binary disk formed in the ideal model (Figure~\ref{fig:toomre_Q_ideal_binary}; also the top panel in Figures~\ref{fig:disk_mass}--\ref{fig:sinks_with_disks}); the single disk formed in the non-ideal OA model (Figure~\ref{fig:toomre_Q_nOA_single}; also the third panel in Figures~\ref{fig:disk_mass}--\ref{fig:sinks_with_disks}); one of the binary disks in the non-ideal OAH model (Figure~\ref{fig:toomre_Q_nOAH_binary_2}; also the disk labeled ``binary 2" in the fifth panel in Figures~\ref{fig:disk_mass}--\ref{fig:sinks_with_disks}); and the binary disk formed in the non-ideal OAH+jets model (Figure~\ref{fig:toomre_Q_nOAHj_binary}; also the bottom panel in Figures~\ref{fig:disk_mass}--\ref{fig:sinks_with_disks}). We also indicate the disk-to-stellar mass ratio, as plotted in Figure~\ref{fig:disk_mass_ratio}, in the upper-left corner of the panels showing the surface density; as discussed in the review by \cite{kratter_lodato_2016}, disks are likely to be unstable for mass ratios $M_{\rm disk}/M_{\rm stellar} \gtrsim 0.1$. 

The binary disk formed in the ideal model appears to be the most stable, with $Q \gg 1$ throughout the disk lifetime. Some spiral structures appear in the surface density shown in Figure~\ref{fig:toomre_Q_ideal_binary}, but they are not well-resolved, as each spiral arm has the width of a single gas cell (visible as spherical blobs in the figure). This disk also has a low disk-to-stellar mass ratio, as seen in Figure~\ref{fig:disk_mass}: the mass ratio briefly approaches $\sim$0.1 soon after the formation of the disk, but then rapidly drops to $\lesssim$0.01.

The disks formed in the non-ideal models, meanwhile, show some evidence of gravitational instability, particularly at early times in their evolution. The bottom-left panels of Figures~\ref{fig:toomre_Q_nOA_single}--\ref{fig:toomre_Q_nOAHj_binary} show regions where $Q \sim 1$, and clear spiral features are visible in the corresponding surface density plots. The single disk formed in the non-ideal OA model shown in Figure~\ref{fig:toomre_Q_nOA_single} is the most massive disk formed out of all models and has a disk-to-stellar mass ratio $\gtrsim$0.1 for approximately 100 kyr during the first part of its evolution, as shown in Figures~\ref{fig:disk_mass} and \ref{fig:disk_mass_ratio}. The spiral structure in the surface density becomes even more evident at $t = 126$ kyr. The binary disk formed in the non-ideal OAH, shown in Figure~\ref{fig:toomre_Q_nOAH_binary_2}, has the most asymmetric structure. This is because this binary is in a higher-order quadruple system with a second binary, as shown in Figure~\ref{fig:proj_density_nOAH_binaries}, and the two binary disks interact and exchange matter as they evolve. The binary disk formed in the non-ideal OAH+jets model, shown in Figure~\ref{fig:toomre_Q_nOAHj_binary}, has a large region where $Q \sim 1$ and $M_{\rm disk}/M_{\rm stellar} \sim 0.1$ at $t = 66$ kyr; however, as seen in Figure~\ref{fig:disk_mass_ratio}, the mass ratio then quickly drops to $\lesssim$0.01. Even thought these three disks show regions that approach the conditions for gravitational instability at early times, we observe no disk fragmentation in any of our models  (though see Appendix \ref{appendix_rotating_envelope} for further discussion of the rotating fragmenting envelope in the non-ideal OA$+$jets model). We discuss this further and compare with observations in Section~\ref{section_observations}.


\subsubsection{Disk radial profiles}\label{section_disk_radial_profiles}
\begin{figure*}
    \includegraphics[width=2\columnwidth]{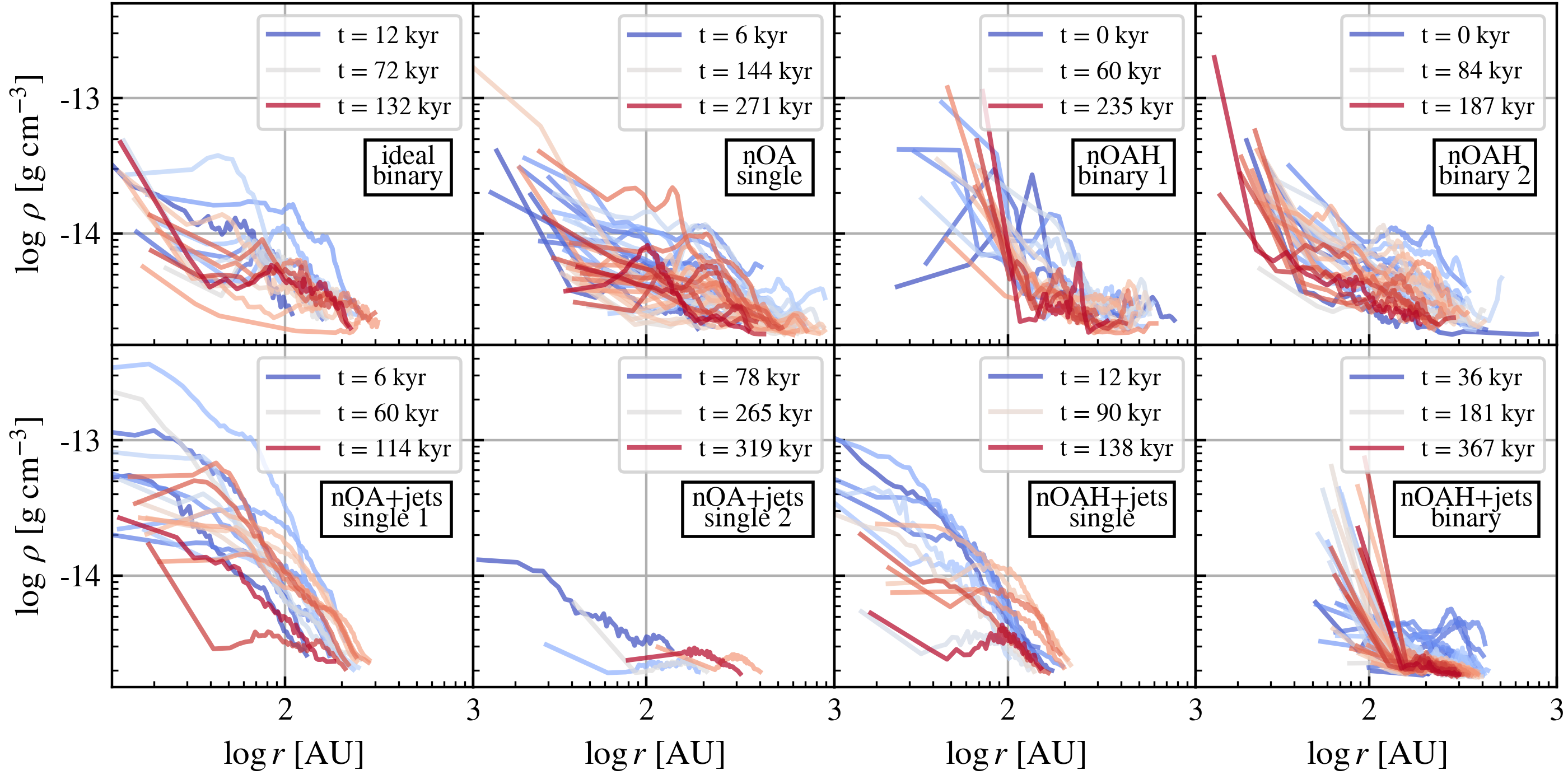}
    \caption{Time evolution of the radial density profiles of all disks formed across all models. Only snapshots with disks containing $\geq 1000$ gas cells are plotted. Blue lines correspond to earlier snapshots; red lines correspond to later snapshots.}
    \label{fig:rad_prof_density}
\end{figure*}
\begin{figure*}
    \includegraphics[width=2\columnwidth]{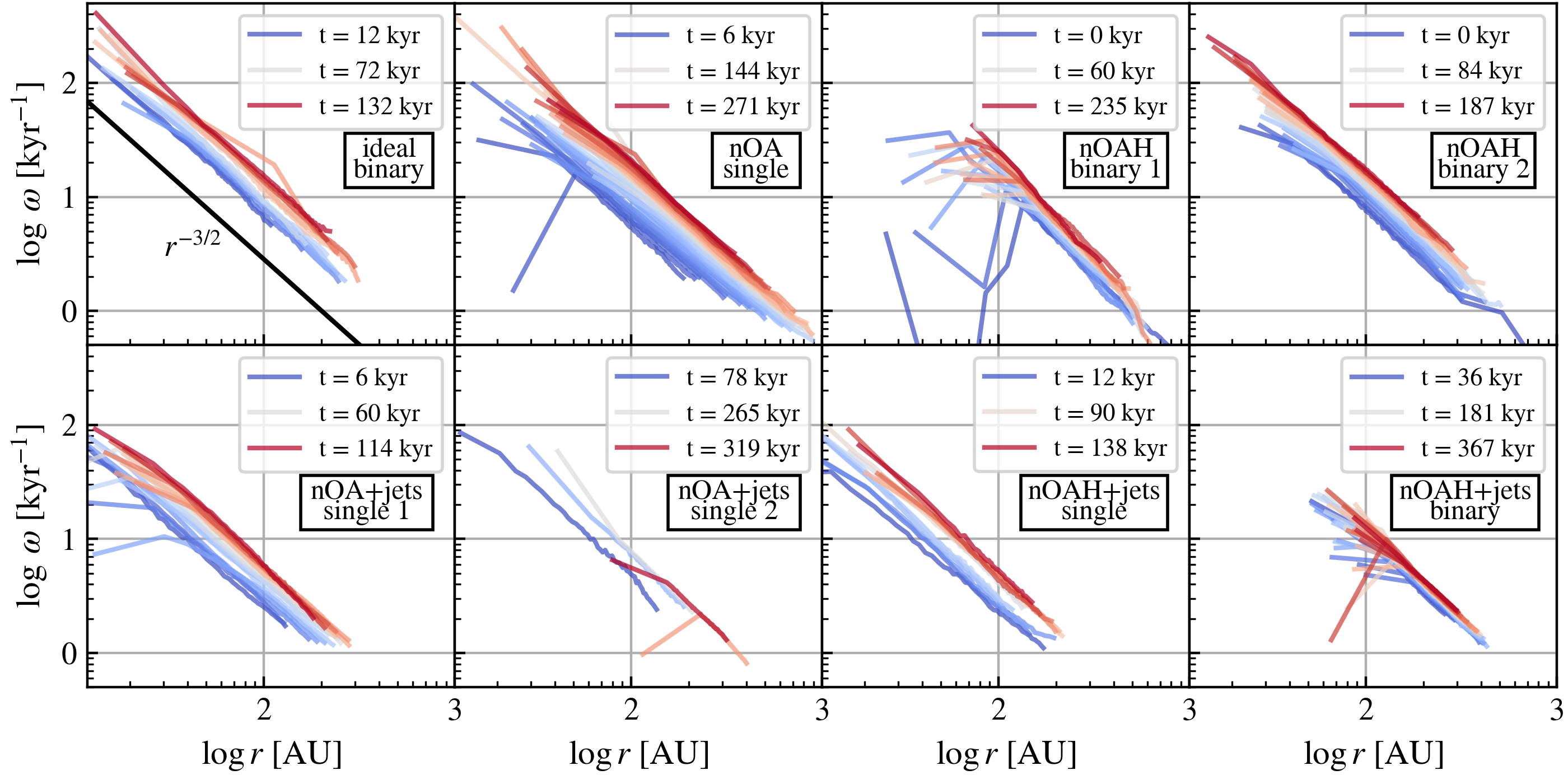}
    \caption{Same as Figure~\ref{fig:rad_prof_density}, but showing radial profiles of the angular velocity.}
    \label{fig:rad_prof_omega}
\end{figure*}
\begin{figure*}
    \includegraphics[width=2\columnwidth]{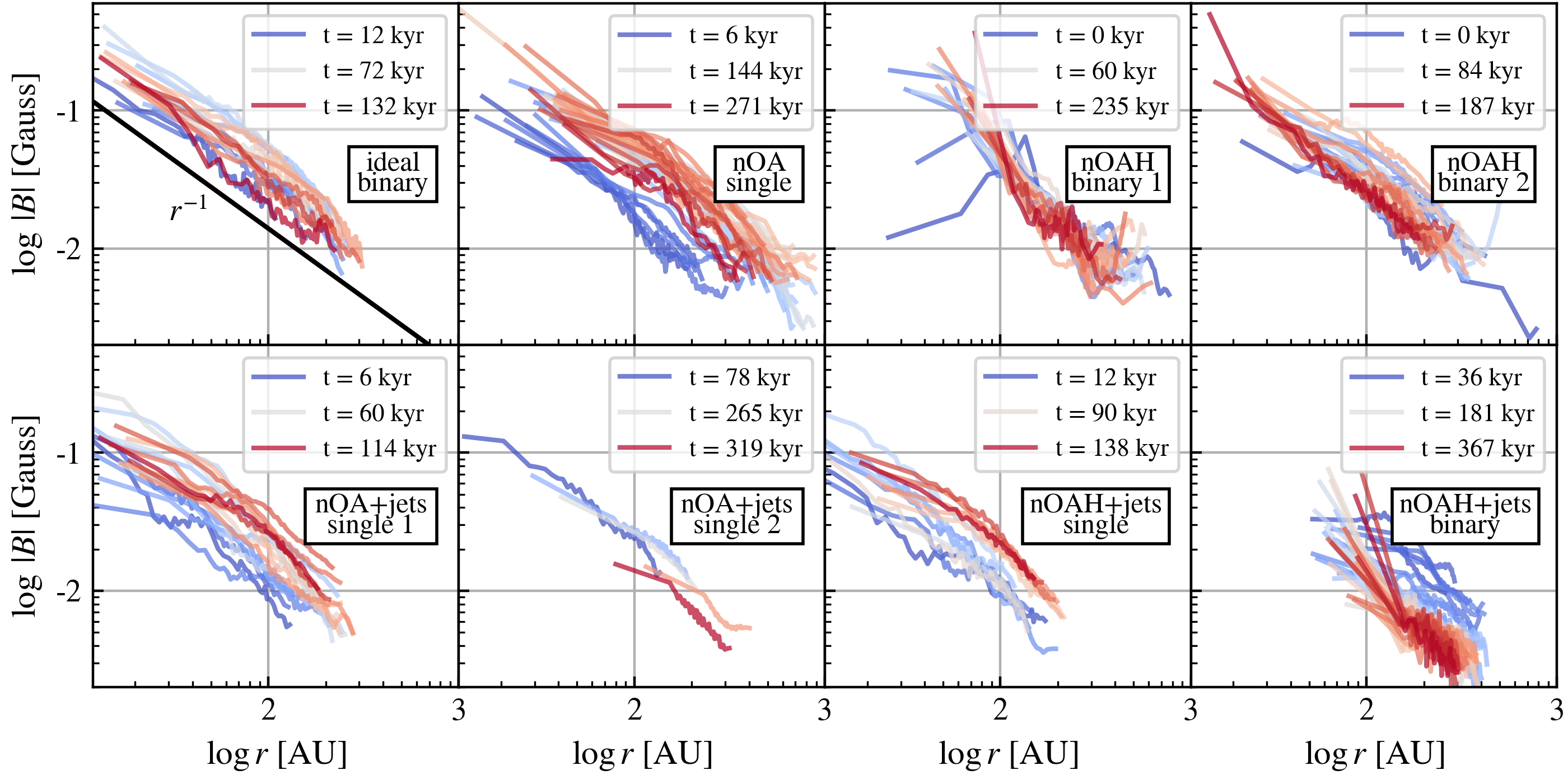}
    \caption{Same as Figure~\ref{fig:rad_prof_density} but showing radial profiles of the magnetic field strength.}
    \label{fig:rad_prof_B_mag}
\end{figure*}
\begin{figure*}
    \includegraphics[width=2\columnwidth]{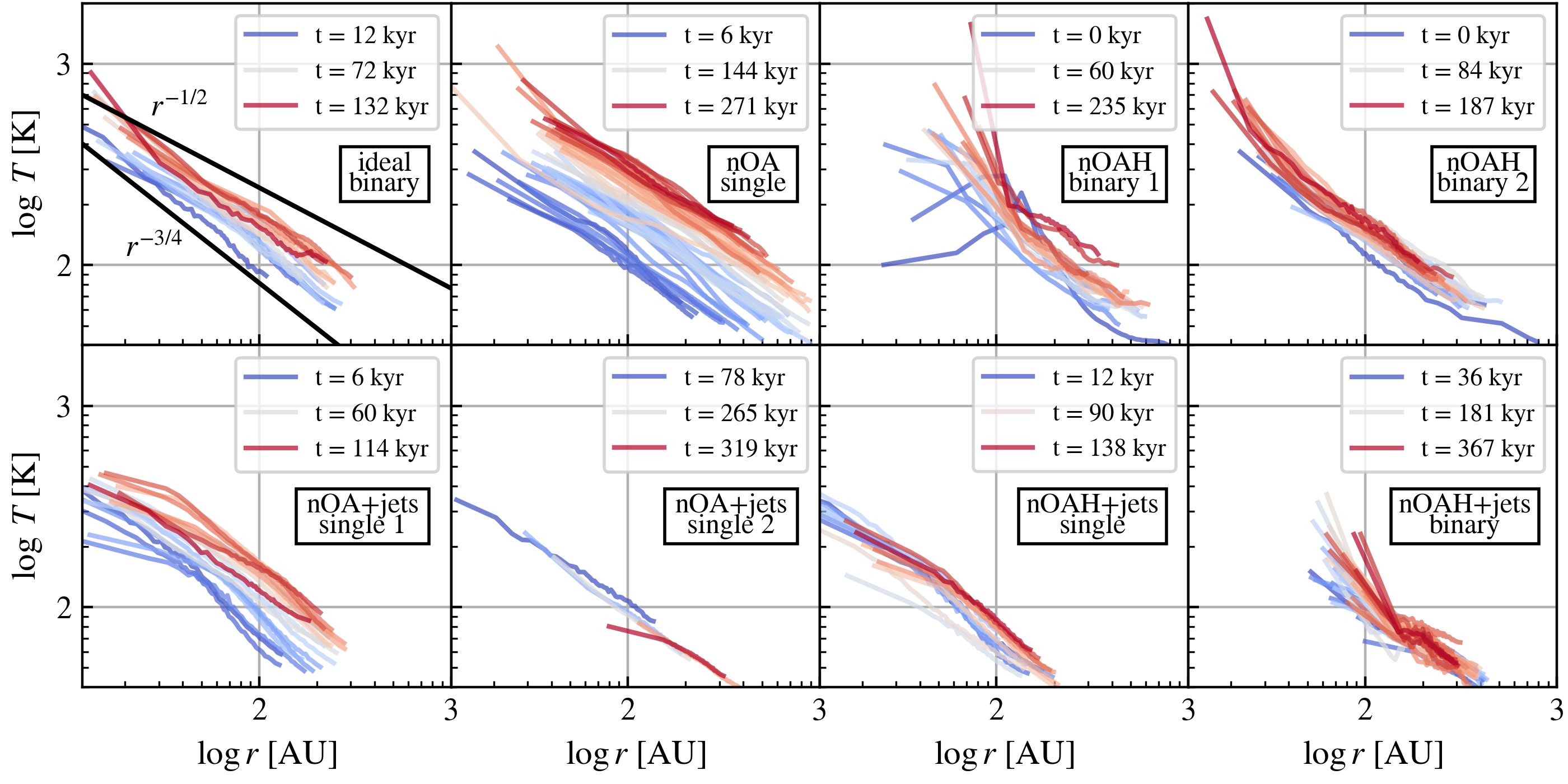}
    \caption{Same as Figure~\ref{fig:rad_prof_density}, but showing radial profiles of the gas temperature.}
    \label{fig:rad_prof_temperature}
\end{figure*}
\begin{figure*}
    \includegraphics[width=2\columnwidth]{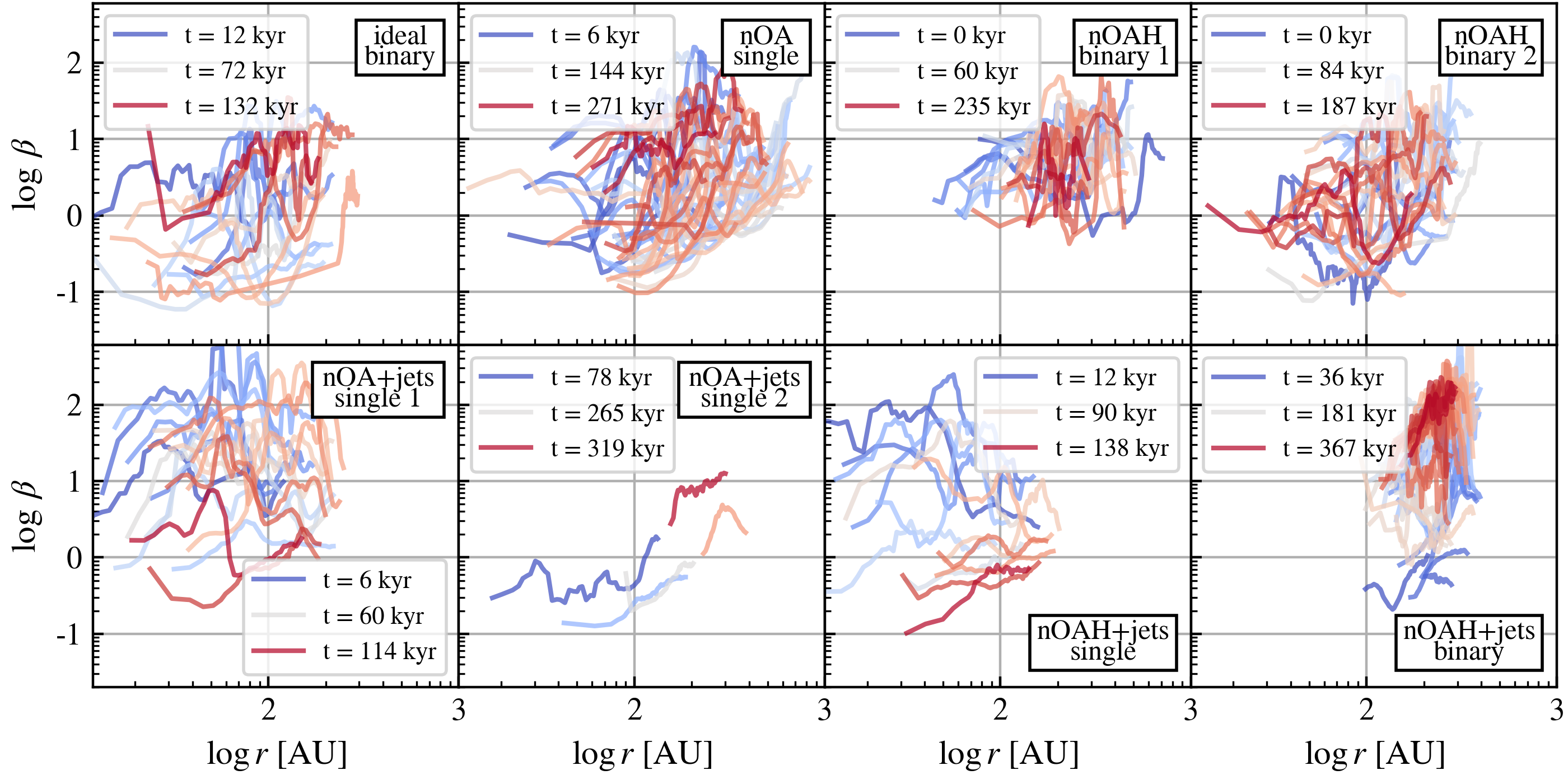}
    \caption{Same as Figure~\ref{fig:rad_prof_density}, but showing radial profiles of the plasma $\beta = P_{\rm th}/P_{\rm mag}$. Here we average over gas cells within $|z| < 5$ AU to isolate the disk midplane.}
    \label{fig:rad_prof_beta}
\end{figure*}
\begin{figure*}
    \includegraphics[width=2\columnwidth]{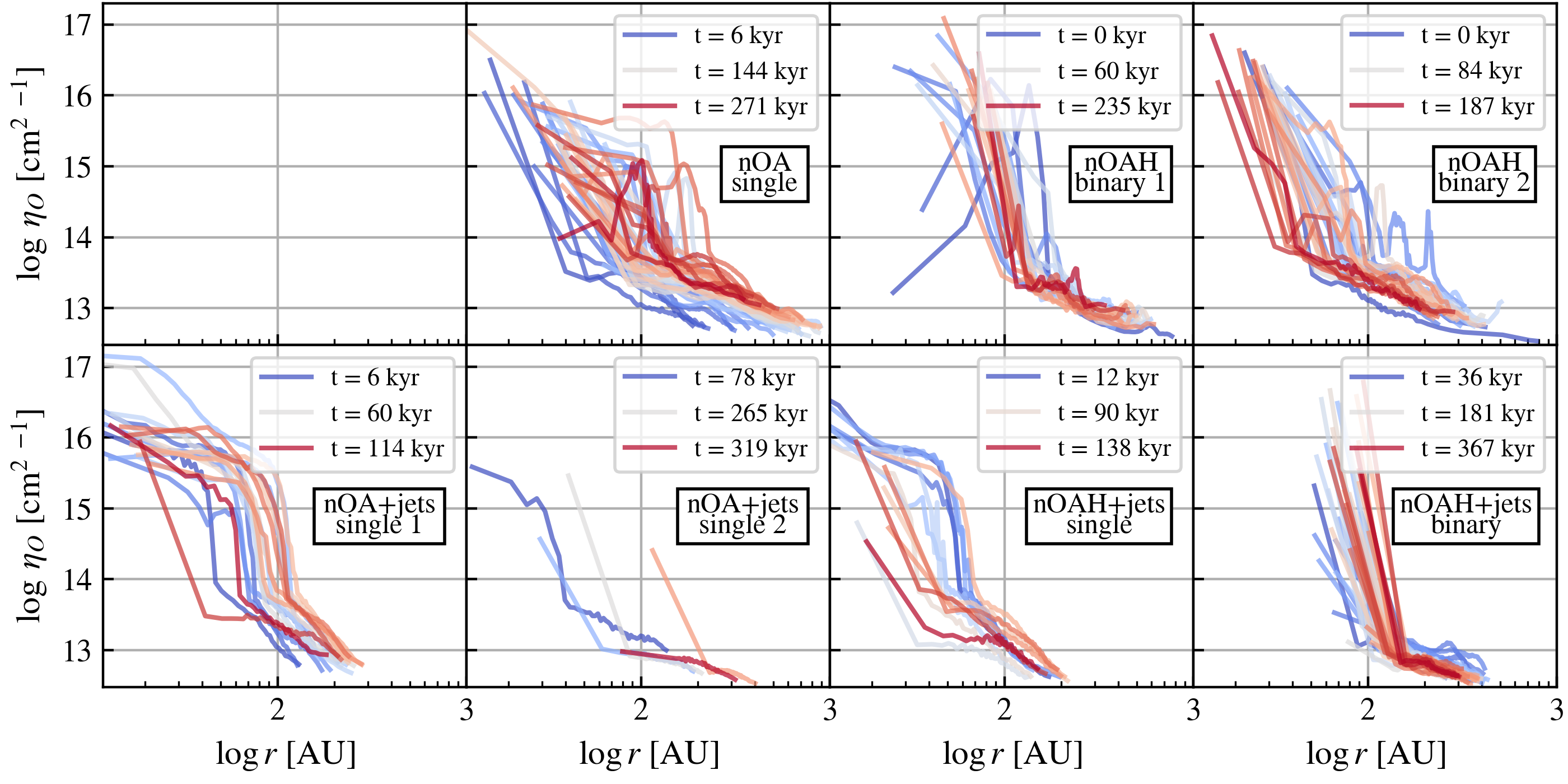}
    \caption{Same as Figure~\ref{fig:rad_prof_density}, but showing radial profiles of the Ohmic resistivity, $\eta_{\rm O}$.}
    \label{fig:rad_prof_eta_O}
\end{figure*}
\begin{figure*}
    \includegraphics[width=2\columnwidth]{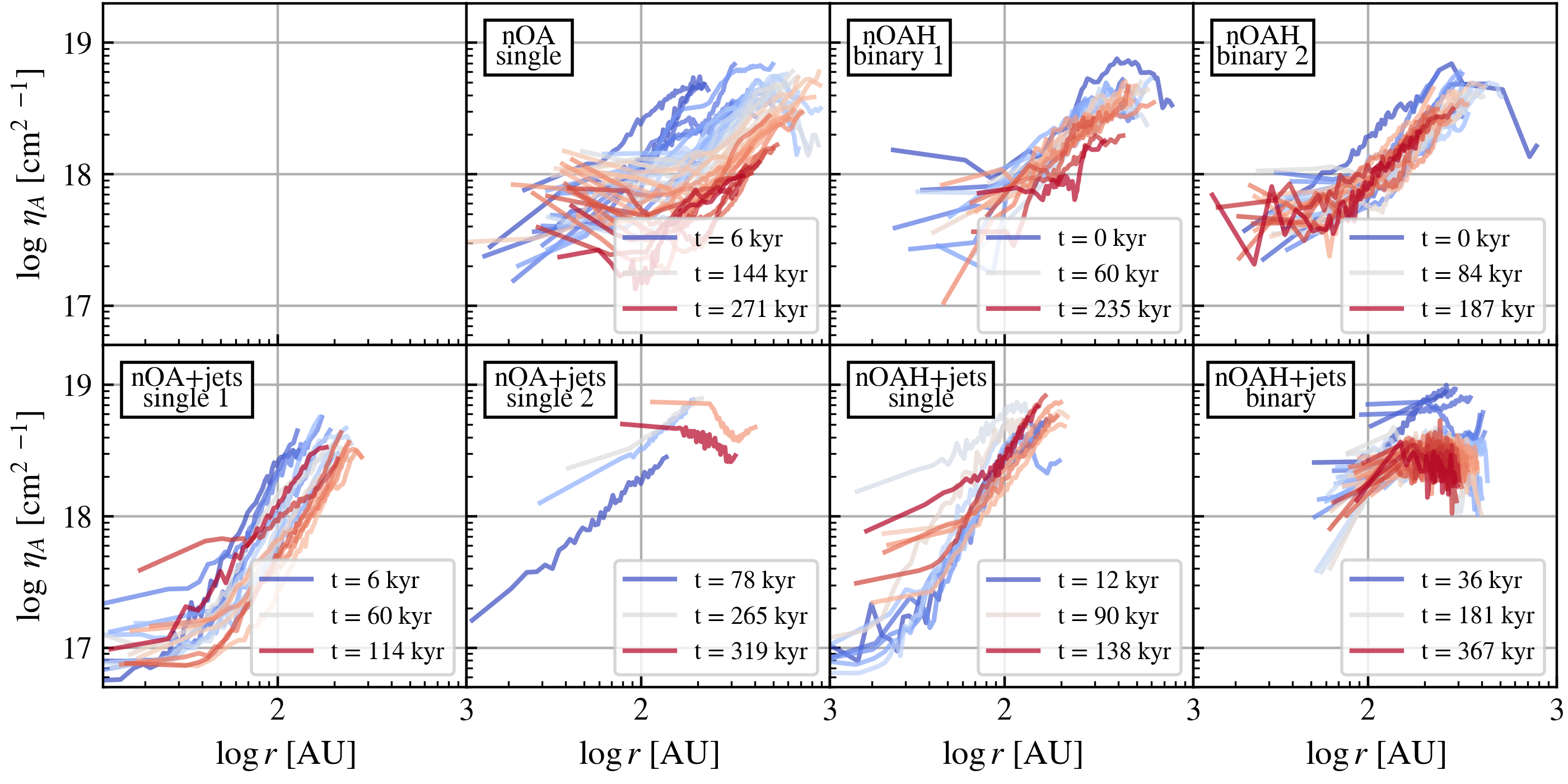}
    \caption{Same as Figure~\ref{fig:rad_prof_density}, but showing radial profiles of the ambipolar diffusion resistivity, $\eta_{\rm A}$.}
    \label{fig:rad_prof_eta_A}
\end{figure*}
\begin{figure}
    \includegraphics[width=\columnwidth]{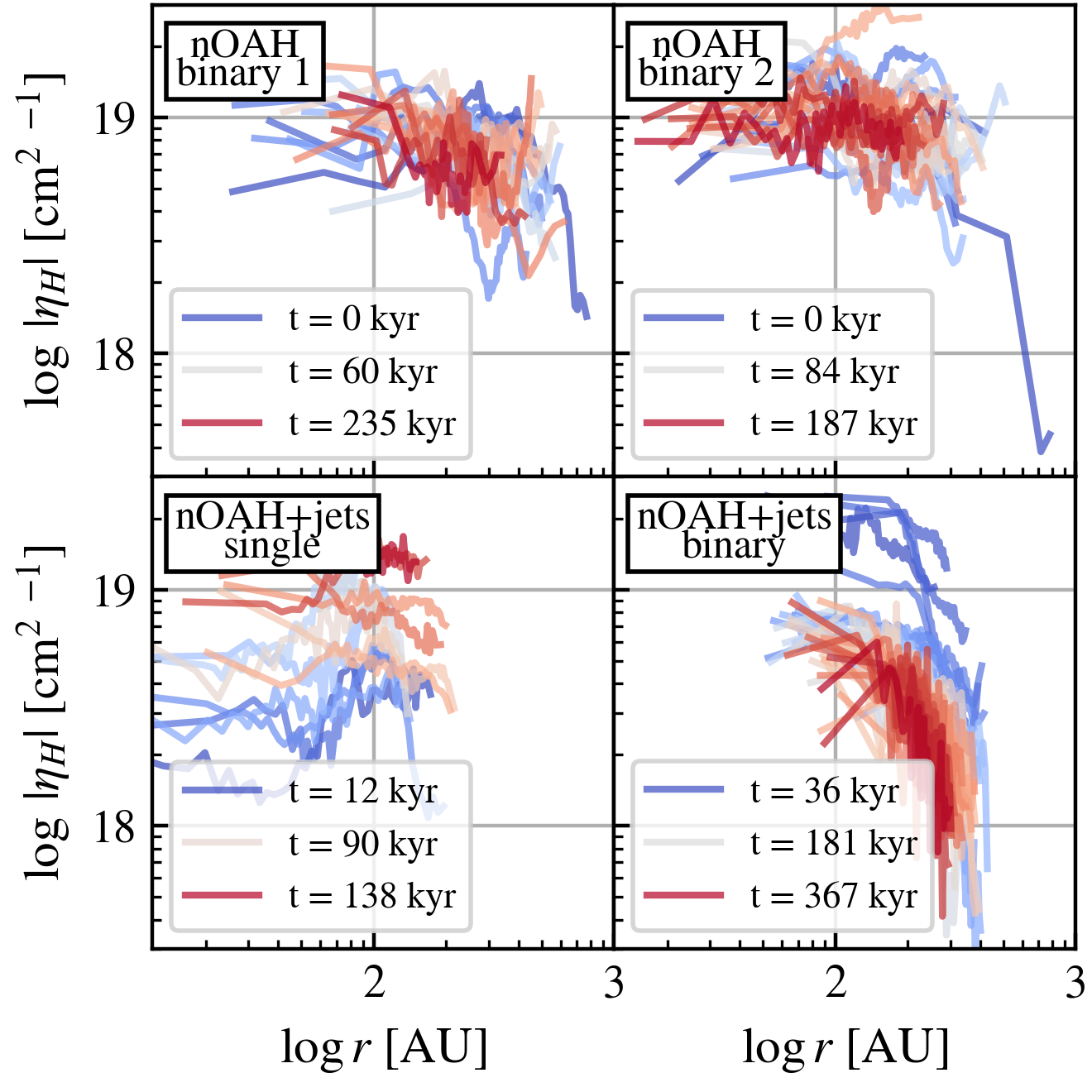}
    \caption{Same as Figure~\ref{fig:rad_prof_density}, but showing radial profiles of the absolute value of the Hall effect resistivity, $|\eta_{\rm H}|$, which is negative throughout the disk.}
    \label{fig:rad_prof_eta_H}
\end{figure}
In addition to examining the evolution of bulk disk properties such as mass and radius, we also construct radial profiles of disk properties such as density, magnetic field strength, angular velocity, temperature, and nonideal resistivities. We create the radial profiles by sorting the gas cells belonging to each disk by distance to the disk center of mass, binning the cells into 50 bins containing approximately equal cell counts, and taking the average value of each bin. For each disk, we plot the radial profile for each snapshot in which the disk contains at least 1000 cells (i.e., $M_{\rm disk} \geq 0.01 M_{\odot}$). 

Figure~\ref{fig:rad_prof_density} shows the evolution of the radial density profiles. It is not clear whether a power-law of the form $\rho(r) \propto r^{a}$ can be used to characterize the density profile of the disks, owing to the the presence of density structures (such as spirals), observed as peaks and valleys in the density profiles. The density generally decreases in time as the disk becomes less massive. For binary disks, the inner cavity observed in the radial profiles tends to correspond with the binary separation: the binary in the ideal MHD model (first row, first column of the figure) and the second binary in the non-ideal OAH model (first row, fourth column) are close binaries with separation $\lesssim$10 AU, while the first binary in the non-ideal OAH model (first row, third column) and the binary in the non-ideal OAH$+$jets model (second row, fourth column) are separated by $\sim$100 AU.

As seen in Figure~\ref{fig:rad_prof_omega}, the angular velocity profile of each disk shows clear Keplerian rotation, $\omega(r) \propto r^{-3/2}$, with the angular velocity increasing with time as the disk continues accreting material and angular momentum. It appears that gravitational torques (seen as spiral structures) and magnetic braking are somewhat inefficient at transporting angular momentum on disk scales.

Figure~\ref{fig:rad_prof_B_mag} shows the evolution of the radial profiles of the disk magnetic field strength, which generally follow a $B(r) \propto r^{-1}$ relation. The magnetic field strength in the ideal MHD disk is comparable to that in the non-ideal MHD models. In some disks, such as the single disk in the non-ideal OA model (first row, second column), the magnetic field strength increases with time; in other disks, such as the binary disk in the non-ideal OAH$+$jets model (second row, fourth column), the field strength decreases with time. We generally observe the magnetic field strength increasing over time in the non-ideal single disks and decreasing in the non-ideal binary disks, though a detailed analysis of this trend requires a larger statistical sample.

Figure~\ref{fig:rad_prof_temperature} shows the evolution of the radial temperature profiles. In general, the disk is heated by stellar irradiation, which acts at the disk surface, and viscous dissipation due to accretion, which is maximal at the disk midplane \citep{ueda_2023}. The disk temperature structure is therefore a function of the vertical coordinate $z$ as well as the radial coordinate $r$, and depends on the optical properties of the disk. A simplified model for disk midplane temperature is \citep[e.g.,][]{sierra_2020, ueda_2023}
\begin{equation}
    T_{\rm disk} = \left(T_{\rm irr}^4 + T_{\rm acc}^4 \right)^{1/4},
\end{equation}
where $T_{\rm irr}$ and $T_{\rm acc}$ denote the disk temperature determined by
stellar irradiation and accretion heating, respectively. For a passively-heated, flared disk in radiative equilibrium, the temperature due to stellar irradiation can be estimated as \citep[e.g.,][]{chiang_goldreich_1997, dullemond_2001, huang_2018}
\begin{equation}
    T_{\rm irr}(r) = \left(\frac{\varphi L_*}{8 \pi \sigma_{SB} r^2}\right)^{1/4} \propto r^{-1/2},
\end{equation}
where $\varphi$ is the flaring angle of the disk ($\varphi \ll 1$), $L_*$ the stellar luminosity, and $\sigma_{SB}$ the Stefan-Boltzmann constant. The temperature due to viscous heating during accretion, meanwhile, can be approximated as \citep[e.g.,][]{nakamoto_nakagawa_1994, sierra_2020}
\begin{equation}
    T_{\rm acc}(r) = \frac{3}{4}\left(\tau_R + \frac{2}{3}\right) \frac{3 \dot{M} \Omega_K}{8 \pi \sigma_{SB}} \propto r^{-3/4},
\end{equation}
where $\tau_R$ is the Rosseland optical depth measured perpendicular to the disk midplane, $\dot{M}$ the accretion rate, and $\Omega_K = \sqrt{GM_*/r^3}$ the Keplerian angular velocity for stellar mass $M_*$. We show both $T(r) \propto r^{-3/4}$ (accretion-dominated heating) and $T(r) \propto r^{-1/2}$ (irradiation-dominated heating) as solid black lines in the top-left panel of Figure~\ref{fig:rad_prof_temperature}. Viscous heating is expected to dominate at early times and small radii, while heating by stellar irradiation is expected to dominate at larger radii and during later stages of disk evolution \citep[e.g.,][]{ueda_2023}. For most disks in our models, $T(r) \propto r^{-3/4}$ appears to be a better fit to the disk radial profiles, consistent with expectations of accretion-dominated heating. However, when we examine the radial temperature profiles of all gas contained within spheres of radius $\sim10^5$ AU centered at the disk center of mass, we observe a better fit to $T(r) \propto r^{-1/2}$ for gas beyond a few 100 AU of the center, suggesting that stellar irradiation dominates the heating in this region.

Figure~\ref{fig:rad_prof_beta} shows the evolution of the disk plasma $\beta = P_{\rm th}/P_{\rm mag}$, the ratio of thermal to magnetic pressure support throughout the disk: $\beta < 1$ implies that the disk is magnetically dominated, while $\beta > 1$ suggests that the disk is thermally dominated. We calculate $\beta$ for a slice of the disk with $|z|\leq5$ AU to better represent the pressure balance at the disk midplane. For nearly all of the disks, there is a weak trend of increasing $\beta$ with radius, suggesting that thermal support becomes more important in the outer disk. Several of the non-ideal disks are almost entirely thermally dominated throughout their evolution, such as the first single disk in the non-ideal OA$+$jets model (second row, first column), the first binary disk in the non-ideal OAH model (first row, third column), and the two disks in the non-ideal OAH$+$jets model (second row, third and fourth columns). However, the other non-ideal disks are more similar to the ideal binary disk (first row, first column) in that, while $\beta > 1$ throughout much of the disk at certain points in its evolution, at other times $\beta < 1$ for large regions of the disk. In general, $\beta$ is comparable between the ideal and non-ideal disks. Thermal pressure support tends to dominate, but the magnetic pressure support is not negligible.

Figures~\ref{fig:rad_prof_eta_O},~\ref{fig:rad_prof_eta_A}, and~\ref{fig:rad_prof_eta_H} show the evolution of the radial profiles of the Ohmic dissipation, ambipolar diffusion, and Hall effect resistivities throughout the disks. Ohmic resistivity is most efficient in the inner disk ($r \lesssim 100$ AU), with a steep drop-off in the outer regions of the disk, but is subdominant to both ambipolar diffusion and the Hall effect. Ambipolar diffusion is more efficient in the outer disk. In models including the Hall effect, the Hall effect generally dominates over ambipolar diffusion and Ohmic resistivity throughout most of the disk. We do not observe a clear systematic trend with time in any of the non-ideal MHD resistivities, which are complex, non-linear functions of the gas density, temperature, magnetic field strength, and properties of charged species (see Eq.~\ref{eq:eta_dependencies}). For example, as seen in the top two panels of Figure~\ref{fig:rad_prof_eta_H}, the Hall resistivity throughout the two binary disks in the non-ideal OAH model fluctuates over a range of values with no clear time dependence; however in the non-ideal OAH$+$jets model, the Hall resistivity increases with disk age for the single disk, but decreases with disk age for the binary disk, as seen in the bottom two panels of Figure~\ref{fig:rad_prof_eta_H}. 

\subsection{Self-consistent outflows}\label{results_outflows}
\begin{figure*}
    \includegraphics[width=2\columnwidth]{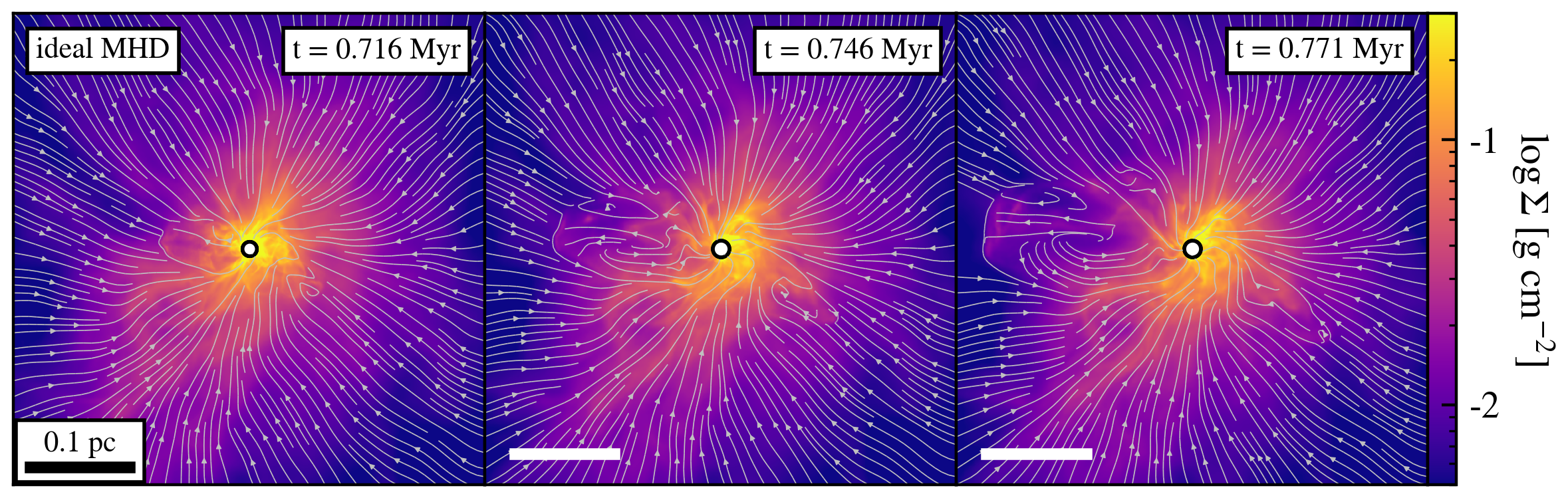}
    \caption{Column density projected along the $x$-axis showing the time evolution of the self-consistently launched outflow in the ideal MHD model. Streamlines trace the density-weighted velocity. The outflow is launched from the central binary and is nearly parallel to the plot $x$-axis in this projection.}
    \label{fig:proj_density_ideal_outflow_time_evolution}
\end{figure*}
\begin{figure}
    \centering\includegraphics[width=0.78\columnwidth]{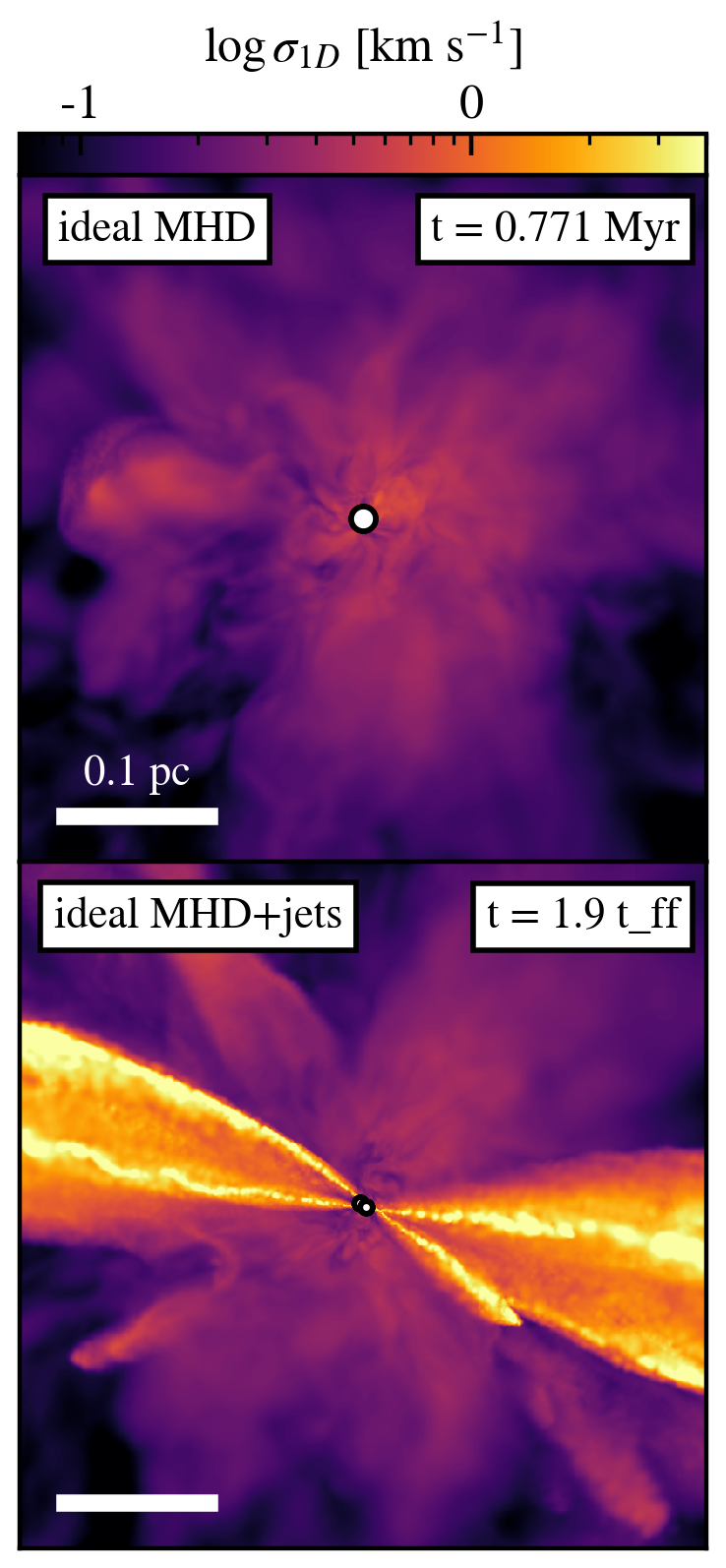}
    \caption{Line-of-sight 1D velocity dispersion ($\sigma_{\rm 1D}$) along the $x$-axis for the ideal MHD (top row) and the ideal$+$jets (bottom row). The top panel shows the same time and viewing angle as the rightmost panel of the projected column density plot in Figure~\ref{fig:proj_density_ideal_outflow_time_evolution}.}
    \label{fig:velocity_dispersion_ideal_outflows}
\end{figure}
Although we include a model with sub-grid protostellar jet feedback for each set of MHD physics in our calculations, we also observe some self-consistently launched outflows in our ideal MHD model with no sub-grid jet feedback. Figure~\ref{fig:proj_density_ideal_outflow_time_evolution} shows the time evolution of this outflow as seen in the column density when projected along the $x$-axis. The outflow is launched from the central binary disk system and is approximately parallel to the plot $x$-axis. A bow shock due to the outflow can be seen in the figure. We note, however, that this outflow is less collimated and has significantly lower velocity than the sub-grid jet feedback. Figure~\ref{fig:velocity_dispersion_ideal_outflows} shows the 1D line-of-sight velocity dispersion $\sigma_{\rm 1D}$ of this outflow in the top panel; for comparison, the bottom panel shows the velocity dispersion of the ideal$+$jets model at the same point in time. The left lobe of the outflow can be distinguished in the top panel of Figure~\ref{fig:velocity_dispersion_ideal_outflows}. While the sub-grid jets approach a velocity dispersion of $\sigma_{\rm 1D} \sim 10$ km s$^{-1}$, the self-consistent outflow in the top model is closer to $\sigma_{\rm 1D} \sim 0.5$ km s$^{-1}$, more than an order of magnitude lower. We observe no self-consistent outflows in any of the non-ideal MHD models.

\section{Discussion}
We have presented the results of a suite of six calculations following the formation of stellar clusters and protostellar disks under various assumptions of ideal and non-ideal MHD and with optional sub-grid protostellar jet feedback. Here, we discuss differences in stellar and disk formation outcomes among our models and compare our results to several previous numerical studies. We also discuss several challenges, which complicate drawing direct comparisons between observed and modeled disk properties, before concluding with a few caveats concerning the validity of our microphysical model used to calculate the non-ideal MHD resistivities.

\subsection{Effects of including non-ideal MHD}\label{discussion_nonideal_mhd}
Prior to initiating the self-gravitating star-forming calculations, we observe smoother cloud structure in the calculations with non-ideal (Ohmic$+$ambipolar) MHD than in the ideal MHD calculations, as shown by the column density plots in Figure~\ref{fig:proj_density_turbsphere}. We further note that the degree of smoothing is correlated with the dust grain size $a_g$ used for calculating the non-ideal MHD resistivities, with larger grains leading to smoother density distributions and smaller grains leading to more structure at smaller ($\lesssim$0.1 pc) scales. This is due to  ambipolar diffusion, which leads to the formation of smooth C-type shocks \citep{draine_1980_C_shock, draine_1983_C_shock} and is highly sensitive to the assumed dust grain size, as shown in Figure~\ref{fig:density_profile_TURBSPHERE_shock_thickness}. Smaller dust grains lead to lower values of $\eta_{\rm AD}$, with the suppression of the ambipolar resistivity becoming more pronounced at higher densities. The density distribution in the non-ideal model with the smallest dust grains ($a_g = 0.001$ $\mu$m) is therefore most similar to the ideal model as the non-ideal effects become dynamically insignificant. However, it is important to note that the dynamical behavior of the gas in the opposite limit of larger dust grains and thereby stronger ambipolar diffusion \textit{does not} resemble the hydrodynamical case, as can be seen by comparing the top middle and bottom left plots in Figure~\ref{fig:proj_density_turbsphere}, due to the effects of ion-neutral interactions under the influence of a magnetic field.

Several other studies of cluster formation with non-ideal MHD, such as \cite{wurster_2019_low_mass_cluster} and \cite{lebreuilly_2024a}, found that the evolution of the large-scale structure ($\gtrsim$0.05 pc) is independent of whether ideal or non-ideal MHD is assumed. These studies, however, did not include an initial turbulent stirring phase equivalent to our \textsc{TURBSPHERE} calculations. Our results are qualitatively similar to the driven turbulent box calculations of \cite{ntormousi_2016}, who compared the effects of assuming ideal MHD versus including ambipolar diffusion on the resulting density and magnetic field distributions. They found that the density and magnetic field morphologies differed between the two sets of MHD physics, with the densest structures becoming broader and more massive under the influence of ambipolar diffusion. They also found that the power spectra of the density and velocity fields showed a steeper cutoff, with a turnover occurring at a smaller wavenumber in the non-ideal than in the ideal MHD case. We observe a similar result in our Figure~\ref{fig:TURBSPHERE_power_spectrum}, and note that the turnover point in the non-ideal power spectra depends on the assumed dust grain size. 

Ion-neutral friction has been previously invoked as a damping mechanism for interstellar turbulence, suggesting the existence of a cutoff in the turbulent energy cascade below a characteristic ambipolar diffusion length scale at which the motions of the two species decouple \citep{zweibel_1983, mouschovias_1991, brandenburg_1994, li_houde_2008}. However, as indicated by Figure~\ref{fig:density_profile_TURBSPHERE_shock_thickness}, this characteristic scale is sensitive to poorly-constrained microphysical parameters such as the dust grain size. Furthermore, rather than modeling an evolving dust grain size distribution, our calculations are limited to assuming a single dust grain radius, taken to be $a_g = 0.1$ $\mu$m in our fiducial model. This value may be representative of dust grains within the dense protostellar disk \citep{pollack_et_al_1994}, but likely overestimates the typical dust grain size in more diffuse regions of the cloud \citep{mathis_1977}, meaning that our fiducial non-ideal TURBSPHERE calculations may overestimate the degree of smoothing caused by ambipolar diffusion. Indeed, recent observations of the nearby star-forming region NGC 1333 by \cite{pineda_2024} found no evidence of dissipation in both the ion and neutral power spectra down to $\sim$4000 AU scales, suggesting that the characteristic ambipolar length scale, if it exists, is below this scale. We further discuss the influence of dust grain size on the non-ideal resistivities in Section \ref{section_caveats}.

We continue to observe differences in structure once self-gravity is turned on and star formation begins. Figure~\ref{fig:proj_density_pre-sf} shows that, just prior to star formation, the ideal MHD model is the most structured, and the densest region is more centrally concentrated. The morphology of the dense gas also differs between the two non-ideal MHD models, which can be attributed to inclusion of the Hall effect. Ambipolar diffusion tends to relax magnetic field bending via the magnetic tension force, acting in the same direction as the bending of the field lines. The Hall effect, meanwhile, induces a Hall drift orthogonal to the magnetic field bending and is not in principle a diffusive term. These two terms may induce fluid motions orthogonal to one another \citep[see, e.g.,][]{zhao_2021}, resulting in different density and magnetic field structures. The onset of star formation, as well as the number of stars formed, differs between the models, likely as a result of these structural differences.

Figure~\ref{fig:B_vs_rho_pre-sf} shows the median value and 95th percentile of the magnetic field strength as a function of density just after the onset of star formation. The magnetic field strength in all models follows the scaling $B \propto \sqrt{\rho}$ between densities $\rho \gtrsim 10^{-21}$ g cm$^{-3}$ and $\rho \lesssim 10^{-12}$ g cm$^{-3}$, and there is little difference between the median value of the magnetic field strength between the ideal and non-ideal MHD models. This is consistent with the results of \cite{wurster_2019_low_mass_cluster}, who compare the mean magnetic field strength between ideal and non-ideal MHD calculations for models with different values of the initial mass-to-magnetic flux ratio $\mu_{\Phi}$ over a similar range of densities. At our current level of resolution, our simulations do not probe past densities appreciably higher than $\rho \sim 10^{-12}$ g cm$^{-3}$. However, at higher densities, the difference between the ideal and non-ideal calculations may become more apparent. \cite{masson_2016} examined the effects of ambipolar diffusion on the properties of the first Larson core and its surroundings, and showed that while non-ideal calculations with both low ($\mu_{\Phi} = 5$) and high ($\mu_{\Phi} = 2$) magnetization reached a plateau in magnetic field strength at about $B \sim 0.1$ G, the magnetic field strength in the ideal models continued to increase with density, following the expected scaling with density for magnetic flux freezing in a contracting sphere and reaching values as high as $B \sim 1$ G. They attribute this plateau to the formation of a magnetic diffusion barrier, and derive an estimate for the saturation limit of the magnetic field strength which is in good agreement with the numerical value they find for the plateau. A similar plateau was found by \cite{vaytet_2018}, who also included Ohmic resistivity, and more recently by \cite{mayer_2025}, who included all three non-ideal MHD effects. The magnetic plateaus found in these works occurred at higher densities and/or magnetic field strengths than are attained by our calculations, so it is not surprising that we do not observe this effect. However, we do observe that the magnetic field strengths in the non-ideal calculations are slightly higher than in the ideal model for densities $\rho \gtrsim 10^{-16}$ g cm$^{-3}$. This may simply be a consequence of the different initial conditions generated by the TURBSPHERE calculations, or else due to the fact that we use equal bins in log density space to compute our profiles and that the highest-density bins ($\rho \gtrsim 10^{-13}$ g cm$^{-3}$) in the non-ideal models contain many fewer particles than in the ideal model. In either case, we do not consider the difference in magnetic field strength between the ideal and non-ideal MHD models to be significant at these densities.

In the case of ideal MHD, \cite{He_Ricotti_2025} observed a universal $B-\rho$ relationship in their zoom-in calculations of disk formation in massive, magnetically critical or super-critical GMC cores, regardless of the initial magnetic intensity or cloud size. They found that each GMC and core has two regimes: a low-density regime where the mean $B$ is approximately constant with $\rho$, and a high-density regime where the scaling follows $B \propto \sqrt{\rho}$. They interpreted the low-density regime as corresponding to the case when the density of the gas can increase or decrease without affecting the value of $B$, which occurs when the gas motion has no preferred direction, as in the case of isotropic turbulence. Motion along the magnetic field lines can compress or decompress the gas without modifying the magnetic field strength. Motion perpendicular to the field lines can increase or decrease the magnetic field strength, but if the motion is isotropic, this simply results in a constant scatter around the mean $B$, where the scatter is smaller for stronger $B$ (lower $\mu_{\Phi}$) as the stronger magnetic tension/pressure suppresses perpendicular compression. In our Figure~\ref{fig:B_vs_rho_pre-sf}, we observe this constant $B$ regime for $\rho \lesssim 10^{-21}$ g cm$^{-3}$. At higher densities, the gas motion is no longer isotropic but rather follows the compression of the core due to self-gravity, and the assumption of flux-freezing implies $B \propto \sqrt{\rho}$. Our results suggest that this relationship is not affected by the assumption of non-ideal MHD, at least prior to the development of the magnetic plateau at even higher densities, as discussed in the preceding paragraph. 

In almost all of our models, disks form shortly ($\sim$100$-$200 kyr) after the onset of star formation, as seen in Figure~\ref{fig:sinks_with_disks}, and persist another $\sim$150$-$400 kyr (see Table~\ref{tab:model_results}). Contrary to the ``magnetic braking catastrophe," disk formation occurs even in the ideal MHD models. However, the quadruple ``disk" in the ideal$+$jets model is notably less massive than the disks formed in other models and is not well resolved given our mass resolution of $\Delta m = 10^{-5}\; M_{\odot}$. As a binary disk forms in the ideal MHD model with no jets, it is possible that our sub-grid jet feedback model, and not the ideal MHD assumption, is more disruptive to disk formation. Aside from the quadruple ``disk" in the ideal$+$jets model, we observe no significant differences in disk properties such as mass, radius, or magnetization between the binary disk in the ideal MHD model and the disks in the non-ideal MHD models, except that the ideal binary disk lifetime is $\sim$200 kyr shorter than the non-ideal binary disk lifetimes. We caution, however, that our sample of disks is too small to draw any conclusions about disk population statistics: our sample effectively consists of one ideal MHD disk and nine non-ideal MHD disks, where the non-ideal disks differ from one another in type (single vs. multiple), included physics (non-ideal MHD effects and sub-grid feedback), and properties such as size and plasma $\beta$ (see Figs.~\ref{fig:disk_mass},~\ref{fig:disk_radius}, and~\ref{fig:rad_prof_beta}). Further studies beginning with a larger initial cloud mass and forming a greater number of disks are necessary to make any definitive statements about differences in disk properties between ideal and non-ideal MHD models.

When we extend our analysis to include the lower-density ($n > 10^8$ cm$^{-3}$) envelopes in which the disks are embedded, however, the differences between the ideal and the non-ideal calculations 
are more apparent. As shown in Figs.~\ref{fig:disk_mass} and \ref{fig:disk_radius}, the envelopes in the two ideal MHD models (with and without jet feedback) are significantly less massive and more compact than the non-ideal MHD envelopes. The envelope in the ideal model is only slightly more massive than the binary disk it encloses and disappears after just under 200 kyr, while the non-ideal envelopes are several orders of magnitude more massive than the embedded disks and persist for $\sim$300-600 kyr. These results suggest that while magnetic braking does not entirely suppress disk formation in the ideal MHD case, it does notably reduce the extent of the surrounding lower-density envelope material in which the disks are embedded and from which they are able to accrete from. This may lead to less massive or shorter-lived disks in the ideal MHD case.

Our disk formation outcomes are in agreement with the study of low-mass cluster formation by \cite{wurster_2019_low_mass_cluster}, who concluded that ``there is no magnetic braking catastrophe." They modeled the collapse of $M_0 = 50\; M_{\odot}$ turbulent molecular cloud cores using smoothed particle hydrodynamics (SPH) calculations containing $N = 5\times10^6$ particles (i.e., using the same mass resolution $\Delta m = 10^{-5} \; M_{\odot}$ as in our calculations) and including radiative transfer and all three non-ideal MHD effects. They compared the results of assuming ideal vs. non-ideal MHD as well as varying the initial mass-to-magnetic flux ratio $\mu_{\Phi} = 3$, 5, 10, and 20. Their models formed 8$-$19 stars by 1.45 $t_{\rm ff}$ (275.5 kyr), likely due to their higher initial cloud density $\rho_0 = 1.22 \times 10^{-19}$ g cm$^{-3}$ compared to our $\rho_0 = 2.77 \times 10^{-20}$ g cm$^{-3}$. They found that single disks of radii $\sim$10$-$80 AU and multiple disks of radii up to $\sim$500 AU formed in all of their models, with no obvious dependence on model parameters. The magnetic field strength in the disks was independent of the initial magnetic field strength of the cloud, although the disk field strengths in the non-ideal MHD models spanned a narrower range of values, suggesting that non-ideal MHD processes may moderate the disk magnetic field. These calculations did not include protostellar feedback.

As the Hall effect is numerically challenging to implement, several works investigating non-ideal MHD in stellar cluster and protostellar disk formation have focused on ambipolar diffusion only. Our non-ideal OA and non-ideal OA$+$jets models can be directly compared with these works. \cite{lebreuilly_2024a} calculated the collapse of $M_0 = 500$$-$1000 $M_{\odot}$ molecular clouds with radiative transfer and ambipolar diffusion using adaptive mesh refinement (AMR), investigating the effects of ideal MHD vs. ambipolar diffusion, varying the initial mass-to-magnetic flux ratio, and using flux-limited diffusion vs. assuming a barotropic equation of state. Their fiducial non-ideal MHD model, which was integrated up to a final star formation efficiency (SFE) of 0.15, had $\mu_{\Phi} = 10$ and formed 88 stars, with approximately 70\% of stellar systems hosting a disk at some point during the calculations. They compared disk properties at birth, 10, and 20 kyr after formation; we note that this is a significantly earlier stage of disk evolution than is reached in our models. They found that the initial mass-to-magnetic flux ratio ($\mu_{\Phi} = 10$, 50) had a greater effect on disk size and radius than including ambipolar diffusion, with typical disk sizes of $\sim$30$-$40 AU in the $\mu_{\Phi} = 10$ models and $\sim$60$-$80 AU in the $\mu_{\Phi} = 50$ models.

In addition to non-ideal MHD effects, misalignment between the magnetic field direction and initial rotation axis of the cloud has been investigated as a possible resolution to the magnetic braking catastrophe. The isolated spherical collapse, $M_0=1 \; M_{\odot}$ ideal MHD calculations of \cite{joos_2012} found that disks formed in all of their models with an initial misalignment $\theta_{B}\gtrsim20^{\circ}$ and mass-to-flux ratio $\mu_{\Phi}\gtrsim2$. The resistive (i.e., Ohmic) non-ideal MHD calculations of \cite{machida_2020} showed that the disk, outflow, and magnetic field axis are rarely aligned, except when $\theta_{B} = 0^{\circ}$, and that outflows tend to be suppressed when $\theta_{B} = 90^{\circ}$. Most of the disks in our models show some tendency to be perpendicular to the local magnetic field, albeit with significant scatter, as seen in Figure~\ref{fig:disk_B_misalignment}; magnetic braking is least effective in this configuration. The disks in the Hall models, however, become increasingly anti-aligned with the local field over time. This is consistent with the 2D axisymmetric calculations of \cite{zhao_2021}, who found that disks formed during Hall-dominated collapse could only have an anti-aligned magnetic field polarity, such that $\ve{\omega}_{\rm disk} \cdot \ve{B} < 0$. Again, a larger statistical sample is needed in order to make any definitive statements on the impact of non-ideal effects on disk orientation.

Turbulence also reduces the effectiveness of magnetic braking. \cite{He_Ricotti_2023, He_Ricotti_2025}, found that in the ideal MHD limit large ($R_{\rm disk} \sim$ 200$-$6000 AU), rotationally-supported, thick ($q \sim$ 0.2$-$0.5) disks formed around massive protostars even for pre-stellar cores with $\mu_{\Phi} \sim$ 1$-$5; only cores with $\mu_{\Phi} < 1$ failed to form disks. The large turbulence resulting from the non-axisymmetric gravitational collapse of the cores reduced the effectiveness of the magnetic braking torque within the disks by roughly an order of magnitude, which they deduced by comparing the magnitudes of the turbulent $B_r$ component of the magnetic field within the disks to the ordered $B_{\phi}$ component. Turbulence also dominated the vertical support within the inner disk and was comparable to the magnetic pressure support in the outer disk; the balance of these forces determined the disk thickness. They commented that this is in contrast to disks around low-mass stars, which tend to be supported instead by thermal pressure. Indeed our Figure~\ref{fig:rad_prof_beta} suggests that thermal pressure generally dominates throughout the disk, though at certain stages of disk evolution the magnetic pressure may dominate. We do not resolve the disk scale height in our calculations and so do not comment on the vertical support of our disks. However, turbulence can also diffuse the magnetic field and modify the magnetic field topology via turbulent magnetic reconnection \citep[e.g.,][]{santos_2012, santos_2016}. Our initial molecular cloud cores are supersonically turbulent, which may contribute to the reduced effectiveness of magnetic braking throughout the collapse.

In summary, disks form regardless of the assumption of ideal or non-ideal MHD. This may be due to the reduced effectiveness of magnetic braking resulting from misalignment between the disk rotation axis and the magnetic field as well as the inclusion of supersonic turbulence, conditions which are relevant even in the ideal MHD case. However, non-ideal MHD disks are generally longer-lived and are embedded within larger, lower-density rotating envelopes, suggesting that non-ideal MHD effects are relevant at these scales and are important for modeling the long-term evolution of disks as they continue to accrete material from these envelopes. Non-ideal MHD effects are also expected to be stronger at higher densities ($\rho \gtrsim 10^{-12}$ g cm$^{-3}$) and smaller scales ($\lesssim$1 AU) than are probed by our chosen resolution \citep{wurster_2016_nicil}.

\subsection{Effects of including protostellar jet feedback}\label{sec:effects_of_jet_feedback}
Our models with jet feedback consistently show increased fragmentation when compared to their counterparts with no feedback, resulting in more stars being formed, as summarized in Table~\ref{tab:model_results}. The final masses of the stars formed in models with jets are also roughly an order of magnitude less massive than those formed in models without jets, as seen in Figure~\ref{fig:sink_mass}. Previous calculations examining the effects of including protostellar feedback, such as radiation and jets, on low-mass star formation by \cite{hansen_2012}, showed that while radiative feedback can suppress fragmentation, outflows reduce protostellar masses and accretion rates and thereby luminosities. When outflows are included, the total stellar luminosity drops by an order of magnitude and radiation is much less effective at suppressing fragmentation. Calculations by \cite{gus_2021} using the \textsc{STARFORGE} numerical framework also showed that including jet feedback dramatically disrupts the accretion flow on $\lesssim$0.1 pc scales and shifts the turnover in the stellar initial mass function (IMF) to smaller scales.

We observe disk formation in our models both with and without jet feedback, although it is possible that our jet feedback prescription is somewhat disruptive to disk formation, particularly when there are many nearby stars. The non-ideal OAH$+$jets model, which forms 10 stars within a $\sim$120 kyr period, only forms one disk which persists for longer than a couple of snapshots ($\gtrsim$18 kyr): a number of other disks formed at this time are destroyed by jet feedback from neighboring protostars a snapshot or two ($\sim$6$-$12 kyr) after formation. The ideal$+$jets model, meanwhile, forms the smallest and least-resolved quadruple ``disk" out of all models. By injecting mass and disturbing the material in the vicinity of existing protostars, however, jet feedback may also promote a ``second generation" of disk formation. As seen in Figures~\ref{fig:disk_mass}--\ref{fig:sinks_with_disks}, the quadruple disk formed in the ideal+jets model (second panel) and the binary disk formed in the non-ideal OAH+jets model (bottom panel) form $\sim$0.4 Myr after the onset of star formation, after the dispersal of the initial envelope, from gas which   has been kicked up by jet feedback. We also observe from Figure~\ref{fig:disk_mass} that jet feedback tends to replenish and stabilize the envelope mass.

As discussed in Section~\ref{results_outflows} and shown in Figure~\ref{fig:proj_density_ideal_outflow_time_evolution}, we observe a loosely-collimated, low-velocity ($\sigma_{\rm 1D} \sim 0.5$ km s$^{-1}$) outflow ejected from the binary disk system in the ideal MHD model. However, we do not observe any highly-collimated, high-velocity ($\sigma_{\rm 1D} \gtrsim 10$ km s$^{-1}$) outflows in any our models without sub-grid jet feedback, necessitating the use of the sub-grid model to capture some of the dynamical effects caused by high-velocity jet feedback. Additionally, our sink particle prescription and mass resolution of $\Delta m = 10^{-5}\; M_{\odot}$ mean that we do not resolve the star-disk interaction boundary. As protostellar jets are thought to be launched due to the winding of the magnetic field lines and infall of material at the protostellar surface \citep{shu_1988, pelletier_1992}, it is likely that our models do not accurately capture coupled jet-disk dynamics, such as the correlation between episodic accretion and outflow events \citep{arce_2007}. Similar studies of protostellar disk formation at comparable resolution \citep[e.g.,][]{wurster_2019_low_mass_cluster, lebreuilly_2024a, lebreuilly_2024b} likewise do not observe self-consistent jet launching and attribute the lack of jets to insufficient numerical resolution, frequent changes in disk orientation due to close encounters, and the lack of coherent magnetic field structure due to turbulent initial conditions. Self-consistent jet launching has been observed in higher-resolution calculations assuming ideal and resistive (Ohmic) non-ideal MHD \citep[e.g.,][]{tomida_2013, machida_2020}, as well as non-ideal calculations with ambipolar diffusion and the Hall effect \citep[e.g.,][]{wurster_lewis_2020}. However, the non-ideal MHD calculations of \cite{wurster_lewis_2020} showed that non-ideal MHD tended to suppress magnetically-launched outflows from the stellar core. Given enough initial rotation or turbulence, thermally-launched outflows were observed in the non-ideal MHD models after the formation of the stellar core.

In a companion study to their investigation of the effects of including ambipolar diffusion in calculations of disk formation in $M_0 = 1000 \; M_{\odot}$ clouds, \cite{lebreuilly_2024b} also included sub-grid jet feedback using an implementation similar to the \textsc{STARFORGE} framework (i.e., a fraction $f_{\rm w} = 0.3$ of accreted material is launched along the sink angular momentum axis at a fraction $f_{\rm K} = 0.3$ of the escape velocity). They found that including outflows did not have a significant effect on the disk masses and radii by the time the two models were evolved up to a SFE of 0.1 ($\sim$30$-$40 kyr after the onset of star formation). However, including outflows significantly increased fragmentation and lowered accretion luminosities by roughly an order of magnitude, reducing the disk temperatures by a factor of about 2.

\subsection{Comparisons with observations}\label{section_observations}
The disks formed in our models have masses of a few 0.001$-$0.1 $M_{\odot}$ and radii of a few 10$-$100 AU, as seen in Figures~\ref{fig:disk_mass} and~\ref{fig:disk_radius}; given the small sample of disks we do not present more detailed statistics. The VANDAM survey towards a sample of 328 protostars in the Orion molecular clouds \citep{tobin_2020} 
measured dust continuum emission and reported average dust disk radii of 45 and 37 AU, as well as average dust disk masses of 26 and 15 $M_{\oplus}$ (or $7.8 \times 10^{-5}$ and $4.5 \times 10^{-5} \; M_{\rm \odot}$), for Class 0 and Class I sources, respectively \citep[with ages of sources in these classes estimated to be roughly 0.2 and 0.5 Myr; see, e.g.,][]{evans_2009, dunham_2015}. These disks are somewhat more compact and significantly less massive than those formed in our numerical models. However, a number of uncertainties make it difficult to draw direct comparisons between observed and modeled disk sizes. The gas disk masses and radii are measured directly from our simulations and are not post-processed to account for optical depth, observational effects, and instrumental resolution. The disk properties reported by the VANDAM survey, meanwhile, describe only the dust content of the disks, which may not be well-coupled to the gaseous component due to both different dynamics as well as biases and limitations in the attainable resolution. While a dust-to-gas ratio $f_{dg} = 0.01$ is commonly used to calculate the mass of the gaseous component, uncertainties in the dust-to-gas ratio, the spatial distribution of dust grains, and the dust size distribution can lead to large errors in the gas properties inferred from the dust emission, particularly when the flux from dust thermal emission is used to estimate the disk mass \citep[see, e.g., the review by][]{tsukamoto_2023_review}. As disks evolve on timescales shorter than the age-spread of star-forming regions, the predicted distribution of disk properties is also likely to depend upon initial conditions and the resulting star cluster age distribution. Given these caveats, our results are broadly in agreement with observations, as well as previous numerical non-ideal MHD studies.

As discussed in Sections~\ref{section_bulk_properties} and \ref{section_disk_stability}, we do not observe disk fragmentation in any of our models, although we do identify a large ($\sim$2000 AU) rotating structure in the non-ideal OA+jets model that is gravitationally unstable and fragments to form a number of stars (see Figure~\ref{fig:proj_density_rotating_core_fragmentation} and the analysis of this structure in Appendix~\ref{appendix_rotating_envelope}). The disks shown in Figures~\ref{fig:toomre_Q_nOA_single}--\ref{fig:toomre_Q_nOAHj_binary} are most likely to be unstable during the first $\sim$50--100 kyr of disk evolution, when the disk-to-stellar mass ratio is $M_{\rm disk}/M_{\rm stellar} \gtrsim 0.1$ and the Toomre Q parameter in parts of the disk is $Q \sim 1$. However, a number of observed protostellar systems have been argued to be situated in young, gravitationally-unstable disks, suggesting evidence for disk fragmentation as a mechanism for multiple system formation. \cite{tobin_2016_triple_system} reported observations of a disk with spiral structure surrounding three Class 0 protostars in the L1448 IRS3B system. They used dust continuum emission to estimate the disk mass and C$^{18}$O molecular line emission to determine the velocity profile of the disk and concluded that the disk around IRS3B is marginally unstable ($Q \sim 1$) at radii between 150 AU and 320 AU for a wide range of accretion rates through the disk, consistent with both the continued presence of spiral arms as well as the formation of the IRS3B-c protostar at a separation of 254 AU from the center of the system. 

More recently, \cite{li_2025_septuple_system} detected a young septuple protostellar system embedded within a disk in the high-mass star-forming region NGC 6334IN, with an average projected separation of 298 AU between nearest-neighbor pairs. They also used gas kinematics and dust continuum emission to estimate a disk-to-stellar mass ratio $M_{\rm disk}/M_{\rm stellar} \sim$ 0.2$-$0.6 and a Toomre Q parameter $Q > 10$ in the central $\sim$500 AU region, suggesting that the disk is in a gravitationally-unstable regime. In contrast to  \citet{tobin_2016_triple_system}, the rotating structure in \cite{li_2025_septuple_system} is quite irregular and does not clearly visually resemble an ordered disk. They use the presence of rotation, proximity of members, and apparent gas instability to argue that the septuple system likely formed from disk fragmentation. 

We stress that here all binary and multiple companions form at wider separations, outside the typical disk scale, via filament and core fragmentation. A similar result for massive, ideal MHD disks was observed by \cite{He_Ricotti_2023}, whose disks were on average Toomre-stable. Consequently, close companion separation, i.e., within $\sim$100 AU, and a marginally unstable disk do not definitively indicate that a particular companion formed from disk fragmentation. We also note that the gas profile of our disks follows Keplerian rotation throughout (see Figure~\ref{fig:rad_prof_omega}), suggesting that disturbances in the disk produced by migration or dynamical interactions may be difficult to detect at low observational resolution.  We find that the rotation signature extends to $\sim$1000 AU in some cases, which suggests that the disks connect to a lower-density rotating envelope. Consequently, a rotation signature does not necessarily indicate an ordered disk. Given the apparent ubiquity of migration of companions predicted by core fragmentation models \citep[see also][]{OffnerDunham2016,LeeOffner2020,GuszejnovRaju2023,KuruwitaHaugbolle2023}, we caution against drawing inferences about companion formation based on their location within a disk.

\subsection{Caveats in calculating non-ideal MHD resistivities}\label{section_caveats}
The efficiency and relative ordering of the non-ideal MHD resistivities depend on microphysical parameters such as the dust grain size distribution and cosmic ray ionization rate. While we adopt a single dust grain radius $a_g = 0.1\;\mu$m and fixed cosmic ray ionization rate $\zeta_{\rm CR} = 1.6 \times 10^{-17}$ s$^{-1}$ for simplicity, several other studies have investigated the effects of modifying the dust grain properties and/or changing the cosmic ray ionization rate on disk formation. \cite{wurster_2021} used models of simple parametrized disks to predict the effects of different grain size assumptions on the values and relative importance of the non-ideal MHD resistivities, comparing models assuming an MRN \citep[Mathis-Rumpl-Nordsieck;][]{mathis_1977} distribution at several different magnetic field strengths and with or without cosmic ray attenuation to those using singly-sized dust grains, as commonly used in numerical studies such as this work. While the Hall resistivity was the dominant term throughout the disk in most of these idealized models, the ratios $\eta_{\rm O}/\eta_{\rm A}$ and $|\eta_{\rm H}|/\eta_{\rm A}$ differ by less than a factor of ten throughout a significant fraction ($\sim$30$-$90\%) of the disk, suggesting that, regardless of which term dominates, all three non-ideal MHD effects are relevant when considering the evolution of the disk. \cite{zhao_2021} performed 2D axisymmetric calculations using a modified MRN grain size distribution, fixing the power law index and maximum grain size but varying the minimum grain size, to investigate the effect on disk formation. They found that removing the smallest nanometer-sized grains ($a_g \lesssim 10$ nm) promoted disk formation, and that truncating the MRN at different minimum grain sizes led to either a Hall-dominated ($a_{\rm min} = 0.03 \; \mu$m) or ambipolar-dominated ($a_{\rm min} = 0.1 \; \mu$m) regime, with the morphologies and kinematics of the envelope and disk showing clear differences between the two types of collapse. While observations suggest that the smallest grains are rapidly depleted due to grain growth in cold dense cores \citep[e.g.,][]{tibbs_2016}, the exact grain size distribution in protostellar environments is still poorly constrained. As dust grains may coagulate and fragment during the gravitational collapse and disk formation phases, however, it is unlikely that the size distribution of grains in the disk resembles that observed in the ISM \citep{dominik_1997, ormel_2009}. Non-ideal MHD studies have only just begun to follow dust grain property evolution \citep{marchand_2023, tsukamoto_2023} but have already shown that these processes can significantly modify the values of the non-ideal MHD resistivities throughout disk formation and evolution. Dust grains are also expected to be distributed differently from the gas in protostellar disks due to processes such as settling and radial drift \citep{dullemond_2004, miotello_2023, birnstiel_2024}. While we assume that the gas and dust are well-coupled and use a constant gas-to-dust ratio $f_{dg} = 0.01$ in order to calculate the non-ideal MHD resistivities, future studies should take into account dust dynamics and evolving dust grain properties in order to more accurately model non-ideal MHD effects.

However, in addition to considering more detailed microphysics, the fundamental assumptions underlying non-ideal MHD effects in protostellar disk regimes may need to be revisited. Recently, \cite{hopkins_2024}  challenged several approximations typically used to derive the standard non-ideal MHD equations in weakly-ionized astrophysical systems such as protostellar disks. They argue that these equations are no longer self-consistent when magnetic gradients become too steep and the implied drift velocities between different charged species exceed the thermal velocities. In the case of superthermal drift velocities, the collision rates used to calculate the non-ideal MHD resistivities, which depend on the thermal velocities but assume negligible relative drift velocities, are incorrect. Superthermal drift velocities also generate micro- and mesoscale plasma instabilities, giving rise to a so-called ``anomalous resistivity." They argue that the leading-order effects of correcting for superthermal drifts amount to modifying the effective collision rates and adding an approximate anomalous resistivity. These corrections reduce the ambipolar diffusivity and greatly enhance the Ohmic resistivity such that the Hall-dominated regime is effectively eliminated. The non-ideal effects then always act diffusively and not dispersively as with the Hall effect, acting to smooth out magnetic gradients and restore subthermal drift velocities. As protostellar disk conditions typically fall in the Hall-dominated regime, this has significant implications for the relative ordering of the non-ideal terms throughout the disk and calls into question several expected disk behaviors ascribed to the Hall effect, such as the preference for anti-alignment between the disk rotation axis and local magnetic field. 

\section{Summary}
We follow the collapse of magnetized, turbulent 50 $M_{\odot}$ molecular cloud cores under different assumptions of (non-)ideal MHD and protostellar feedback through the the formation of stellar systems and disks using the MHD code \textsc{GIZMO}. Our results on disk formation and evolution can be summarized as follows:
\begin{itemize}
    \item Inclusion of ambipolar diffusion leads to the development of C-type shocks during the turbulent driving phase, modifying the resulting density distribution. This phenomenon is strongly sensitive to the assumed dust grain size. 
    \item Disks form in all models, with ideal MHD and non-ideal MHD, with and without sub-grid jet feedback. Disruption by jet feedback causes certain disks to be short-lived ($\lesssim$12 kyr); however, jet feedback also drives a second generation of disk formation by replenishing the available mass reservoir after the natal envelope has been dispersed and/or accreted. The disk population is diverse.
    \item Disk masses of $\mathcal{O}(0.001-0.1\, M_{\odot})$ and radii of $\mathcal{O}(10-100$ AU$)$ are broadly in agreement with observed protostellar disks, given the large uncertainties in dust models and flux-to-mass conversions.
    \item Disks tend to belong to the most massive stars; smaller disks are likely unresolved in these simulations.
    \item No disk fragmentation is observed. Disks are stable according to the Toomre $Q$ criterion. Multiple systems form due to core and filament fragmentation.
    \item Due to our small yet diverse disk sample, we do not observe any significant differences in disk properties between the ideal and non-ideal MHD models. Non-ideal MHD disks may be more massive and somewhat longer-lived. A larger statistical study is necessary to determine whether these differences are significant.
    \item The differences between the ideal and non-ideal MHD models are more pronounced when considering the lower-density ($n > 10^8$ cm$^{-3}$) rotating envelopes in which the disks are embedded. The ideal MHD envelopes are notably less massive and more compact, implying a smaller mass reservoir available to be accreted by the disk as it evolves.
    \item All disks tend to be relatively well-aligned with the stellar angular momentum axis ($\theta_{\rm rel} \lesssim 20^{\circ}$) and have an aspect ratio $q \lesssim 0.2$.
    \item Disk angular momenta in models without the Hall effect show some tendency to be perpendicular to the local magnetic field. The angular momenta of disks in models with the Hall effect tend to become anti-aligned with the local field with time. Again, larger samples are needed to make any definitive claims.
    \item Sub-grid protostellar jet feedback increases core fragmentation and reduces the final stellar masses. 
    \item Low-velocity ($\sigma_{\rm 1D} \sim 0.5$ km s$^{-1}$), loosely-collimated bipolar outflows are observed in the ideal MHD model. We do not observe any self-consistently launched high-velocity, highly-collimated protostellar jets in any of our models.
\end{itemize}

\section*{Acknowledgements}
NF acknowledges support from the U.S. Department of Energy, Office of Science, Office of Advanced Scientific Computing Research under Award Number DE-SC0023112. SSRO acknowledges support by NSF 2407522 and a Peter O'Donnell Research Fellowship. PFH was supported by a Simons Investigator Grant. Calculations and analyses for this paper were performed on the Frontera supercomputer at the Texas Advanced Computing Center (TACC\footnote{http://www.tacc.utexas.edu}) under the allocations AST23007 and AST23034. The column density and velocity dispersion figures were made using \texttt{yt}\footnote{https://yt-project.org}.

\section*{Data Availability}
The data supporting the plots within this article is available by request to the corresponding authors. A public version of the \textsc{GIZMO} code is available\footnote{http://www.tapir.caltech.edu/~phopkins/Site/GIZMO.html}.



\bibliographystyle{mnras}
\bibliography{example} 




\appendix

\section{Fragmenting rotating envelope} \label{appendix_rotating_envelope}
\begin{figure}
	\includegraphics[width=\columnwidth]{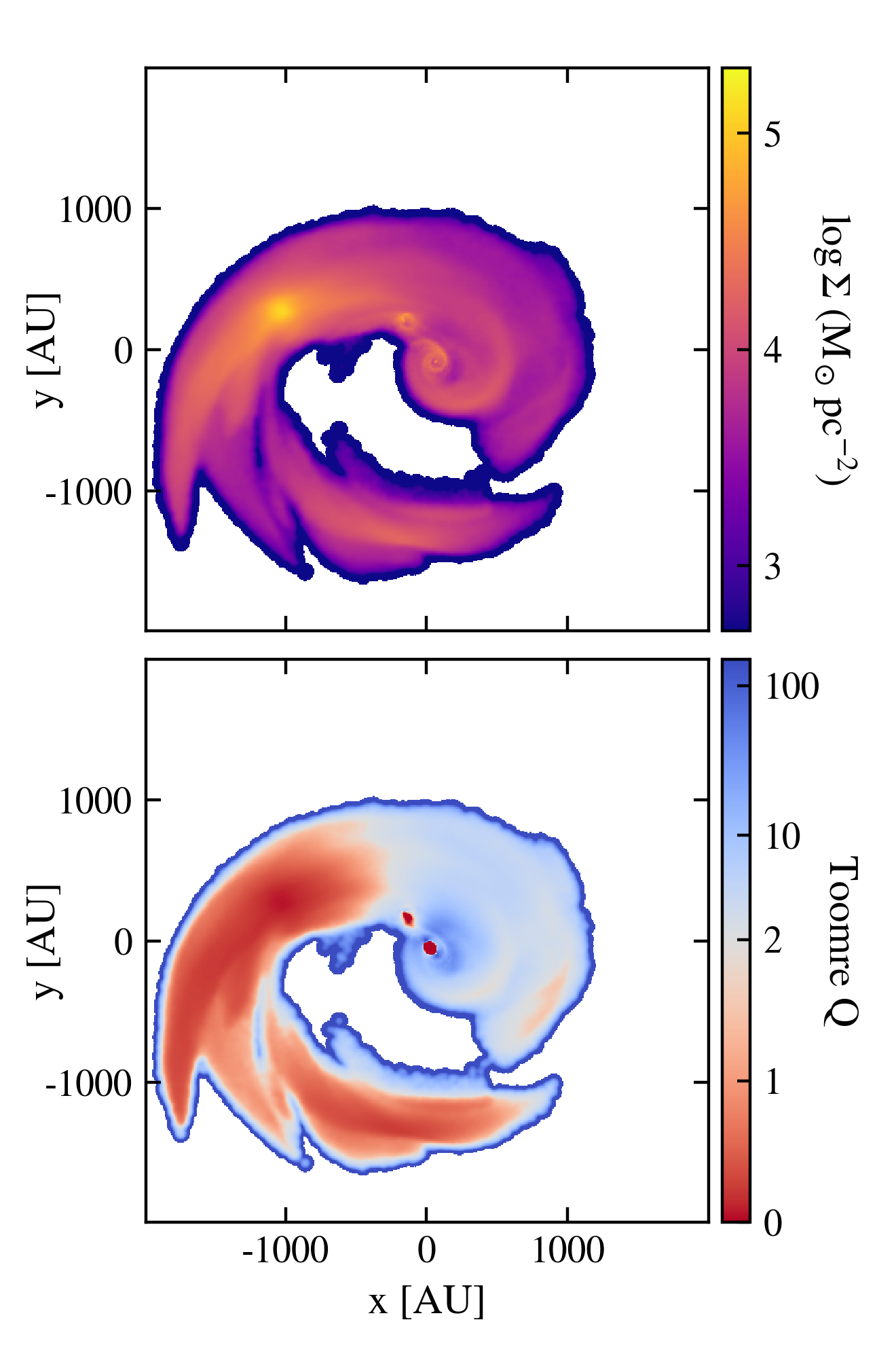}
    \caption{Projected surface density (top) and Toomre $Q$ (bottom) values of the gas with density $n \geq 10^8$ cm$^{-3}$ within 2000 AU of the binary system in the nOA$+$jets model just prior to the formation of the third star, as shown in the top row of the column density projections in Figure~\ref{fig:proj_density_rotating_core_fragmentation}. We project along the direction of the net angular momentum of the gas, and use a linear colormap normalization between $Q=0$ and $Q=2$ and a logarithmic normalization for $Q > 2$.}
    \label{fig:rotating_core_toomre_Q}
\end{figure}

\begin{figure}
	\includegraphics[width=\columnwidth]{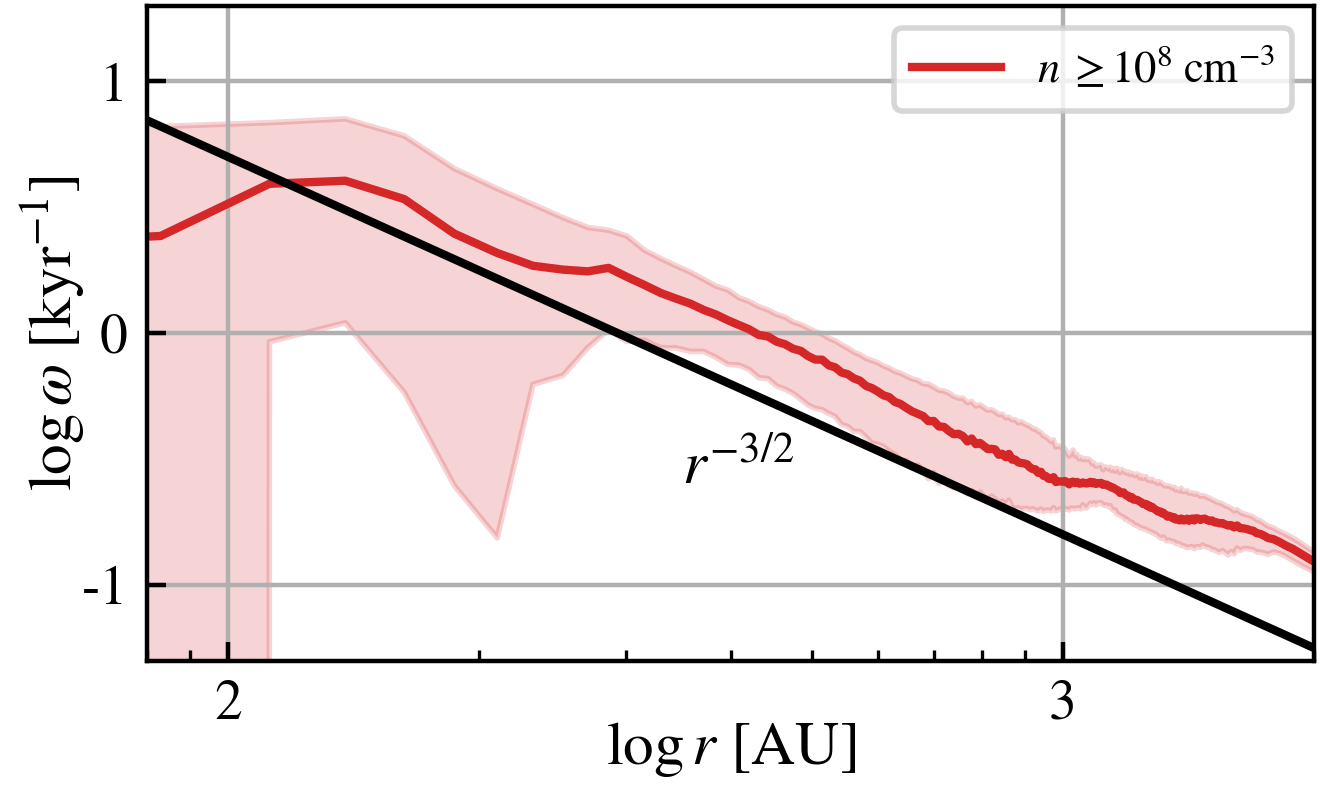}
    \caption{Angular velocity profile of the gas with density $n \geq 10^8$ cm$^{-3}$ within 2000 AU of the binary system in the nOA$+$jets model just prior to the formation of the third star, as shown in the top row of the column density projections in Figure~\ref{fig:rotating_core_velocity_profile}. The shaded region encloses the 5th and 95th percentiles. The black line indicates Keplerian rotation, $\omega(r) \propto r^{-3/2}$.}
    \label{fig:rotating_core_velocity_profile}
\end{figure}

While we do not observe disk fragmentation in any of our models, we identify a rotating structure extending out to $\sim$2000 AU in the non-ideal OA$+$jets model. This structure is centered around the first two stars to form in this model, and then later fragments to form an additional 3 stars. This structure is not picked out as a disk by our disk identification algorithm when we use a density threshold $n \geq 10^9$ cm$^{-3}$, as in \cite{joos_2012}. However, adopting a lower density threshold $n \geq 10^8$ cm$^{-3}$ leads to the identification of this structure as a disk in Keplerian rotation. The top row of Figure~\ref{fig:rotating_core_toomre_Q} shows the surface density of this lower density ``disk", while the bottom row of the figure shows that it is quite gravitationally unstable beyond the inner $\sim$1000 AU according to the Toomre $Q$ criterion. Figure~\ref{fig:rotating_core_velocity_profile} shows that the angular velocity profile of this structure is in Keplerian rotation. We identify this structure as part of the rotating envelope material; however, its presence suggests that the boundary between the disk and the envelope might not be as distinct as is typically assumed.

\section{Oblique isothermal C-shock test}\label{appendix_C_shock}
The steady-state oblique isothermal C-shock, which permits a semi-analytic solution, is a well-known test problem for ambipolar diffusion \citep[e.g., ][]{mac_low_1995_C_shock, duffin_pudritz_2008_C_shock, masson_2012_C_shock, wurster_2014_C_shock}. Here we present the results of the test in 1D, following the problem setup and initial conditions of \cite{mac_low_1995_C_shock} and \cite{wurster_2014_C_shock}. In this test, an isothermal gas in a magnetic field is reflected off of a wall, with the initial uniform magnetic field $\ve{B}_0 = B_{x0} \uve{x} + B_{y0} \uve{y}$ oriented at some angle $\theta = \tan^{-1}(B_{y0}/B_{x0})$ with respect to the velocity $\ve{v}_0 = v_{x0} \uve{x}$. We include Ohmic resistivity and ambipolar diffusion, but ignore the Hall effect for this test.

The semi-analytic solution to the problem can be derived from the non-ideal MHD equations by setting $v_z = B_z = 0$ and assuming that all quantities are independent of the $y$- and $z$-coordinates as well as time. The divergence constraint $\grad \cdot \ve{B} = 0$ then implies that $B_x$ is constant. The mass, momentum, and magnetic induction equations can be written in conservative form as
\begin{equation}
    \frac{\partial \ve{U}}{\partial t} + \grad \cdot \ve{F}_{\rm ideal}(\ve{U}) + \grad \cdot \ve{F}_{\rm non-ideal}(\ve{U}) = 0,
\end{equation}
where
\begin{equation}
\begin{split}
    &\ve{U} = 
    \begin{pmatrix}
        \rho \\
        \rho \ve{v} \\
        \ve{B}
    \end{pmatrix}, \\ &\ve{F}_{\rm ideal} = 
    \begin{pmatrix}
        \rho \ve{v} \\
        \rho \ve{v}\ve{v}^T + P - \ve{B}\ve{B}^T \\
        \ve{v} \ve{B}^T - \ve{B}\ve{v}^T
    \end{pmatrix}, \\
    &\ve{F}_{\rm non-ideal} = 
    \begin{pmatrix}
        0 \\
        0 \\
        \eta_{\rm OR} \grad \ve{B} + \eta_{\rm AD} [(\ve{J} \times \ve{b}) \ve{b}^T - \ve{b}(\ve{J} \times \ve{b})^T]
    \end{pmatrix}.
\end{split} 
\end{equation}
Here $P = \rho c_s^2$ is the isothermal pressure, $\ve{J} = \grad \times \ve{B}$, and $\ve{b} = \ve{B}/|\ve{B}|$. We assume that $\eta_{\rm OR}$ is constant and that $\eta_{\rm AD} = v_{\rm A}^2/(\gamma_{\rm AD} \rho_i)$, where $\gamma_{\rm AD}$ is the collisional coupling constant between the ions and neutrals, $\rho_i$ is the ion density (assumed to be constant), and $v_{\rm A} = B/\sqrt{\rho}$ is the Alfv\'{e}n speed. The equations describing the equilibrium C-shock in the $x$-direction are then
\begin{align}
    \frac{d}{dx}(\rho v_x) &= 0, \\
    \frac{d}{dx}\left(\rho v_x^2 + \rho c_s^2 + \frac{1}{2} B_y^2\right) &= 0, \\
    \frac{d}{dx} \left(\rho v_x v_y - B_x B_y \right) &= 0, \\
    \frac{d}{dx} \left(B_y v_x - B_x v_y - \eta_{\rm OR} \frac{dB_y}{dx} - \frac{(B_x^2 + B_y^2)}{\gamma_{\rm AD} \rho_i \rho} \frac{dB_y}{dx} \right) &= 0.
\end{align}
This set of differential equations can be reduced to a single ordinary differential equation for $B_y(x)$,
\begin{multline}
    \frac{dB_y}{dx} = \frac{1}{\left(\eta_{\rm OR} + \frac{(B_x^2 + B_y^2)}{\gamma_{\rm AD} \rho_i \rho}\right)} \bigg[(B_y v_x - B_x v_y) - (B_{y0} v_{x0} - B_xv_{y0}) + \\ \left(\frac{dB_y}{dx}\right) \Bigg|_{x = x_0} \left(\eta_{\rm OR} + \frac{(B_x^2 + B_{y0}^2)}{\gamma_{\rm AD} \rho_i \rho_0}\right)\Bigg],
\end{multline}
where quantities with the subscript 0 denote the initial state at $x = x_0$. We set the wall at $x = 0$ and let $x_0 = 30 L_{\rm AD}$, where $L_{\rm AD}$ is the characteristic ambipolar diffusion length scale, $L_{\rm AD} = \eta_{\rm AD} / v_{\rm A}$. To obtain the steady-state analytic solution, we begin with the pre-shock quantities at the initial state and use a fourth-order Runge-Kutta integrator to calculate $B_y$ at the new position, triggering the shock by setting $\left(\frac{dB_y}{dx}\right)\big|_{x = x_0} = \epsilon$, where $\epsilon < 0$ is small. We then update the other unknowns at this position,
\begin{align}
    v_x &= \frac{1}{2C_1} \Bigg[\left(C_2 - \frac{1}{2}B_y^2\right) + \sqrt{\left(C_2 - \frac{1}{2}B_y^2\right)^2 - 4 C_1^2 c_s^2} \; \Bigg], \\
    v_y &= v_{y0} + \frac{B_x}{C_1}(B_y - B_{y0}), \\
    \rho &= \frac{C_1}{v_x},
\end{align}
where $C_1$ and $C_2$ are constants derived from the initial conditions,
\begin{align}
    C_1 &= \rho_0 v_{x0}, \\
    C_2 &= \rho_0 v_{x0}^2 + \rho_0 c_s^2 + \frac{1}{2}B_{y0}^2.
\end{align}
Our pre-shock initial conditions at $x = x_0$ are $\rho_0 = 1$, $v_{x0} = -5$, and $B_{x0} = B_{y0} = 1/\sqrt{2}$, with constants $c_s = 0.1$, $\rho_i = 10^{-5}$, $\gamma_{\rm AD} = 1$, and $\eta_{\rm OR} = 10^{-5}$.

As in \cite{wurster_2014_C_shock}, instead of implementing a reflective boundary at $x=0$, we initialize two inflows that meet at the center of the domain, with $\ve{v}_0 = v_s \uve{x}$ for $x < 0$ and $\ve{v}_0 = -v_s \uve{x}$ for $x > 0$. Since the shocked gas is at rest at $x = 0$ in the computational problem, the necessary shock speed $v_s = 4.45$ is determined from the semi-analytic solution by comparing the change in velocity across the shock front. We use a large computational domain of $-60 L_{\rm AD} < x < 60 L_{\rm AD}$, where $L_{\rm AD} = 10^5$ for our chosen parameters, to ensure that boundary effects at the edge of the domain do not propagate into the center of the domain. To further enforce stable boundary conditions, we `freeze' the evolution of gas cells with $|x - x_0| > 30 L_{\rm AD}$ by zeroing out their fluxes, forcing their magnetic and hydrodynamical properties to remain constant. We evolve the solution for a long time, $t_{\rm max} = 40 \tau_{\rm AD}$, where $\tau_{AD} = \eta_{\rm AD}/v_{\rm A}^2 = 10^5$, so that the shock reaches a steady state solution before comparing the results to the semi-analytic solution. We use $N = 3000$ gas cells for this test.

\begin{figure}
	\includegraphics[width=\columnwidth]{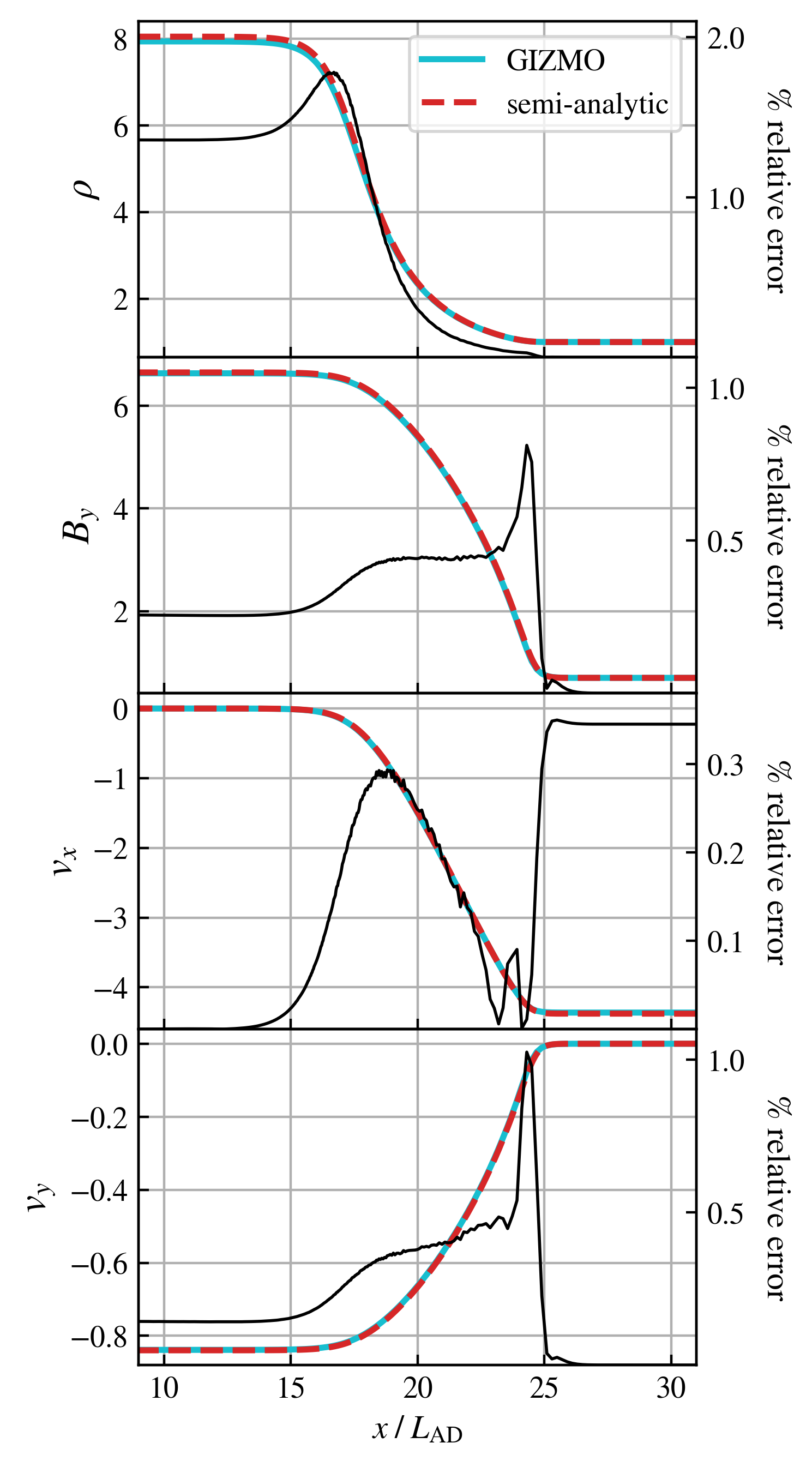}
    \caption{Density, $y$-component of the magnetic field, and $x$- and $y$-components of the velocity in the oblique isothermal C-shock problem at $t = 40 \tau_{\rm AD}$. The cyan line shows the numerical solution obtained with GIZMO, while the dashed red line shows the semi-analytic solution. The thin black line shows the percent relative error between the numeric and semi-analytic solutions, as defined in Eq.~\ref{eq:rel_err}.}
    \label{fig:C_shock_results}
\end{figure}

Figure~\ref{fig:C_shock_results} shows the density, $y$-component of the magnetic field, and $x$- and $y$-components of the velocity across the shock front, comparing the numerical solution obtained with GIZMO at $t = t_{\rm max}$ to the semi-analytic solution. We also plot the percent relative error for each quantity $q$, which we define as in, e.g., \cite{mac_low_1995_C_shock}, \cite{duffin_pudritz_2008_C_shock}:
\begin{equation}\label{eq:rel_err}
    \% \; \text{relative error} \equiv 100 \; \bigg| \frac{q_{\rm analytical} - q_{\rm numerical}}{\max(q_{\rm analytical})} \bigg|
\end{equation}
As can be seen in Figure~\ref{fig:C_shock_results}, the numerical and semi-analytic solutions show good agreement, with the maximum percent relative error at most a few percent, verifying our implementation of ambipolar diffusion.

\section{Whistler wave propagation test}\label{appendix_whistler}
\begin{figure}
	\includegraphics[width=\columnwidth]{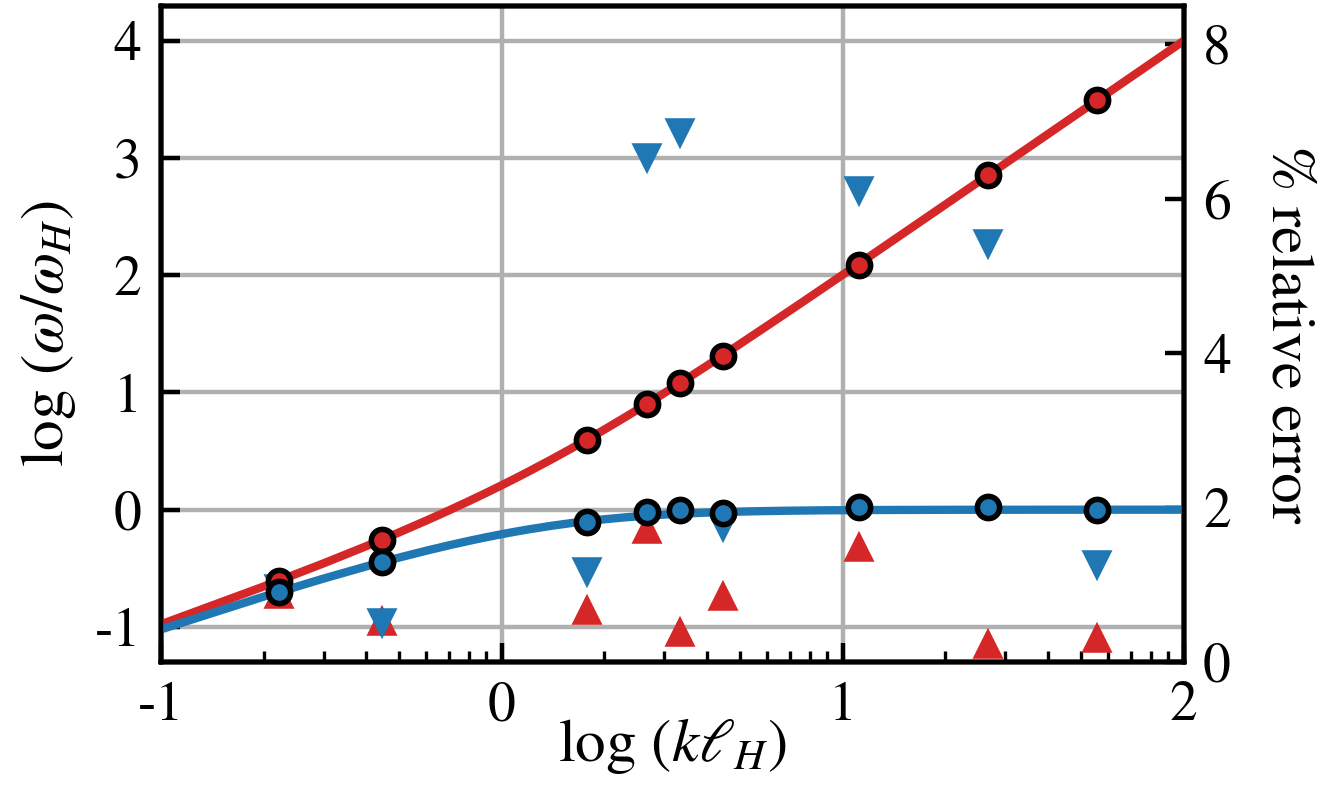}
    \caption{Theoretical (solid lines) and measured (circles) whistler dispersion relation. Downward-pointing triangles mark the percent relative error for the low-frequency ion-cyclotron wave (blue); upward-pointing triangles mark the relative error for the high-frequency whistler wave (red).}
    \label{fig:whistler_results}
\end{figure}

The Hall effect modifies the dispersion relation of linear MHD waves, leading to the generation of ion-cyclotron waves and electron-cyclotron (i.e., whistler) waves. Perturbation analysis gives the following dispersion relation \citep[e.g.,][]{balbus_terquem_2001}:
\begin{equation}
    \frac{\omega}{\omega_{\rm H}} = k \ell_{\rm H} \Bigg[\sqrt{1 + \left(\frac{k \ell_{\rm H}}{2}\right)^2} \pm \frac{k \ell_{\rm H}}{2} \Bigg],
\end{equation}
where $\omega_{\rm H} = v_{\rm A}^2/\eta_{\rm H}$ is the Hall frequency and $\ell_{\rm H} = \eta_{\rm H}/v_{\rm A}$ is the characteristic Hall length. When $k \ell_{\rm H} \ll 1$, the waves are circularly-polarized Alfv\'{e}n waves; when $k \ell_{\rm H} \gg 1$, the right-hand waves ($+$) are the high-frequency whistler waves while the left-hand waves ($-$) are the low-frequency ion-cyclotron waves, which are bounded by $\omega_{\rm H}$. A common test for numerical implementations of the Hall effect is the measurement of this dispersion relation \citep[e.g.,][]{kunz_lesur_2013,wurster_et_al_2016,marchand_et_al_2018,zier_2024_hall}. We set up a 1D box of length $L = 1$ with uniform density and pressure and resolution $N = 160$ and propagate a polarized traveling wave with wave number $k$ in the $x$-direction along a guide field $B_x$. If $B_{y, z} \ll B_x$, the Hall resistivity $\eta_{\rm H}$ is constant. The initial conditions follow the setup of \cite{zier_2024_hall}: $\rho = 1$, $P = 1$, $B_x = B_0$, $B_y = \delta B \cos(kx)$, $B_z = 0$, $v_x = 0$, $v_y = \delta v \cos(kx)$, and $v_z = \delta v (\sigma / \omega) \sin (kx)$, where $B_0 = 10^{-1}/\sqrt{4\pi}$, $\delta B = 10^{-3}/\sqrt{4 \pi}$, $\delta v = (k/\rho)B_0 \delta B[\omega(\sigma^2 - \omega^2)]$ is the initial velocity perturbation, $\sigma = (\eta_{\rm H}/2) k^2$ modulates the amplitude of the magnetic field oscillations, and $\omega = \sqrt{v_{\rm A}^2 k^2 + \sigma^2}$ modulates the frequency of the oscillations. We use periodic boundary conditions (which restricts $k$ to be an integer times $2\pi$) and an adiabatic equation of state with $\gamma = 5/3$. The analytic solution is
\begin{align}
    B_y &= \delta B \cos(\sigma t) \cos(\omega t - kx), \\
    B_z &= \delta B \sin(\sigma t) \cos(\omega t - kx), \\
    v_y &= \delta v \bigg[\cos(\sigma t) \cos(\omega t - kx) + \frac{\sigma}{\omega}\sin(\sigma t) \sin(\omega t - kx) \bigg], \\
    v_z &= \delta v \bigg[\sin(\sigma t) \cos(\omega t - kx) - \frac{\sigma}{\omega}\cos(\sigma t) \sin(\omega t - kx) \bigg].
\end{align}
We measure the dispersion relation for nine different values of the product $k \ell_{\rm H}$. Table~\ref{tab:whistler_params} summarizes the parameters used for the different tests. We integrate until $t = t_{\rm max}$ and measure $B_y(x_{\rm mid}, t)$ with sampling frequency $\Delta t$, where $x_{\rm mid} = L/2$. We then compute the Fourier transform of $B_y$ with respect to $t$ and identify the positions of the two peaks corresponding to the frequencies at which the left- and right-hand waves propagate. Figure~\ref{fig:whistler_results} shows the results. We find very good agreement between the theoretical and the measured dispersion relation, with a relative error of at most $\sim$7\%.

Numerically implementing the Hall effect is challenging because whistler waves can propagate with effectively unbound speed for very small perturbations (i.e., large $k$); grid-scale disturbances thus propagate fastest and can destabilize numerical schemes. Methods which are second-order in time typically require some kind of dissipation mechanism for damping away small-scale whistler waves \citep{zier_2024_hall,iwasaki_tomida_2025}. In addition to the numerical resistivity inherent to the scheme, our tests with $k \ell_{\rm H} \gtrsim 10$ also require a small amount of Ohmic resistivity in order to stabilize the tests. We find $\eta_{\rm O}/\eta_{\rm H} \geq 10^{-2}$ is sufficient for tests with resolution $N = 160$. While we do not explicitly enforce that the Ohmic resistivity be larger than some threshold value times the Hall coefficient, as in, e.g., \cite{zier_2024_hall}, in practice, we find that $\eta_{\rm O}/|\eta_{\rm H}| \gtrsim 10^{-2}$ for densities $\rho \gtrsim 10^{-14}$ g cm$^{-3}$ and that $(\eta_{\rm O} + \eta_{\rm A})/|\eta_{\rm H}| \gtrsim 10^{-1}$ throughout the entire range of densities present in the 50 $M_{\odot}$, $\Delta m = 10^{-5}\;M_{\odot}$ turbulent collapse calculations. We also compare the whistler test parameters to the sizes of the smallest resolvable structures in our disk formation calculations by estimating $L_{\rm small} \sim C \Delta x_{\rm cell}$, where $\Delta x_{\rm cell} = (\Delta m / \rho)^{1/3}$ is the volume-equivalent cell size and $C$ is a constant on the order of a few. Based on Figures~\ref{fig:B_vs_rho_pre-sf}-\ref{fig:eta_vs_rho_nOAH}, we assume a simple power law for the magnetic field $B \propto \rho^{1/2}$ and assume constant $\eta_{\rm H}$ in order to estimate the product $k_{\rm small} \ell_{\rm H}$, where $k_{\rm small} \sim 1/L_{\rm small}$. We find that in the regime $\rho \lesssim 10^{-14}$ g cm$^{-3}$ where $\eta_{\rm O}/|\eta_{\rm H}| \lesssim 10^{-2}$, the product $k_{\rm small} \ell_{\rm H} \lesssim 1$ if we assume $|\eta_{\rm H}| = 10^{19}$ cm$^2$ s$^{-1}$ and $k_{\rm small} \ell_{\rm H} \lesssim 0.1$ if we assume $|\eta_{\rm H}| = 10^{18}$ cm$^2$ s$^{-1}$, well within the regime where the whistler tests are stable even as $\eta_{\rm O} \rightarrow 0$. We see no signs of obvious numerical instabilities \citep[e.g., destroying the global solution with noise originating from high-density regions, as discussed in the Hall-only calculations of][]{zier_2024_hall}, in our full-physics calculations. Even in these idealized wave propagation tests, in this limit we do not see evidence for numerical instability of the sort that would corrupt the large-scale solutions; we observe only deterioration of the analytic solution into noise-dominated solutions at sufficiently high $k$ when run without sufficient numerical or Ohmic resistivity. Nor do we see such damaging instabilities in either the non-linear Hall discontinuity test problem, nor the linear and non-linear Hall magnetorotational instability problems presented and studied in \cite{hopkins_2017_gizmo_nonideal_mhd_diffusion}.

\begin{table}
    \centering
    \begin{tabular}{ |c|c|c|c|c|c|  }
        \hline
        $k/(2\pi)$ & $\eta_{\rm H}$ & $\eta_{\rm O}/\eta_{\rm H}$ & $k \ell_{\rm H}$ & $t_{\rm max}$ & $\Delta t$ \\
        \hline
        2 & $5 \times 10^{-4}$ & $-$ & 0.22 & 400 & 0.0500 \\
        2 & $1 \times 10^{-3}$ & $-$ & 0.45 & 200 & 0.0250 \\
        4 & $2 \times 10^{-3}$ & $-$ & 1.78 & 100 & 0.0125 \\
        6 & $2 \times 10^{-3}$ & $-$ & 2.67 & 80  & 0.0100 \\
        3 & $5 \times 10^{-3}$ & $-$ & 3.34 & 200 & 0.0250 \\
        4 & $5 \times 10^{-3}$ & $-$ & 4.45 & 200 & 0.0125 \\
        5 & $1 \times 10^{-2}$ & $1 \times 10^{-2}$ & 11.14 & 300 & 0.0200 \\
        6 & $2 \times 10^{-2}$ & $1 \times 10^{-2}$ & 26.73 & 600 & 0.0025 \\
        5 & $5 \times 10^{-2}$ & $1 \times 10^{-2}$ & 55.68 & 800 & 0.0050 \\
        \hline
    \end{tabular}
    \caption{Parameters specifying the tests used to measure the whistler dispersion relation: the wave number $k$, Hall resistivity $\eta_{\rm H}$, minimum Ohmic resistivity $\eta_{\rm O}$ required for stability, product $k \ell_{\rm H}$, integration time $t_{\rm max}$, and sample spacing $\Delta t$ required to measure the frequencies.}
    \label{tab:whistler_params}
\end{table}


\bsp	
\label{lastpage}
\end{document}